\documentclass{aa}  

\usepackage{filecontents}
\usepackage{xcolor}
\usepackage{graphicx}
\usepackage[varg]{txfonts}
\usepackage[colorlinks,allcolors=blue]{hyperref}
\usepackage[english]{babel}
\usepackage{float}
\usepackage{subcaption}
\bibpunct{(}{)}{;}{a}{}{,} 

\begin{document} 
\newcommand{\IoII}{SEDCSJ1411528$-$1154286}
\newcommand{\IoIV}{SEDCSJ1409277$-$1149267}
\newcommand{\IoV}{EDCSNJ1411028$-$1147006}
\newcommand{\IoVI}{SEDCSJ1410429$-$1144385}
\newcommand{\IoVII}{SEDCSJ1411478$-$1140389}
\newcommand{\IoVIII}{SEDCSJ1410518$-$1139195}
\newcommand{\IoIX}{SEDCSJ1410089$-$1138578}
\newcommand{\IoX}{SEDCSJ1410532$-$1137091}
\newcommand{\IoXI}{SEDCSJ1411416$-$1134421}
\newcommand{\IoXII}{SEDCSJ1411319$-$1133048}
\newcommand{\IoXIII}{SEDCSJ1411348$-$1132522}
\newcommand{\IoXV}{SEDCSJ1411480$-$1148562}
\newcommand{\IoXVI}{SEDCSJ1410349$-$1146140}
\newcommand{\IoXIX}{EDCSNJ1411036$-$1148506}
\newcommand{\IIoI}{SEDCSJ1411449$-$1158184}
\newcommand{\IIoII}{SEDCSJ1411112$-$1154452}
\newcommand{\IIoIV}{SEDCSJ1410214$-$1148167}
\newcommand{\IIoVb}{SEDCSJ1411096$-$1147245}
\newcommand{\IIoV}{SEDCSJ1411098$-$1147242}
\newcommand{\IIoVI}{SEDCSJ1410204$-$1142155}
\newcommand{\IIoVII}{SEDCSJ1411275$-$1139433}
\newcommand{\IIoVIII}{SEDCSJ1410249$-$1138157}
\newcommand{\IIoIX}{SEDCSJ1410568$-$1131594}
\newcommand{\IIoX}{SEDCSJ1411033$-$1146028}
\newcommand{\IIoXI}{SEDCSJ1411431$-$1143589}
\newcommand{\IIoXII}{SEDCSJ1411243$-$1140510}
\newcommand{\IIoXIII}{SEDCSJ1410463$-$1145508}
\newcommand{\IIoXIV}{SEDCSJ1411296$-$1137130}
\newcommand{\MS}{main-sequence}
\newcommand{\muhh}{$\mu_\mathrm{H_2}$}
\newcommand{\DV}{$\Delta V_{\rm baseline}$}

\title{SEEDisCS I. Molecular gas in galaxy clusters and their large-scale structure}
\subtitle{The case of CL1411.1$-$1148 at $z\sim0.5$}
\titlerunning{Cold gas content of galaxies in and around
  CL1411.1$-$1148}
\author{D. Sp\'{e}rone-Longin\inst{\ref{inst1},}\thanks{e-mail: damien.sperone-longin@epfl.ch} 
\and P. Jablonka\inst{\ref{inst1}, \ref{inst2}} \and
  F. Combes\inst{\ref{inst3}, \ref{inst4}} \and
  G. Castignani\inst{\ref{inst1}} \and 
  M. Krips\inst{\ref{inst5}} \and \\
  G. Rudnick\inst{\ref{inst6}} \and
  D. Zaritsky\inst{\ref{inst7}} \and 
  R. A. Finn\inst{\ref{inst8}} \and
  G. De Lucia\inst{\ref{inst9}} \and 
  V. Desai\inst{\ref{inst10}} }
\authorrunning{Sp\'{e}rone-Longin et al.}  
\institute{Laboratoire d'astrophysique, \'{E}cole Polytechnique F\'{e}d\'{e}rale de
  Lausanne (EPFL), Observatoire de Sauverny, 1290 Versoix,
  Switzerland \label{inst1} \and 
  GEPI, Observatoire de Paris, PSL University, CNRS, 5 Place Jules Janssen, 92190 Meudon,
  France \label{inst2} \and 
  Observatoire de Paris, LERMA, CNRS, Sorbonne University, PSL Research University, 75014 Paris,
  France \label{inst3} \and 
  Coll\`{e}ge de France, 11 Place Marcelin
  Berthelot, 75231 Paris, France \label{inst4} \and 
  IRAM, Domaine Universitaire, 300 rue de la Piscine, 38406
  Saint-Martin-d’H\`{e}res, France \label{inst5} \and 
  Department of Physics and Astronomy, The University of Kansas, Lawrence, KS,
  USA \label{inst6} \and 
  Steward Observatory and Department of
  Astronomy, University of Arizona, Tucson, AZ, USA \label{inst7} \and 
  Department of Physics and Astronomy, Siena College, Loudonville, NY,
  USA \label{inst8} \and INAF – Osservatorio Astronomico di Trieste,
  Via G. B. Tiepolo 11, 34143 Trieste, Italy \label{inst9} \and
  IPAC, Mail Code 100-22, Caltech, 1200 E. California Boulevard,
  Pasadena, CA, USA \label{inst10} }

\date{}

\abstract {
  We investigate how the galaxy reservoirs of molecular gas fuelling star
  formation are transformed while the host galaxies infall  onto
  galaxy cluster cores. As part of the Spatially Extended ESO Distant Cluster
  Survey (SEEDisCS), we present CO(3-2) observations of 27 star-forming
  galaxies obtained with the Atacama Large Millimeter Array (ALMA). These
  sources are located inside and around CL1411.1$-$1148 at $z=0.5195$,
  within five times the cluster virial radius. These targets were selected to have
  stellar masses (M$_{\rm star}$), colours, and magnitudes similar to those
  of a field comparison sample at similar redshift drawn from the Plateau de Bure
  high-$z$ Blue Sequence Survey (PHIBSS2). We compare the cold
  gas fraction ($\mu_{\rm H_2}=$ M$_{\rm H_2}$/M$_{\rm star}$), 
  specific star formation rates (SFR/M$_{\rm star}$) and depletion timescales 
  ($t_{\rm depl}=$ M$_{\rm H_2}$/SFR) of our main-sequence galaxies to the
  PHIBSS2 subsample. While the most of our galaxies (63\%) are
  consistent with PHIBSS2, the remainder fall below
  the relation between \muhh\ and M$_{\rm star}$ of the PHIBSS2 galaxies at
  $z\sim0.5$. These low-\muhh\ galaxies are not compatible with the tail of
  a Gaussian distribution, hence they correspond to a new population of
  galaxies with normal SFRs but low gas content and low depletion times ($\lesssim 1$ Gyr), 
  absent from previous surveys. We suggest that the star formation activity of these galaxies has
  not yet been diminished by their low fraction of cold molecular gas.
}

\maketitle


\section{Introduction\label{section1}}

Galaxy surveys have revealed a strong bimodality of the galaxy population in
colour, star formation rates (SFRs), and morphology \citep[e.g.,
SDSS,][]{Strateva2001}. Galaxies can indeed be broadly described as either
red predominantly early-type galaxies with little or no star formation
or blue predominantly late-type galaxies with active star formation
\cite[e.g.][]{Driver2006,Brammer2009, Muzzin2013}. A major thrust of the ongoing
research is to understand how the quenching of star formation starts
and works in galaxies, which leads ultimately to the build-up of the
passively evolving population. The fraction of star-forming galaxies is the
lowest inside galaxy clusters, while at the same time the fraction of early-type
morphologies (lenticulars, ellipticals) is the highest in
the field \citep{Dressler1980,Blanton2009}. There is no shortage of
proposed physical mechanisms to explain how galaxies stop forming stars at a
higher frequency in clusters relative to the field: tidal stripping
\citep{Gnedin2003}, ram-pressure stripping \citep{Gunn1972}, thermal evaporation
\citep{Cowie1977}, encounters with other satellites
\citep[`harassment',][]{Moore1996}, and removal of the diffuse gas reservoir of
galaxies \citep[`strangulation',][]{Larson1980, Zhang2019}. However, we are
still lacking the observational evidence that will distinguish between the
relative importance of the different mechanisms put forward and set their sphere
of influence.

Interestingly, the removal of HI gas and suppression of star formation seems to
occur at large distances from the cluster cores \citep[$\sim$2-4 virial
radii;][]{Solanes2002, Gomez2003, Haines2015}. The implication is that
galaxies are possibly pre-processed over cosmic time before they fall into the
cluster cores \citep[e.g.][]{Einasto2018,Olave-Rojas2018,Salerno2020};
our current understanding is that the largest gravitationally bound
overdensities in the initial $\Lambda$ cold dark matter ($\Lambda$CDM) 
density field collapse and gradually
merge to form increasingly more massive clusters connected by filaments
\citep{Springel2018}. This network of matter, called the cosmic web, is
observed up to a redshift of $z\sim1$ \citep{Pimbblet2004,
Kitaura2009, Guzzo2018} and is a potential site for pre-processing. One piece
of evidence for this is that massive red galaxies have preferentially been
found close to the filament axes \citep{Malavasi2016, Laigle2018, Kraljic2018,
Gouin2020}. Another possibility is that the cluster environment, and in
particular the hot intracluster medium, actually extends beyond the
cluster virial radius \citep{Zinger2018}.

Unfortunately, the cluster infall regions still remain poorly explored around galaxy 
clusters due to the dearth of deep imaging and accompanying spectroscopy 
in these extended regions.
The first and seminal wide-field investigations at intermediate redshift ($z
\sim 0.3 - 0.9$) focused on individual very massive systems \citep[$\sigma
\ge$ 900 km\,s$^{-1}$; e.g.][]{Kodama2001, Moran2007,Koyama2008, Patel2009,
Tanaka2009}, or even superclusters \citep{Lemaux2012}. They highlighted
the need for a variety of physical quenching processes acting well beyond
the cluster virial radii. Larger surveys followed such as the CLASH-VLT
survey \citep[][]{Biviano2013}, the ORELSE survey\citep{Lubin2009}, the PRIMUS
survey \citep{Berti2019}, and the IMACS cluster building survey \citep{Dressler2013},
leading to improved sampling of datasets and analyses.

To date most studies have focused only on the consequences of quenching (i.e.
the properties of the stellar populations). The gas that fuels star formation,
which is ultimately what must be affected to stop star formation, has
barely been explored in dense environments. We have undertaken a new approach,
which has the significant advantage of allowing us to link the galaxy stellar
mass build-up and the cosmic evolution of the galaxy molecular gas reservoir.
In other words, it allows us to establish how molecular gas is fuelling 
star formation, and how it is modified when star formation is on its way to
quenching \citep[e.g.][]{Castignani2020b}.

In order to shed light on the above issues, we are conducting a survey of
the large-scale structures (LSS) around two spectroscopically well-characterised,
intermediate-redshift, medium-mass clusters. They are selected from the ESO
Distant Cluster Survey \citep[EDisCS;][]{White2005}. This paper presents our
results for CL1411.1$-$1148 and the analysis of our ALMA dataset. It is
organised as follows. In Sect. \ref{sample} we present the sample selection 
and the observations with the Atacama Large Millimeter Array
(ALMA). In Sect. \ref{results}, we present our results and make a comparison
with the field population. We discuss our results in Sect. \ref{discussion},
and summarise our conclusions in Sect. \ref{conclusion}. In the following we
assume a flat $\Lambda$CDM cosmology with $\Omega_{\rm m} = 0.3$,
$\Omega_\Lambda = 0.7$ and $H_{0} = 70$\,km\,s$^{-1}$\,Mpc$^{-1}$
\citep[see][]{Riess2019, Planck2018}, and we use a Chabrier initial mass
function (IMF) \citep{Chabrier2003}. All magnitudes are in the AB system.



\section{Sample and observations\label{sample}}

EDisCS contains 18 systems at $0.4<z<0.8$ spanning the mass range from
groups to massive clusters (velocity dispersions between $\sim$200 and 1200
km\,s$^{-1}$), each with $\sim$20 to 70 spectroscopically confirmed members
\citep{Halliday2004, Milvang-Jensen2008}. Multi-band optical $B, V, I$, and $R$
photometry and spectroscopy were obtained with VLT/FORS2 and $J$, and $K_s$
bands gathered with the SOFI instrument on the NTT. \textit{Spitzer} MIPS 24-micron observations were
also obtained for a subset of clusters \citep{Finn2010}.

The Spatially Extended EDisCS survey (SEEDisCS) focuses on CL1301.7$-$1139 and
CL1411.1$-$1148 at redshifts $z_{\rm cl} = 0.4828$ and 0.5195 and
velocity dispersions $\sigma_{\rm cl} = 681$ and 710 km\,s$^{-1}$, respectively. Their
intermediate masses make them close analogues to the progenitors of typical
local clusters, whose velocity dispersions peak at around 500 km\,s$^{-1}$
\citep{Milvang-Jensen2008}. Deep $u$, $g$, $r$, $i$, $z$ and $K_{\rm s}$
images were taken with CFHT/MEGACAM and WIRCam. They cover a region that
extends up to $\sim 10 \times$ $R_{200}$, with $R_{200}$ corresponding to the
cluster virial radius. Our observational strategy follows three main steps:
\textit{\textbf{ i)}} identifying the LSS around the two clusters using accurate 
photometric redshifts (normalised median absolute deviation 
$\sigma_{\rm NMAD}=0.036$; Rerat et al. in prep.);
\textit{\textbf{ ii)}} spectroscopically following up these LSS to characterise
them precisely and to study the properties of the galaxy stellar populations;
\textit{\textbf{ iii)}} using ALMA CO observations to reveal the status of the
galaxy cold gas reservoirs.

\subsection{Sample selection}

Our ALMA targets were selected in the LSS around CL1411.1$-$1148 
and using three criteria. 
First, targets were chosen to fall within three times the cluster 
velocity dispersion (3$\times \sigma_{cl}$), corresponding to a 
redshift interval $\Delta z = \pm 0.010$ around the cluster
redshift. This is measured from the galaxy spectroscopic redshifts obtained
with VLT/FORS2, VLT/VIMOS, or MMT/Hectospec, or from a robust
redshift estimate from the IMACS Low Dispersion Prism (LDP). 
Second and with only one exception, the selected targets are located at 
a projected cluster centric distance smaller than 5$\times R_{200}$. 
Third, the targets span the same range of stellar
masses; $u$, $g$ and $i$ magnitudes; and colours from the combination of 
these bands as our initial comparison sample of normal star-forming 
field galaxies with CO information, the Plateau de Bure high-$z$ 
Blue Sequence Survey 2 \citep[PHIBSS2;][]{Freundlich2019}. 
This means that stellar masses were between log(M$_{\rm
star}$/M$_{\odot}$)=10 and 11, $i \le 22$, $g-i$ between $\sim$\,1 and 2.2,
and $u-g$ between $\sim$\,0.6 and 1.5. Two galaxies from the original
EDisCS spectroscopic sample were detected by \textit{Spitzer} at 24\,$\mu$m in the
central $\sim $1.8 $\times$ 1.8 Mpc region of CL1411.1$-$1148 ($R_{200}=1.27$ Mpc), 
above the 97 $\mu$Jy 80\% completeness flux limit of the EDisCS Spitzer observations
\citep{Finn2010}. They are identified by a bold black circle in all figures.

\begin{figure}[h]
	\resizebox{\hsize}{!}{\includegraphics{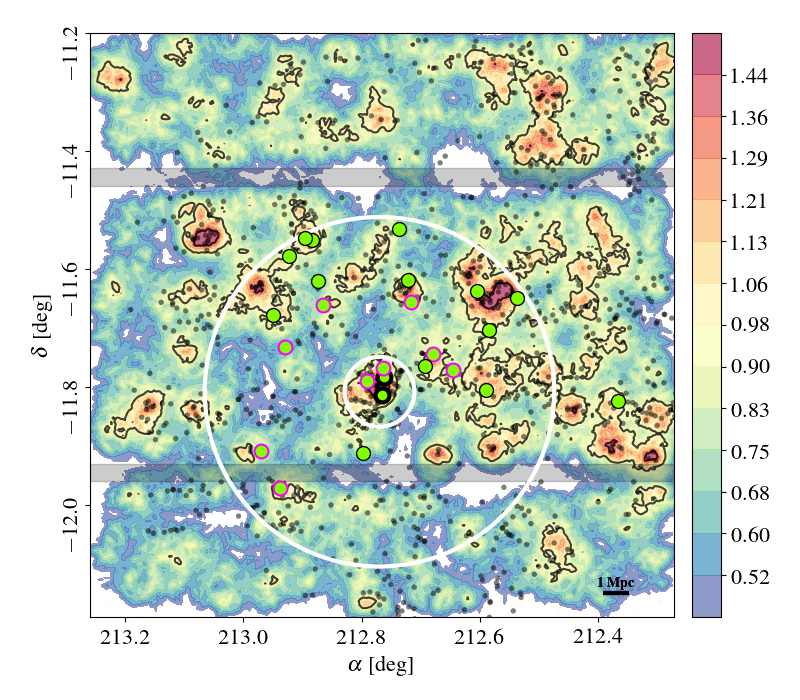}}
    \caption{Density map of the CFHT/MEGACAM 1$\degr\times$1$\degr$ field around
      CL1411.1$-$1148. The colour-coding indicates the $\log_{10}$ of the density of galaxies
      averaged over the ten nearest neighbours. Black contours are at 1 and
      $3\sigma$ above the field mean density. The grey points identify all
      galaxies with a $z_\mathrm{spec}$ within $\mathrm{5\sigma_{cl}}$ of the
      cluster redshift. The inner and outer white circles are positioned at $R_{200}$ and
      $5R_{200}$ radius, respectively. The grey bands indicate the gaps between the
      MEGACAM CCDs. The green circles show our ALMA targets. The thick black
      outline identifies the two cluster members detected at 24\,$\mu$m
      by \textit{Spitzer}. The pink outer rings show the position of the
      galaxies with low gas fraction (see Fig. \ref{Ms_fg} and
      Sect. \ref{gas_fractions}).}
    \label{targets}
\end{figure}

Figure \ref{targets} shows the galaxy density map in the
1$\degr\times$1$\degr$ region centred on CL1411.1$-$1148. Densities are
calculated within a photometric redshift slice of $ \pm
(1+z_{cl})\times \sigma_{\rm NMAD}$ = 0.0547 around the cluster redshift. Within
this photometric redshift slice, we use a `nearest neighbour' approach, in
which for any point $(x,y)$ the distance $r_N(x,y)$ to the $N{th}$ nearest
neighbour is estimated. The galaxy density is thus the ratio between 
$N$ (fixed) and the surface defined by the adaptive distance: $\rho_N(x,y)=\frac{N}{\pi
r^2_N(x,y)}$. We chose $N=10$, which corresponds to an average spatial
scale (i.e. the mean distance between the ten galaxies) of about 0.8 Mpc, with
90\% of the values being smaller than $\sim$1.5 Mpc. We selected 27
star-forming galaxies, satisfying the three criteria detailed
above, and  mapping the variety of local densities encountered
inside and around the cluster as they appear from the photometric redshift
estimates. 

\begin{figure}[h]
	\resizebox{\hsize}{!}{\includegraphics{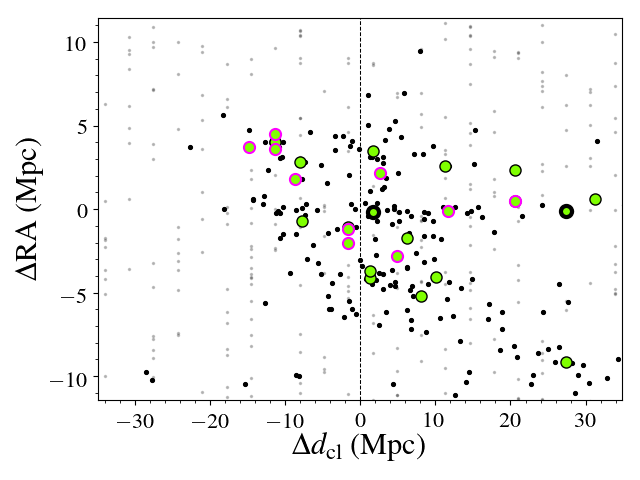}}
    \caption{Light cone centred on $z_{\rm cl}=0.5195$ and extending to 
    $\pm3\sigma_{\rm cl}$ in redshift. In right ascension, $1\degr$ is covered.
    The vertical line corresponds to the cluster redshift.
    The grey points indicate the galaxies with a photometric redshift. Galaxies 
    with spectroscopic redshifts are in black. Our sample is in green, 
    lower \muhh\ galaxies (see Sect. \ref{gas_fractions}) are outlined in pink, and 
    \textit{Spitzer} observed galaxies are outlined in thick black. Distances 
    are expressed relative to the brightest cluster galaxy (BCG).} 
    \label{light_cone}
\end{figure}

Figure \ref{light_cone} provides another 2D view of the spatial distribution of
our ALMA targets over the same 1$\degr\times$1$\degr$ MEGACAM field of view. The
galaxy positions are calculated relative to the position of the brightest
cluster galaxy (BCG) in redshift and right ascension (RA). The galaxy relative
position in redshift, $\Delta d_{\rm cl}$, is computed by taking the difference
between the comoving distances of the galaxy and the BCG. The relative
position in RA, $\Delta$RA, is obtained by transforming the angular separation
between the BCG and the galaxy into a distance, using the angular distance at
the redshift of the galaxy. Our full spectroscopic sample within $\pm$
$\mathrm{3\times\sigma_{cl}}$ of $z_{\rm cl}$ is presented, as are the
photometric redshift cluster member candidates. The finger-of-God
structure due to the relative velocities of the CL1411.1$-$1148 galaxies is
clearly seen along the $\Delta d_{\rm cl}$-axis. Many of our targets are located
in LSS related to CL1411.1$-$1148, such as the one extending
westward from the cluster centre and up to 30 Mpc behind it; a few are in more
isolated (lower density) regions. The information on our targets are summarised in
Table \ref{obs_table}.

PHIBSS2 encompasses 60 galaxies with CO(2-1) detections at $0.49\le z \le0.8$,
with stellar masses (M$_{\rm star}$) higher than $10^{10.1}$ M$_\sun$ and SFRs 
above 3.5 M$_\sun$yr$^{-1}$ selected from the COSMOS, AEGIS
and GOODS-North deep fields. A subsample of 19 systems falls at $0.49 \leq z
\leq 0.6$ and is used as comparison sample for our study.

\begin{figure}[h]
  \resizebox{\hsize}{!}{\includegraphics{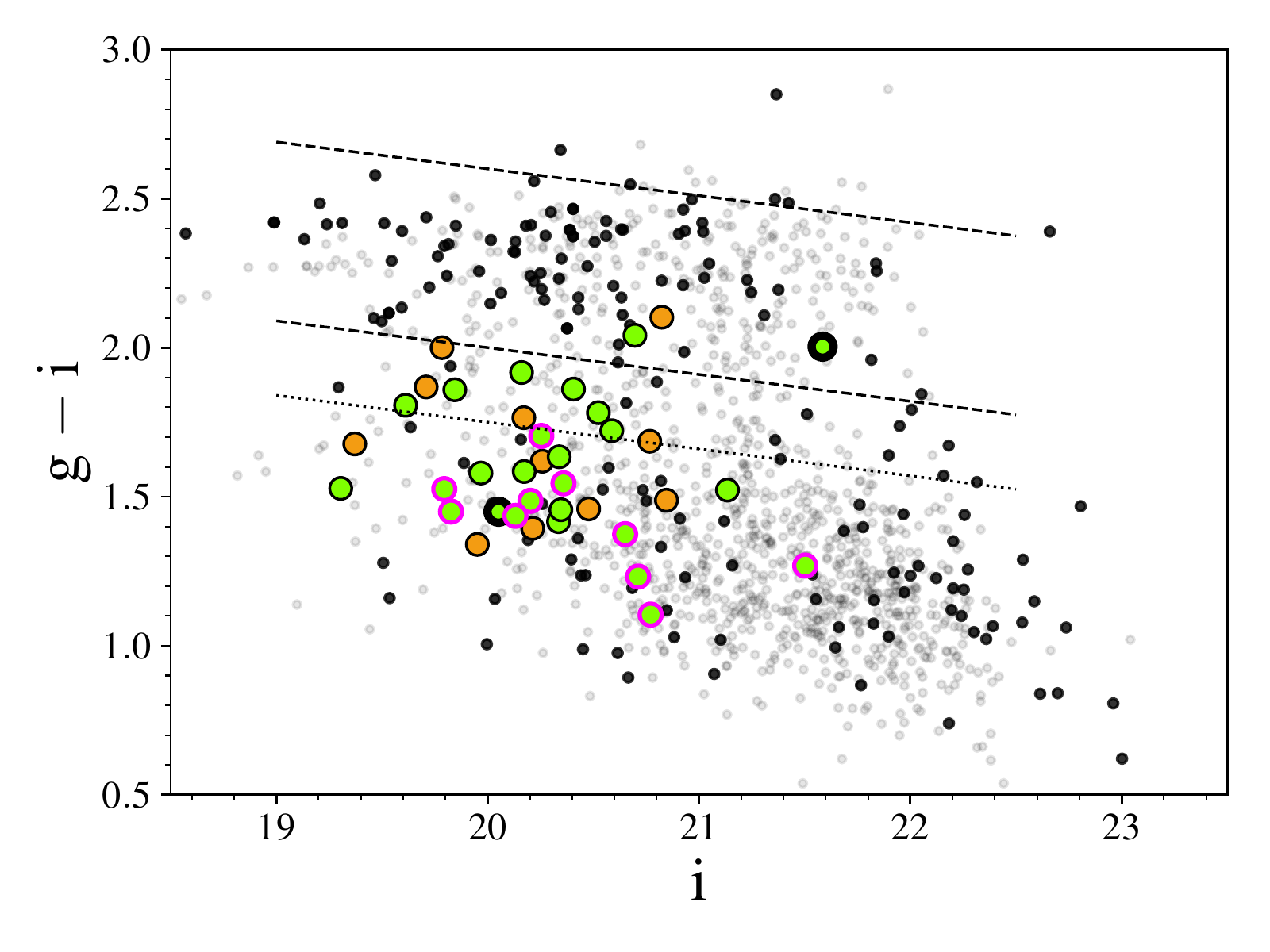}}
   \caption{Observed colour--magnitude diagram, $g-i$ as a function of
      $i$ for the CL1411 galaxies. Our ALMA sample is in green. 
      The dots with the thick black borders are our \textit{Spitzer}-observed galaxies,
      and those with the pink borders are our low-\muhh\ sample (see
      Sect. \ref{lowmuh2}). The orange dots are the PHIBSS2 galaxies. 
      The grey points are the photometric redshift members. 
      The small black dots are the spectroscopic redshift
      galaxies, within $3\sigma_{\rm cl}$ of the cluster
      redshift. The dashed lines delimit the red sequence and its $\pm$0.3
      mag dispersion. The dotted
      line delimits the transition zone between the blue clump and the
      red sequence, 0.25 mag below the lower boundary of the red
      sequence.}
    \label{CMD}
\end{figure}

Figure \ref{CMD} presents the distribution of the $0.49 \leq z \leq 0.6$ PHIBSS2 field
galaxies and our ALMA sample in the $g-i$ versus $i$ colour--magnitude diagram
(CMD). The position of the red sequence of CL1411.1$-$1148 is derived by
considering the initial galaxy sample of EDisCS in the centre of
CL1411.1$-$1148, for which we have $V$- and $I$- as well as $g$- and $i$-band
photometry. We first identify the passive galaxies in the ($V$, $I$) CMD from
\citet{DeLucia2007}. This provides us with their positions in the ($g$, $i$)
plane and allows us to fit the corresponding mean locus of the red sequence, and
place its $\pm0.3$ mag dispersion.

The $g$- and $i$-band photometry for the PHIBSS2 galaxies comes from the
original CFHT Legacy survey catalogue for the COSMOS and AEGIS fields
\citep{Erben2009}, while they were derived from $B$, $V$, and $I$ for the
galaxies in the GOODS-North field from the 3D-HST catalogues
\citep{Capak2004}. The latter may suffer from some uncertainties as no proper
photometric calibration between these bands and $g$ and $i$ exists for
galaxies.

\begin{figure}[htbp]
	\resizebox{\hsize}{!}{\includegraphics{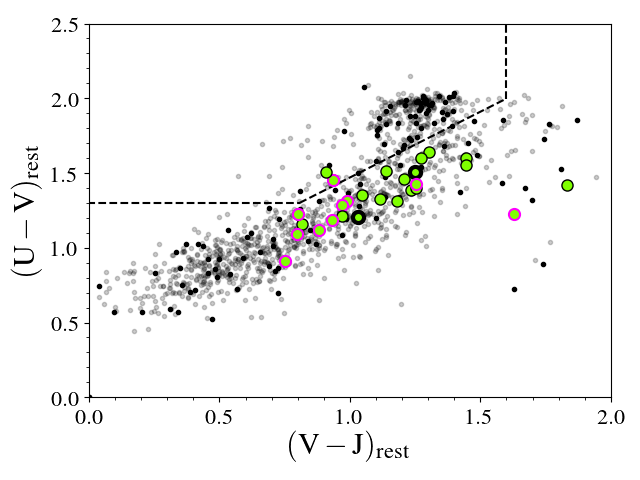}}
 	\caption{Rest-frame UVJ colour--colour diagram. The dashed lines separate passive galaxies, 
    in the upper part, from star-forming galaxies \citep{Williams2009}. Colours and symbols are the same as in Fig. \ref{CMD}}
    \label{rf_CCD}
\end{figure}

Figure \ref{rf_CCD} presents the rest-frame $U-V$ versus $V-J$ colour--colour diagram (CCD)
that helps discriminate between passive and star-forming galaxies
\citep{Williams2009}. The rest-frame colours were derived with EAZY
\citep{Brammer2008}. We used the Johnson-Cousins $U$ and $V$ bands, and the
2MASS $J$ band \citep{Skrutskie2006}, together with a set of six templates: five 
main component templates obtained following the \citet{Blanton2007a} algorithm
and one for dusty galaxies \citep{Brammer2008}. 

As expected, most of our targets fall in the star-forming region. Two systems,
\IIoVIII\ and \IoVIII, are formally located within the passive region, however
close to the boundary between the two regimes. None of them is located in the
red sequence in Fig. \ref{CMD}, but rather  in or close to the green
valley, hence they are most likely transitioning to a quenched
regime. On the other hand, the CO targets within the red sequences of
the ($g$, $i$) CMD in Fig. \ref{CMD} are not located in the passive region of
the $UVJ$ plane, meaning that they are dusty.

\begin{table*}[htbp]
\caption{Sample properties: Galaxy ID, coordinates, optical spectroscopic redshift, and SED-based
  estimates of the galaxy  $M_{\rm star}$ and SFRs}
\label{obs_table}
\centering
\begin{tabular}{l c c l r r}
\hline \hline \\
IDs & R.A. (J2000) & Dec (J2000) & ${z_{\rm spec}}$ 
& \multicolumn{1}{p{1.5cm}}{\centering M$_{\mathrm{star}}$ \ ($\mathrm{10^{10}\,M_{\odot}}$)} 
& \multicolumn{1}{p{1.5cm}}{\centering $\mathrm{SFR}_{SED}$ \ ($\mathrm{M_{\odot}\,yr^{-1}}$)} \\
\hline \\
\IoIV\ & 14:09:27.7553 & $-$11:49:26.734 & 0.5275 & $7.08^{+1.47}_{-1.47}$ &  $19.14^{+10.58}_{-11.90}$ \\
\IoIX\ & 14:10:08.9192 & $-$11:38:57.846 & 0.5217 & $2.95^{+0.61}_{-0.68}$ & $5.09^{+3.81}_{-3.52}$ \\
\IIoVI\ & 14:10:20.3750 & $-$11:42:15.477 & 0.5199 & $14.13^{+2.93}_{-3.58}$ & $10.52^{+2.18}_{-7.63}$ \\
\IIoIV\ & 14:10:21.4016 & $-$11:48:16.687 & 0.5226 & $12.02^{+2.49}_{-2.21}$ & $18.92^{+3.92}_{-9.59}$ \\
\IIoVIII\ & 14:10:24.9817 & $-$11:38:15.703 & 0.5199 & $5.62^{+1.68}_{-2.07}$ & $3.82^{+1.14}_{-2.99}$ \\
\IoXVI $^\dagger$ & 14:10:34.9274 & $-$11:46:14.071 & 0.5210 & $3.09^{+0.64}_{-0.64}$ & $5.71^{+2.70}_{-2.89}$ \\
\IoVI  $^\dagger$ & 14:10:42.8737 & $-$11:44:38.509 & 0.5308 & $5.89^{+1.08}_{-1.08}$ & $9.82^{+5.99}_{-4.18}$ \\
\IIoXIII\ & 14:10:46.3146 & $-$11:45:50.845 & 0.5214 & $13.8^{+3.18}_{-2.22}$ & $44.87^{+10.33}_{-40.30}$ \\
\IoVIII $^\dagger$ & 14:10:51.8133 & $-$11:39:19.548 & 0.5185 & $4.9^{+0.79}_{-0.90}$ & $3.33^{+1.65}_{-1.57}$ \\
\IoX\ & 14:10:53.2482 & $-$11:37:09.091 & 0.5193 & $4.68^{+0.97}_{-0.86}$ & $15.38^{+10.63}_{-11.16}$ \\
\IIoIX\ & 14:10:56.8242 & $-$11:31:59.398 & 0.5171 & $5.5^{+1.64}_{-1.52}$ & $6.27^{+1.88}_{-3.68}$ \\
\IoV $^\star$ & 14:11:02.8248 & $-$11:47:01.302 & 0.5202 & $4.79^{+1.10}_{-0.99}$ & $15.92^{+6.60}_{-7.88}$ \\
\IIoX $^\dagger$ & 14:11:03.2799 & $-$11:46:02.789 & 0.5231 & $1.82^{+0.50}_{-0.42}$ & $4.59^{+1.27}_{-2.27}$ \\
\IoXIX $^\star$ & 14:11:03.5909 & $-$11:48:50.573 & 0.5282 & $2.29^{+0.63}_{-0.58}$ & $5.58^{+3.41}_{-3.28}$ \\
\IIoVb $^\dagger$ & 14:11:09.6219 & $-$11:47:24.523 & 0.5002\tablefootmark{a} & $0.74^{+0.22}_{-0.21}$ & $2.86^{+0.84}_{-1.42}$ \\ 
\IIoII\ & 14:11:11.2342 & $-$11:54:45.236 & 0.5292 & $10.72^{+2.22}_{-2.22}$ & $16.29^{+3.38}_{-11.82}$ \\
\IIoXII $^\dagger$ & 14:11:24.3255 & $-$11:40:51.064 & 0.5171 & $1.15^{+0.32}_{-0.27}$ & $5.71^{+1.58}_{-1.84}$ \\
\IIoVII $^\dagger$ & 14:11:27.5630 & $-$11:39:43.290 & 0.5203 & $3.39^{+0.70}_{-0.70}$ & $10.89^{+2.26}_{-4.51}$ \\
\IIoXIV\ & 14:11:29.6609 & $-$11:37:13.061 & 0.5259 & $5.5^{+1.39}_{-1.64}$ & $4.65^{+1.18}_{-4.22}$ \\
\IoXII\ & 14:11:31.9412 & $-$11:33:04.781 & 0.5231 & $3.31^{+0.76}_{-0.61}$ & $7.62^{+4.30}_{-3.33}$ \\
\IoXIII\ & 14:11:34.7740 & $-$11:32:52.216 & 0.5172 & $4.9^{+1.01}_{-0.90}$ & $8.85^{+8.66}_{-7.54}$ \\
\IoXI\ & 14:11:41.6397 & $-$11:34:42.092 & 0.5198 & $5.13^{+0.94}_{-1.06}$ & $7.53^{+4.94}_{-3.56}$ \\
\IIoXI $^\dagger$ & 14:11:43.0675 & $-$11:43:58.969 & 0.5156 & $1.82^{+0.42}_{-0.34}$ & $6.19^{+1.43}_{-2.71}$ \\
\IIoI $^\dagger$ & 14:11:44.9883 & $-$11:58:18.447 & 0.5149 & $5.75^{+1.59}_{-1.72}$ & $12.79^{+3.54}_{-7.51}$ \\
\IoVII\ & 14:11:47.7871 & $-$11:40:38.956 & 0.5159 & $2.51^{+0.46}_{-0.35}$ & $10.64^{+10.78}_{-6.98}$ \\
\IoXV\ & 14:11:47.9664 & $-$11:48:56.199 & 0.5156 & $11.22^{+2.07}_{-2.33}$ & $9.38^{+8.1}_{-5.5}$ \\
\IoII $^\dagger$ & 14:11:52.8004 & $-$11:54:28.643 & 0.5160 & $3.63^{+0.75}_{-0.67}$ & $6.12^{+4.23}_{-2.96}$ \\
\hline
\end{tabular}
\tablefoot{
Central galaxies detected by \textit{Spitzer} are indicated with$^\star$. Galaxies with low \muhh\ are identified with $^\dagger$. \tablefoottext{a}{Galaxy with $z_{\rm LDP}$ as $z_{\rm spec}$.}
}
\end{table*}


\subsection{ALMA observations \label{alma_obs}}

Fluxes in the CO(3-2) line, falling at $\sim$226 GHz in the ALMA Band 6 for $z\sim$
0.52, were acquired during the ALMA Cycles 3 and 5 (programs 2015.1.01324.S,
2017.1.00257.S).  The observations were
conducted in the compact configurations C36$-$2 and C36$-$3, with 38 to 42
antennas, and C43$-$2, with 45 to 50 antennas, in Cycle 3 and 5,
respectively. This led to beam sizes of 0.94\arcsec$\times$ 0.89\arcsec\ and
1.2\arcsec$\times$ 0.95\arcsec\ for Cycle 3 and 5, respectively.
The integration times were of 7.5 hours (11 hours with
overheads) in the 225.51--228.86 GHz spectral window. The resulting
rms noise ranges from 0.09 to 0.25 mJy/beam in both cycles, and the spectral
resolution is 50.7 km\,s$^{-1}$ for Cycle 3 and between 10.3 and 41
km\,s$^{-1}$ for Cycle 5, depending on the binning applied to reach sufficient
signal-to-noise ratio.

A standard data reduction was performed with the CASA ALMA Science Pipeline
\citep{McMullin2007}.The problematic antennas and runs were flagged. The
continuum was fitted over the entire spectral window, except for
the channels corresponding to the CO line, and subtracted.  The final
datacubes were created with the \texttt{tclean} routine using a Briggs
weighting and a robustness parameter of 0.5, which is a trade-off between
uniform and natural weighting. Finally we performed a primary beam correction,
with the \texttt{impbcor} routine to obtain an astronomically correct image of
the sky.

The final continuum-subtracted and primary-beam-corrected maps were exported to be analysed using
GILDAS\footnote{\url{http://www.iram.fr/IRAMFR/GILDAS}}. The \textit{i}-band
images of our targets, the CO maps and spectra are shown in Fig. \ref{maps}.


\section{Derived parameters\label{results}}


\subsection{CO flux and molecular gas mass}

Fluxes, $S_\mathrm{CO}\,\Delta V$, 
were obtained by selecting the velocity window centred on the peak emission
and maximising the flux over it and the spatial extent of the source.

Following \cite{Lamperti2020}, the error on the flux is defined as
\begin{equation}
\epsilon_{\rm CO} = \frac{\sigma_{\rm CO}\Delta V}{\sqrt{\Delta V\Delta w_{ch}^{-1}}},
\end{equation}
where $\sigma_{\rm CO}$ is the rms noise (in Jy) calculated in units of spectral resolution $\Delta w_{ch}$, and
$\Delta V$ (in km\,s$^{-1}$) is the width of the spectral window in which the line flux is calculated,
$\Delta w_{ch} = 50.7$\,km\,s$^{-1}$ for Cycle 3 and $\Delta w_{ch} = \{ 10.3; 20.6; 41 \} $\,km\,s$^{-1}$, 
depending on the binning applied to the spectrum, for Cycle 5. All intensity maps and
integrated spectra are shown in Fig \ref{maps} of the Appendix. A few of our targets show double-peaked 
emission lines, which is an indication of rotation. This will be
analysed in a forthcoming paper.

The intrinsic CO luminosity associated with a transition between the levels $J$
and $J-1$ is expressed as
\begin{equation} \label{eq1}
    L'_\mathrm{CO(J\rightarrow J-1)}=3.25 \times10^{7} S_\mathrm{CO(J\rightarrow
      J-1)} \Delta V \, \nu_\mathrm{obs}^{-2} \, D_\mathrm{L}^{2} \, (1+z)^{-3},
\end{equation}
where $L'_\mathrm{CO(J\rightarrow J-1)}$ is the line luminosity expressed in
units of ${\rm K\,km\,s\textsuperscript{-1} pc^2}$; $S_\mathrm{CO(J\rightarrow
  J-1)} \Delta V$ is the velocity-integrated flux in ${\rm Jy\,km\,s^{-1}}$;
$\nu_\mathrm{obs}$ is the observed frequency in GHz; $D_\mathrm{L}$ is the
luminosity distance in Mpc; and $z$ is the redshift of the observed galaxy
\citep{Solomon1997, Solomon2005}.

The total cold molecular gas mass (M$_\mathrm{H_2}$) is then estimated as
\begin{equation} \label{eq2}
    M_\mathrm{H_2}=\alpha_\mathrm{CO} \frac{L'_\mathrm{CO(J\rightarrow J-1)}}{r_{J1}}, 
\end{equation}
where $\alpha_\mathrm{CO}$ is the CO(1-0) luminosity-to-molecular-gas-mass
conversion factor, considering a 36\% correction to account for interstellar
helium, and $r_{J1}=L^{\prime}_\mathrm{CO(J\rightarrow J-1)} / L^{\prime}_\mathrm{CO(1-0)}$ the
corresponding line luminosity ratio.

The $\alpha_\mathrm{CO}$ factor depends on different parameters: the average
cloud density, the Rayleigh-Jeans brightness temperature of the CO transition,
and the metallicity of the giant molecular clouds (GMCs) of the galaxy
\citep{Leroy2011, Genzel2012, Bolatto2013, Sandstrom2013}. In the Milky Way,
in nearby main-sequence (MS) star-forming galaxies, and in low-metallicity galaxies
different methods are used to estimate this conversion factor. They converge to 
$\alpha_\mathrm{CO}=4.36 \pm 0.9\,M_\odot\,{\rm
 (K\,km\,s^{-1}\,pc^2)^{-1}}$, including the correction for helium, as a good
estimate for normal star-forming galaxies \citep{Dame2001, Grenier2005,
 Abdo2010, Leroy2011, Bolatto2013, Carleton2017}.

The values of $r_{31}$ have been measured in a number of ways in nearby galaxies, and
range from $\sim$ 0.2 to 2 (rarely reached however) \citep{Mauersberger1999,Mao2010,Wilson2012}. 
\citet{Dumke2001} found that $r_{31}$ could vary within a galaxy from 
$r_{31}\sim0.8$ in the bulge to $r_{31}\sim0.4$ in the disk for local galaxies without 
enhanced star formation.
More recently \citet{Lamperti2020} identified a trend of $r_{31}$ with star
formation efficiency, from $\sim$ 0.2 to 1.2 (with a mean value around 0.5), and
inferred from modelling that the gas density is the main parameter responsible
for this variation. At intermediate ($z\sim0.5$) and high redshifts ($z\sim1.5$),
several studies assumed $r_{31}=0.5\pm0.05$, as we do here as a fair
compromise between all studies \citep{Bauermeister2013, Genzel2015, Chapman2015, 
Carleton2017,Tacconi2018}. We discuss the impact of the choice of
$\alpha_\mathrm{CO}$ and $r_{31}$ on the cold molecular gas masses of our
galaxies in Sect. \ref{caveats}.

The full widths at half maximum (FWHMs) are derived from single or double
Gaussian fits of the CO emission lines. We obtain a median FWHM of
224\,km\,s$^{-1}$ with a standard deviation of 101\,km\,s$^{-1}$ for our entire
ALMA sample, similarly to what is found for our range of stellar masses by \citet{Freundlich2019}.

The intrinsic CO(3-2) luminosity $L^{\prime}_\mathrm{CO(3-2)}$, the line FWHM, the cold
molecular gas mass M$_\mathrm{H_2}$, the corresponding gas-to-stellar-mass ratio
$\mu_{\rm H_2}={\rm M}_\mathrm{H_2}/$M$_\mathrm{star}$, and the redshift of the CO
emission of our sample galaxies are listed in Table \ref{properties_table}. One
galaxy, \IIoVb, exhibits a large difference between its optical and CO
redshifts. This is due to its optical redshift being estimated from
IMACS-LDP, with a precision $\sigma_z = 0.007$ \citep{Just2015}. \IIoVb\ was not
in our initial list of targets for ALMA. It turned out that while our primary
target was not detected (\IIoV\ at $z_{\rm spec} = 0.5260$), 
its companion galaxy within 3\arcsec, \IIoVb\ was. This
is shown in Fig. \ref{maps} in the $i$-band image; the original target
is shown on the left.

\begin{table*}[htbp]
\caption{CO redshift, line-integrated flux, line width, luminosity of the CO(3-2) emission, 
cold molecular gas masses and cold molecular gas-to-stellar mass ratios of the ALMA targets.}
\label{properties_table}
\centering
\begin{tabular}{l l c r r r c}
\hline \hline \\
IDs & $z_{\mathrm{CO}}$ 
& \multicolumn{1}{p{1.5cm}}{\centering $S_{\mathrm{CO(3-2)}}\Delta V$ \\ ($\mathrm{Jy\,km\,s^{-1}}$)} 
& \multicolumn{1}{p{1.5cm}}{\centering FWHM \\ ($\mathrm{km\,s^{-1}}$)} 
& \multicolumn{1}{p{1.5cm}}{\centering $L^{'}_{\mathrm{CO(3-2)}}$ \\ ($\mathrm{10^{8}\,L_{\odot}}$)} 
& \multicolumn{1}{p{1.5cm}}{\centering M$_{\mathrm{H_{2}}}$ \\ ($\mathrm{10^{9}\,M_{\odot}}$)} 
& $\mu_{\mathrm{H_{2}}}$ \\
\hline \\
\IoIV\ & 0.5287 & 0.943 $\pm$ 0.052 & 568 $\pm$ 45 & 15.36 $\pm$ 0.853 & 13.39 $\pm$ 3.16 & $0.189^{+0.084}_{-0.084}$ \\
\IoIX\ & 0.5227 & 0.494 $\pm$ 0.029 & 236 $\pm$ 19 & 7.859 $\pm$ 0.457 & 6.85 $\pm$ 1.62 & $0.232^{+0.103}_{-0.108}$ \\
\IIoVI\ & 0.5200 & 0.655 $\pm$ 0.021 & 110 $\pm$ 45 & 10.326 $\pm$ 0.329 & 9.0 $\pm$ 2.09 & $0.064^{+0.028}_{-0.031}$ \\
\IIoIV\ & 0.5226 & 0.500 $\pm$ 0.008 & 93 $\pm$ 2 & 7.966 $\pm$ 0.132 & 6.95 $\pm$ 1.6 & $0.058^{+0.025}_{-0.024}$ \\
\IIoVIII\ & 0.5199 & 0.327 $\pm$ 0.015 & 262 $\pm$ 10 & 5.154 $\pm$ 0.243 & 4.49 $\pm$ 1.05 & $0.08^{+0.043}_{-0.048}$ \\
\IoXVI $^\dagger$ & 0.5213 & 0.164 $\pm$ 0.014 & 224 $\pm$ 21 & 2.597 $\pm$ 0.225 & 2.26 $\pm$ 0.56 & $0.073^{+0.033}_{-0.033}$ \\
\IoVI $^\dagger$ & 0.5220 & 0.214 $\pm$ 0.018 & 118 $\pm$ 20 & 3.374 $\pm$ 0.288 & 2.94 $\pm$ 0.72 & $0.050^{+0.021}_{-0.021}$ \\
\IIoXIII\ & 0.5215 & 2.150 $\pm$ 0.010 & 206 $\pm$ 2 & 34.094 $\pm$ 0.199 & 29.73 $\pm$ 6.82 & $0.215^{+0.099}_{-0.084}$ \\
\IoVIII $^\dagger$ & 0.5193 & 0.175 $\pm$ 0.025 & 166 $\pm$ 31 & 2.750 $\pm$ 0.400 & 2.4 $\pm$ 0.65 & $0.049^{+0.021}_{-0.022}$ \\
\IoX\ & 0.5213 & 1.398 $\pm$ 0.038 & 404 $\pm$ 12 & 22.03 $\pm$ 0.596 & 19.21 $\pm$ 4.44 & $0.411^{+0.18}_{-0.17}$ \\
\IIoIX\ & 0.5172 & 0.483 $\pm$ 0.014 & 273 $\pm$ 9 & 7.53 $\pm$ 0.225 & 6.57 $\pm$ 1.52 & $0.119^{+0.063}_{-0.061}$ \\
\IoV\ & 0.5207 & 0.884 $\pm$ 0.020 & 148 $\pm$ 6 & 13.95 $\pm$ 0.325 & 12.17 $\pm$ 2.80 & $0.254^{+0.117}_{-0.111}$ \\
\IIoX $^\dagger$ & 0.5231 & 0.108 $\pm$ 0.010 & 243 $\pm$ 22 & 1.724 $\pm$ 0.155 & 1.5 $\pm$ 0.37 & $0.083^{+0.043}_{-0.039}$ \\
\IoXIX\ & 0.5287 & 0.436 $\pm$ 0.049 & 148 $\pm$ 13 & 7.102 $\pm$ 0.799 & 6.19 $\pm$ 1.58 & $0.270^{+0.144}_{-0.138}$ \\
\IIoVb $^\dagger$ & 0.5259\tablefootmark{a} & 0.144 $\pm$ 0.017 & 272 $\pm$ 72 & 2.306 $\pm$ 0.278 & 2.01 $\pm$ 0.52 & $0.270^{+0.150}_{-0.145}$ \\
\IIoII\ & 0.5292 & 0.602 $\pm$ 0.014 & 183 $\pm$ 11 & 9.843 $\pm$ 0.237 & 8.58 $\pm$ 1.98 & $0.080^{+0.035}_{-0.035}$ \\
\IIoXII $^\dagger$ & 0.5168 & 0.106 $\pm$ 0.006 & 183 $\pm$ 14 & 1.65 $\pm$ 0.098 & 1.44 $\pm$ 0.34 & $0.125^{+0.064}_{-0.059}$ \\
\IIoVII $^\dagger$ & 0.5203 & 0.154 $\pm$ 0.006 & 127 $\pm$ 5 & 2.431 $\pm$ 0.101 & 2.12 $\pm$ 0.49 & $0.063^{+0.028}_{-0.028}$ \\
\IIoXIV\ & 0.5259 & 0.411 $\pm$ 0.016 & 243 $\pm$ 27 & 6.634 $\pm$ 0.255 & 5.78 $\pm$ 1.35 & $0.105^{+0.051}_{-0.056}$ \\
\IoXII\ & 0.5233 & 0.736 $\pm$ 0.024 & 266 $\pm$ 13 & 11.75 $\pm$ 0.390 & 10.25 $\pm$ 2.37 & $0.309^{+0.143}_{-0.129}$ \\
\IoXIII\ & 0.5173 & 0.645 $\pm$ 0.034 & 306 $\pm$ 27 & 10.05 $\pm$ 0.525 & 8.77 $\pm$ 2.06 & $0.179^{+0.079}_{-0.075}$ \\
\IoXI\ & 0.5200 & 0.759 $\pm$ 0.027 & 163 $\pm$ 6 & 11.97 $\pm$ 0.424 & 10.44 $\pm$ 2.42 & $0.203^{+0.085}_{-0.089}$ \\
\IIoXI $^\dagger$ & 0.5166 & 0.198 $\pm$ 0.008 & 124 $\pm$ 8 & 3.075 $\pm$ 0.124 & 2.68 $\pm$ 0.62 & $0.147^{+0.068}_{-0.061}$\\
\IIoI $^\dagger$ & 0.5149 & 0.259 $\pm$ 0.012 & 361 $\pm$ 17 & 4.002 $\pm$ 0.189 & 3.49 $\pm$ 0.82 & $0.061^{+0.031}_{-0.032}$ \\
\IoVII\ & 0.5160 & 0.450 $\pm$ 0.020 & 259 $\pm$ 29 & 6.984 $\pm$ 0.312 & 6.09 $\pm$ 1.42 & $0.242^{+0.101}_{-0.090}$ \\
\IoXV\ & 0.5160 & 0.388 $\pm$ 0.024 & 265 $\pm$ 34 & 6.021 $\pm$ 0.378 & 5.25 $\pm$ 1.25 & $0.047^{+0.020}_{-0.021}$ \\
\IoII $^\dagger$ & 0.5166 & 0.102 $\pm$ 0.023 & 166 $\pm$ 58 & 1.584 $\pm$ 0.362 & 1.38 $\pm$ 0.45 & $0.038^{+0.020}_{-0.019}$ \\
\hline
\end{tabular}
\tablefoot{
Galaxies with low \muhh\ are identified with $^\dagger$. \tablefoottext{a}{Galaxy with $z_{\rm LDP}$ as $z_{\rm spec}$.}
}
\end{table*}


\subsection{Stellar masses and star formation rates \label{MS_SFR}}


The stellar masses and SFRs were derived with
MAGPHYS\footnote{\url{http://www.iap.fr/magphys/index.html}}
\citep{daCunha2008} using the $u$, $g$, $r$, $i$, $z$, and \textit{Ks} bands, as
well as the 24 $\mu$m flux when available. The stellar populations and
dust extinction models are those of \citet{Bruzual2003} and
\citet{Charlot2000}. MAGPHYS provides probability density functions
(PDFs) for each parameter (i.e. SFR, M$_{\rm star}$, dust mass, dust
temperature). For each quantity, we considered the peak value of
the PDFs and our uncertainties correspond to the 68\% confidence interval of
the same PDFs. The wavelength coverage of our photometric bands does not allow
the identification of AGNs, which could affect the SFR and M$_{\rm
star}$ estimates. However, the analysis of the [OII]-to-H$\beta$ line ratio
indicates that most likely the emission lines in our galaxy spectra are not
typical of AGNs \citep{SanchezBlazquez2009}. This adds to our current
understanding that there is a smaller percentage of AGNs in clusters (less than 3\%) than in the
field \citep{Miller2003, Kauffmann2004, Mishra2020}.

\begin{figure}[htbp]
	\begin{subfigure}[t]{.5\textwidth}
		\resizebox{\hsize}{!}{\includegraphics{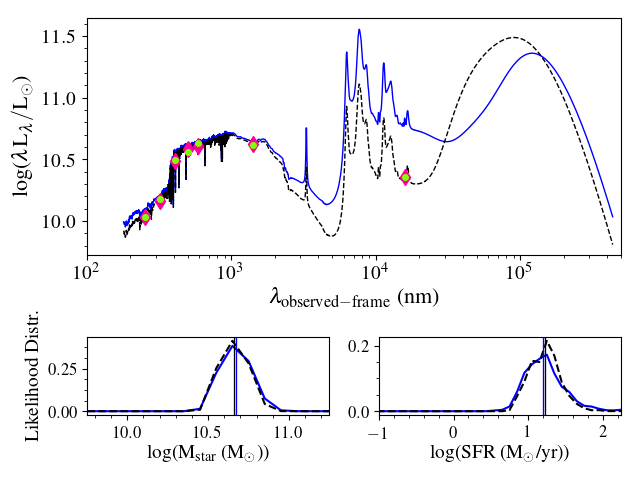}}
		\caption{\IoV .}
	\end{subfigure}
	\begin{subfigure}[t]{.5\textwidth}
		\resizebox{\hsize}{!}{\includegraphics{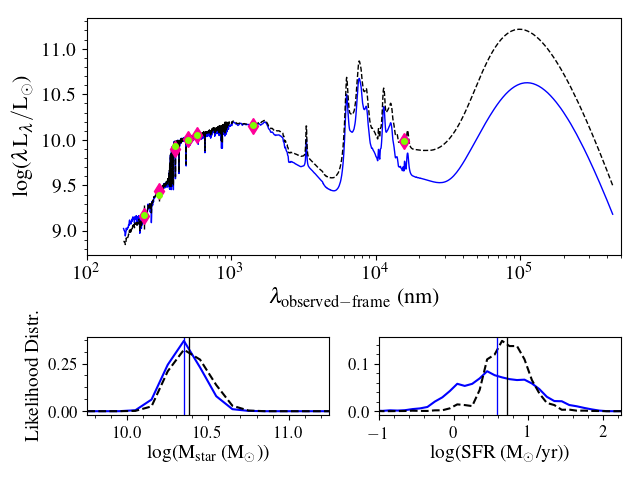}}
		\caption{\IoXIX .}
	\end{subfigure}
    \caption{SEDs and likelihood distributions for the two 
    \textit{Spitzer}-observed galaxies: \IoV\ (a) and \IoXIX\ (b). 
    The blue curves are without the MIPS 24\,$\mu$m fluxes and the 
    black dashed curves are with the MIPS 24\,$\mu$m fluxes. The top panels 
    show the fitted SEDs in black and blue, the observed fluxes in green, 
    and the model fluxes in pink. The bottom panels of (a) and (b) show 
    the likelihood distributions for M$_{\rm star}$ and SFR, 
    and the medians for the M$_{\rm star}$ and SFR, 
    as the blue and black vertical lines, derived with MAGPHYS.
    \label{Spitzer}}
\end{figure}

We used the two galaxies in the core of CL1411.1$-$1148 which were
detected by \textit{Spitzer} at 24 $\mu$m to evaluate the robustness
of our stellar mass and SFR estimates. The 24 $\mu$m flux
(corresponding to $\sim$15$\mu$m at $z\sim0.52$) 
allows us to better constrain the dust
emission and judge its impact on the derived quantities. Figure
\ref{Spitzer} presents the spectral energy distribution (SED) fits and the 
corresponding likelihood distribution of M$_{\rm star}$ and SFR for \IoV\ and \IoXIX. 
For both galaxies, M$_{\rm star}$ stays identical with or without the
24\,$\mu$m flux. As to the SFR, the PDFs of \IoV\ are essentially
identical with or without the 24\,$\mu$m flux point. The case of
\IoXIX\ is different. Its SFR PDF is wider without the 24\,$\mu$m flux
point (peak value at $\log({\rm SFR})=0.5$) than when calculated with
the 24\,$\mu$m point (peak value at $\log({\rm SFR})=0.8$).
The two PDFs have medians within 0.3 dex and 
are highly overlapping, hence the
SFR estimates are consistent with each other. \IoXIX\ is the faintest
galaxy in our sample in the $i$ band ($>$ 21.5, Fig. \ref{CMD}) and is
probably observed edge-on, as seen in Fig. \ref{maps}. 
Its UV flux is very low, with the deepest
rest-frame 4000\,\AA\ break, hence representing the most
challenging and dusty case in our sample.

In summary, our stellar mass estimates are robustly derived from $u$ to
$K_{\rm s}$ photometry. As for the SFRs, missing the \textit{Spitzer} 24\,$\mu$m
flux could lead to underestimated values, but our error bars are realistic
enough to take this possibility into account.

\begin{figure}[htbp]
\centering
\resizebox{\hsize}{!}{\includegraphics{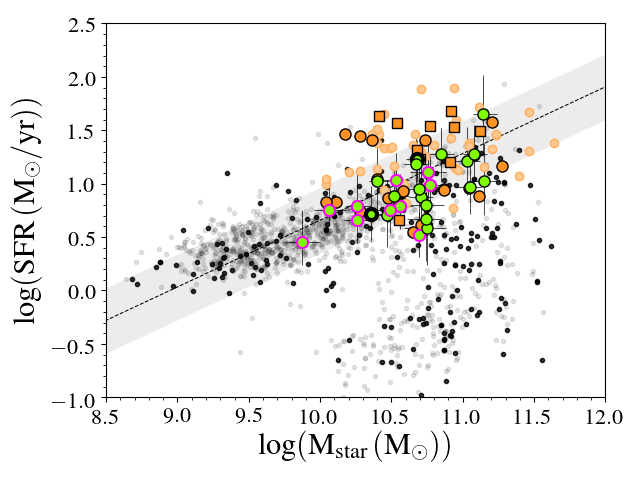}}
\caption{Location of the CL1411 (grey) and ALMA (green) galaxies in the
  stellar mass--SFR plane. The dots with thick black borders represent the
  \textit{Spitzer}-observed galaxies and those with the pink outline
  indicate  the low-\muhh\ galaxies. Galaxies in black and grey are the
  rest of the spectroscopic and photometric samples, respectively, at the
  cluster redshift. The plain orange circles are the PHIBSS2 galaxies
  with $0.49 \leq z \leq 0.6$, while the light orange circles trace the PHIBSS2
  sample at $0.6 < z \leq 0.8$; the orange squares show the cluster LIRGs
  of \citet{Castignani2020b} at $z\sim0.54$. The dashed black line is the
  \citet{Speagle2014} MS corrected for a Chabrier IMF, 
  at our cluster redshift with the
  corresponding $\pm$0.3dex scatter as the grey shaded area.
    \label{MScl1411}}
\end{figure}

Figure \ref{MScl1411} presents the position of our sample galaxies relative to
the MS of normal star-forming galaxies at the same redshift
\citep{Speagle2014}, corrected for a Chabrier IMF.
Our spectroscopic and photometric datasets are both
displayed. More than three-quarters (78\%) of our ALMA targets fall within the
$\pm$ 0.3\,dex dispersion of the MS. Three of our ALMA targets are located just
below the $-0.3$\,dex limit; however, they are still compatible with the MS
considering the uncertainties on the SFRs. Three systems fall in between the MS
and the red sequence. These are systems in the
transition region between star-forming and passive systems. \citet{Mancini2019}
show that this region of the stellar mass--SFR plane contains galaxies that are
quenching, but also galaxies that are undergoing a rejuvenation of star
formation. With the exception of three PHIBSS2 galaxies, which stand clearly
above the MS, PHIBSS2 $z\sim0.55$ systems and our ALMA targets cover the same
SFR--M$_{\rm star}$ space. Their cold gas reservoirs can therefore be compared.


\section{Discussion\label{discussion}}


\subsection{Comparison sample \label{comparison}}

To place our results in a global context, our datasets are compared to the
other CO-line observations available to date. They are listed below in order of
increasing redshift.

 \begin{itemize} 
\item[1.] 53 detections of CO(1-0) for local ($0.001<z<0.05$) IR luminous galaxies \citep{Gao2004};
\item[2.] 19 detections of CO(1-0) for LIRGs with $z\sim0.01$ \citep{Garcia-Burillo2012};
\item[3.] 333 detections of CO(1-0) for galaxies from xCOLD GASS \citep{Saintonge2017} with
 M$_{\rm star}>10^{10}$M$_\odot$ and $z$ between 0.01 and 0.05;
\item[4.] 46 detections of CO(3-2) for galaxies selected from the BASS survey 
\citep{Baumgartner2013} with $z<0.04$ and studied in \citet{Lamperti2020};
\item[5.] 27 detections of CO(1-0) and CO(3-2) for star-forming galaxies with $z$ from 0.06 to 0.3 from the 
EGNoG survey \citep{Bauermeister2013a,Bauermeister2013};
\item[6.] 8 detections of CO(1-0) emission for galaxies selected based on their 4000\,\AA\ 
emission strength with redshifts from 0.1 to 0.23 \citep{Morokuma-Matsui2015};
\item[7.] 9 detections of CO(1-0) for star-forming galaxies inside and in the foreground and 
background of two Abell clusters, A2192 and A963, with $z$ between 0.13 and 0.23, from the 
COOL BUDHIES survey \citep{Cybulski2016};
\item[8.] 8 CO(1-0) and 12 CO(2-1) observations of LIRGs inside clusters with 
redshift between 0.21 and 0.56 \citep{Castignani2020b};
\item[9.] 2 CO(2-1) and 1 CO(1-0) detections of LIRGS inside two clusters at $z=0.397$ and 
$0.489$ \citep{Jablonka2013};
\item[10.] 5 CO(1-0) detections from 24$\mu$m-selected galaxies at $z=0.4$ 
\citep{Geach2009a,Geach2011};
\item[11.] 46 detections of CO(2-1) for star-forming galaxies with $0.5 \leq z \leq 0.8$, as part of the 
PHIBSS2 survey \citep{Tacconi2018,Freundlich2019};
\item[12.] 4 CO(2-1) detections for massive and passive galaxies from the LEGA-C 
survey with $0.6\leq z \leq 0.73$ \citep{Spilker2018};
\item[13.] 52 detections of CO(3-2) for star-forming galaxies with redshifts 
ranging from 1 to 2.3, as part of the PHIBSS1 survey \citep{Tacconi2010,Tacconi2013};
\item[14.] 17 CO(2-1) detections of main-sequence galaxies inside the XMMXCS 
J2215.9$-$1738 cluster at $z=1.46$ \citep{Hayashi2018};
\item[15.] 5 detections of CO(2-1) for near-IR selected galaxies at $z\sim1.5$ from \citet{Daddi2010};
\item[16.] 11 detections of CO(2-1) emission for cluster galaxies at $z\sim1.6$ from \citet{Noble2017, Noble2019};
\item[17.] 2 detections of CO(1-0) emission of massive cluster galaxies at $z\sim1.62$ from \citet{Rudnick2017}
 \end{itemize} 

\begin{figure}[htbp]
  \resizebox{\hsize}{!}{\includegraphics{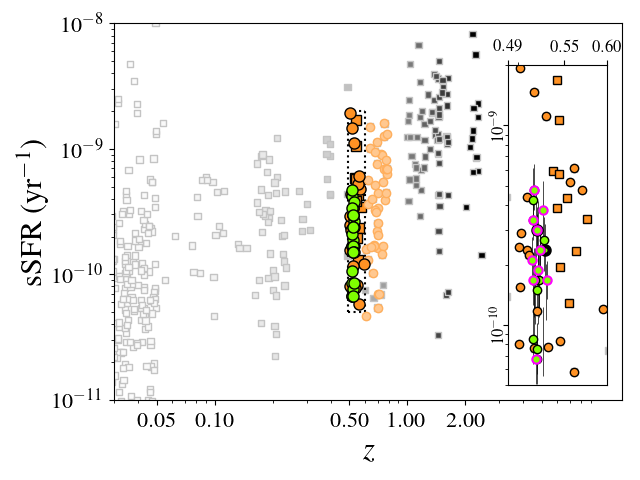}}
    \caption{Specific star formation rates as a function of redshift.
      The green dots identify our ALMA targets; dots with the thick
      black borders are the galaxies detected at 24\,$\mu$m,
      and dots with the pink outlines galaxies with low gas fraction. The orange
      circles stand for the PHIBSS2 galaxies, with the darker shade for
      the systems at $0.49 \leq z \leq 0.6$ and the lighter for the
      galaxies at $0.6 < z \leq 0.8$. The orange squares indicate
      the M0717 LIRGs. 
      The symbols in shades of grey are for the samples we pulled from the
	  literature at different redshifts.
      We provide a zoom-in of the region delineated
      by the dotted lines (see inset), around the redshift of
      CL1411.1$-$1148. While three of the PHIBSS2 sources have sSFRs 
      well above those of our ALMA targets, the sSFRs of the rest of the PHIBSS2 
      sources are in perfect agreement with those from our sample.
      \label{z_sSFR}}
\end{figure}

It is noteworthy that most of these datasets are made of field galaxies, with
the exception of \citet{Geach2009a,Geach2011}, \citet{Jablonka2013},
\citet{Cybulski2016}, \citet{Noble2017,Noble2019} and \citet{Castignani2020b}.
Depending on the focus of the discussions below, we include all or only parts of
this comparison sample. The PHIBSS2 galaxies at $0.49\leq z \leq 0.60$ are the best
field counterparts to our study in terms of redshift range, and even more
importantly because most of the galaxies are forming stars at a normal rate for
their stellar masses. The ten galaxies of the MACS J0717.5+3745 cluster
observed by \citet{Castignani2020b} are, in redshift, the closest cluster galaxy
counterparts to our study. However, they were selected differently,
specifically as LIRGs, and consequently probe on average higher specific SFRs (sSFR = SFR/M$_{\rm star}$)
than our sample, and do not extend down to the  lowest values as our sample does,
as can be seen in Fig. \ref{z_sSFR}. The three field PHIBSS2 galaxies
with the highest gas fraction are systems very clearly above the MS,
hence they do not have counterparts in our dataset. They are
nonetheless included in our analysis.

\subsection{Gas fractions \label{gas_fractions}}

\begin{figure}[htbp]
	\resizebox{\hsize}{!}{\includegraphics{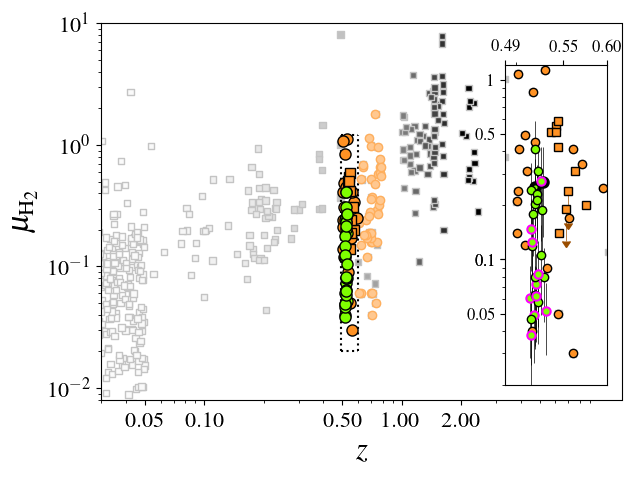}}
	\caption{Molecular gas-to-stellar mass ratio as a function of redshift.
          The colour-coding is as in Fig. \ref{z_sSFR}. We provide a
          zoom-in of the region delineated by the dotted lines (see inset), 
          around the redshift of CL1411.1$-$1148.
    \label{z_fg}}
\end{figure}

Figure \ref{z_fg} shows the variation of the galaxy gas fraction, \muhh $=
{\rm M}_{\rm H_2}/$M$_{\rm star}$, with redshift for both our targets and other
 CO-line measurements published to date. Our sample is the largest sample of galaxies
with direct cold gas measurements at a single intermediate redshift ($0.5000 < z
< 0.5375$) and the only one with galaxies in interconnected cosmic structures
around a given galaxy cluster. The cluster galaxy sample of
\citet{Castignani2020b} at $z\sim0.54$ extends to $\sim$ 1.6 times the virial
radius of M0717, hence stays closer to the cluster centre than the
present study.

We probe a wide range of \muhh\ values, from $\sim$0.04 to 0.30; 44\%
of our galaxies have \muhh\ lower than 0.1. 
This contrasts with the bulk of other datasets at $z>0.05$. Of these, only the PHIBSS2 
sample at $0.49 \leq z \leq 0.8$ has gas fractions that are as low as ours, and even then only $\sim$ 20\% 
of the coeval PHIBSS2 galaxies have $\mu_{H_2}$ below 0.1. The low gas fractions we see 
in our sample compared to those in the field cannot simply be due to cosmic evolution in 
\muhh, as samples at both lower and higher redshift have increased gas fractions. 
It is more likely linked to how we selected our galaxies as many early CO studies that 
dominate the literature values tended to select LIRGs rather than normal star-forming galaxies.
This could impact the derivation of the scaling relations using different studies covering a wide
range of redshifts \citep[e.g.][]{Tacconi2018}.

\begin{figure}[htbp]
	\resizebox{\hsize}{!}{\includegraphics{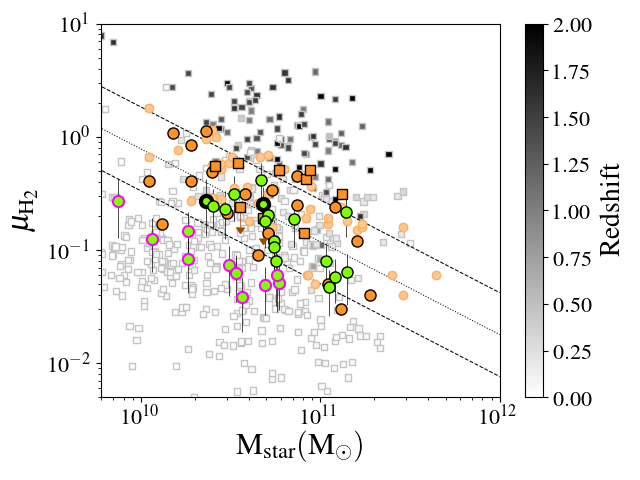}}
	\caption{Fraction of cold molecular gas as a function of the galaxy stellar masses. The colours and shapes of the 
	symbols are the same as in Fig. \ref{z_sSFR}. The dotted line is the fit of the $M_{\rm star}$--\muhh\ 
	relation for the PHIBSS2 galaxies at $0.49<z\leq0.6$, with its variance being represented by the two dashed 
	lines. The pink outlined green symbols are for our low-\muhh\ galaxies, which are located below 
	the $1\times$ $\sigma_{\rm H_2}$ line of the M$_{\rm star}$--\muhh\ relation for the PHIBSS2 field galaxies. }
    \label{Ms_fg}
\end{figure}

Figure \ref{Ms_fg} presents the galaxy cold gas fractions as a function of their
stellar masses. It constitutes the main result of our analysis. At redshifts
similar to those of our sample, $0.49 \leq z \leq 0.6$, the relation between \muhh\ and
M$_{\rm star}$ for the PHIBSS2 subsample has a slope of $\sim-0.82$ and a
variance $\sigma_{\rm H_2}=0.37$ dex. A significant fraction of our targets
fall below this 1 $\times$ $\sigma_{\rm H_2}$ line of the M$_{\rm
star}$--\muhh\ relation for the field galaxies. This means that while 63\% of the 
galaxies in the LSS of CL1411.1$-$1148 have gas mass fractions comparable to
their field counterparts, 37\% lie below the locus defined by field galaxies
at the same stellar mass.  We refer to these ten galaxies as low-\muhh\ systems. 
In order to quantify the significance of this low-\muhh\ population,
we randomly extracted, 100 000 times, 27 galaxies from a normal
distribution of sources with the same mean \muhh\ and same standard deviation
as PHIBSS2. The probability of getting 37\% of the galaxies below
1$\sigma$ is less than 1\%. Therefore, this excess to one side of the field
relation deviates significantly from the expected tail of sources for a
Gaussian distribution, and reveals a population that was absent from previous
surveys. These galaxies are identified with the dagger symbol ($^{\dagger}$) in the tables and they are
highlighted in pink in all figures. Combining our sample with the comparison
field PHIBSS2 subsample with $z \leq 0.6$, the relation between \muhh\ and M$_{\rm
star}$ becomes slightly shallower, with a slope of $-0.51$.

\begin{figure}[htbp]
	\resizebox{\hsize}{!}{\includegraphics{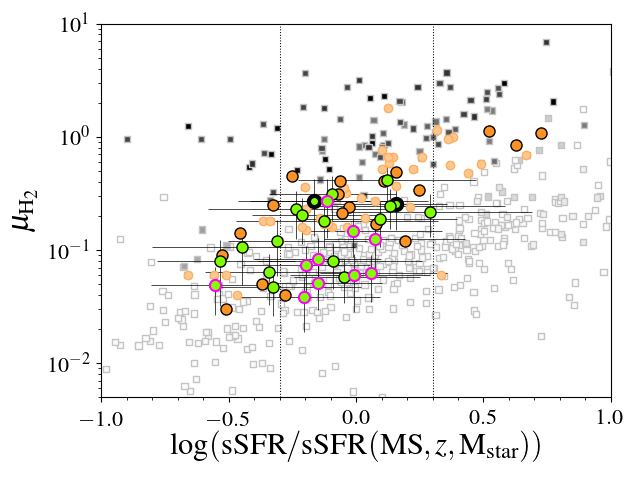}}
	\caption{Fraction of cold molecular gas as a function of the normalised specific 
	star formation rates. The colours and shapes of the symbols are the same as 
	in Fig. \ref{z_sSFR}. The dotted lines represent the $\pm0.3$ dex scatter of the MS.}
    \label{sSFR_fg}
\end{figure}

Interestingly, the SFRs of all but one (\IoVIII) of the low-\muhh\ galaxies are
normal for their stellar masses, indicating that their molecular gas reservoir,
either in mass or in physical properties, is modified before their star
formation activity is impacted. This possibility was also suggested by
\citet{Jablonka2013} for LIRGs in clusters and by \citet{Alatalo2015} for local
elliptical galaxies, who find that their diffuse gas reservoir could potentially
be stripped before the dense gas, which is more closely related to star
formation. This disconnection between \muhh\ and SFR is further illustrated
in Fig. \ref{sSFR_fg} which presents \muhh\ as a function of the galaxy sSFR, 
normalised to their position on the MS, following \citet{Genzel2015}.

Although our galaxies have normal SFRs for their stellar masses, they have a
significantly different distribution of gas fractions than the field
samples. Our low-\muhh\ targets populate a region that to date has remained uncovered at
similar redshift, and reveal that there is a much larger scatter in \muhh\ at
fixed sSFR (nearly twice as much) than previously encountered in other
studies at similar redshifts. To quantify this result, we performed an
Anderson--Darling (A-D) test \citep{Scholz1987} between the
\muhh\ distributions of PHIBSS2 galaxies and ALMA targets, both within the
MS.
The A-D test is more sensitive to differences in the tails of the distributions 
than a Kolmogorov--Smirnov test.
When all MS galaxies are considered, the A-D test results in p = 0.027, 
meaning that there is only a 2.7\% chance that the two
samples come from the same distribution in \muhh . When we restrict the comparison to the 
low-\muhh\ MS galaxies only and the full MS PHIBSS2 galaxies, we find p =
$1.3\times10^{-3}$, which clearly shows that this low-\muhh\ tail of the ALMA
sample comes from a significantly different \muhh\ distribution ($>$99\%) from
that of the PHIBSS2 MS galaxies.

\begin{figure}[htbp]
	\resizebox{\hsize}{!}{\includegraphics{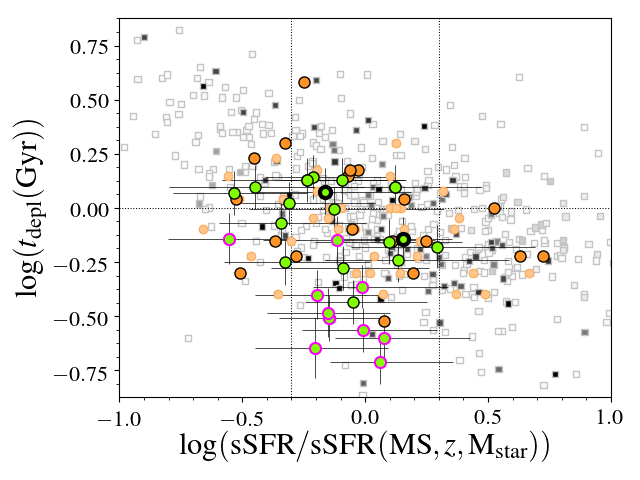}}
	\caption{Dependence of the depletion timescale (in Gyr) as a function of the normalised 
	specific star formation rates. The colours and shapes of the markers are the same as in
	Fig. \ref{z_sSFR}. The horizontal dotted line is located at 1 Gyr 
	and the vertical dotted lines represent the $\pm$ 0.3 dex scatter of the MS.}
    \label{tdepl}
\end{figure}

Figure \ref{tdepl} shows the depletion timescales (M$_{\rm gas}$/SFR) as a
function of the normalised sSFR. Our low-\muhh\ galaxies have short depletion
times $\sim300$ Myr compared to the bulk of the population known so far. This
implies that they should consume their gas and quench more rapidly.


\subsection{Link with local density \label{lowmuh2}}

As a first attempt to link the spatial location of our sample galaxies
and their gas masses, we searched for a possible correlations between
the galaxy gas fraction and cluster-centric distance, but did not find any.

We then looked into the possibility that low gas fractions correlate with local
(over)densities. Figures \ref{targets} and \ref{light_cone} suggest that while the
low-\muhh\ galaxies are embedded in coherent LSS, their
relation with a local and/or small-scale environment does not stand out. Hence, there must be
more than one parameter explaining the status of the galaxy gas reservoir that
we are witnessing.

Similarly, we have looked into the environment of the
PHIBSS2 galaxies. For the galaxies in the COSMOS field, we used the G10-COSMOS
catalogue \citep{Davies2015} for their position and redshift, and the zCOSMOS
20k group catalogue \citep{Knobel2012}. The
3D-HST survey catalogue \citep{Brammer2012, Skelton2014, Momcheva2015} was used
for the PHIBSS2 galaxies within the AEGIS and GOODS-North fields. We used the DEEP2
group catalogue \citep{Gerke2012} for AEGIS. No equivalent group catalogue was
found for GOODS-North.
Among the 19 PHIBSS2 galaxies, only 5 belong to a group or are close to
one. However, we did not find any correlation between their gas content and the
density of their local environment.


\subsection{Caveats \label{caveats}}

The derivation of cold gas masses involves two parameters, $\alpha_\mathrm{CO}$
and $r_{31}$. This raises the question of whether the low-\muhh\
galaxies could arise from our choices of these parameters.

The PHIBSS2 CO conversion factor, $\alpha_\mathrm{CO}$, decreases with
increasing metallicity, from 4.7 down to 3.8 $\,M_\odot\,{\rm
 (K\,km\,s^{-1}\,pc^2)^{-1}}$. Applying the same definition and estimating the
metallicity from its relation with the galaxy stellar mass as in
\citet{Genzel2012}, $\alpha_\mathrm{CO}$ could in principle vary from
$\alpha_{\rm CO}=3.76$ to $4.91\,M_\sun\,{\rm (K\,km\,s^{-1}\,pc^2)^{-1}}$ over
our range of masses. This would increase by 12\% the cold molecular gas mass of the
two lowest mass galaxies (M$_{\rm star}\lesssim10^{10}\,M_\sun$) in our sample,
keep galaxies at M$_{\rm star} \sim 10^{10}\,M_\sun$ at the same positions, and
decrease by 13\% the gas masses of the most massive of our target. None of these
shifts would change the identification of the low-\muhh\ galaxies.

As to the flux ratios, the observations of PHIBSS2 were conducted in \mbox{CO(2-1)}, with $r_{21}=0.77$
as a trade off between the values found in earlier studies, which range from
1 to 0.6 \citep{Freundlich2019}. As seen from Eq. \ref{eq2}, increasing
$r_{31}$ decreases the galaxy gas mass, hence increasing the
gas mass in our low-\muhh\ galaxies is not an option. As shown in previous studies of nearby
galaxies \citep{Mauersberger1999,Mao2010,Lamperti2020}, there is a large scatter
in $r_{31}$ ($\sim$0.1) with any fixed parameter involving the galaxy SFRs. It should be noted,
however, that these nearby samples are mostly composed of galaxies above the main
sequence unlike our targets. The value of $r_{31}$ would need to be decreased by at least a
factor 2 in order to reconcile our sample with the PHIBSS2 galaxies in
Fig. \ref{Ms_fg}. At the lowest tail of the distribution $r_{31}=0.2$ is rarely
encountered \citep{Mao2010}. Moreover, if $r_{31}$ does vary from one galaxy to another, 
some normal systems could have $r_{31}$ higher than 0.5. Hence, they would potentially 
move into the low-\muhh\ region. This issue definitely needs further observations directly in CO(1-0).


\section{Conclusion\label{conclusion}}

We have presented the CO(3-2) emission line fluxes obtained with ALMA for a
sample of 27 galaxies located within $5 \times R_{200}$ of the centre of the
EDisCS cluster CL1411.1$-$1148 at $z=0.5195$. This constitutes the largest sample
of galaxies with direct cold gas measurements at a single intermediate redshift
($0.5000 < z < 0.5375$), and the only sample of galaxies in interconnected cosmic
structures around a galaxy cluster.

Unlike most of the previous studies which targeted galaxies based on their SFRs, our
selection is based on stellar masses and on ground-based photometry in the
$u,\,g$, and $i$ bands only, with the requirement that galaxies are in the blue
cloud of the cluster colour--magnitude diagrams, and have available spectroscopic
redshifts. The derivation of the galaxy stellar masses and star formation rates
placed all but two of our targets on the MS of the normal star-forming galaxies,
with stellar masses between log(M$_{\rm star}$/M$_\sun)=9.8$ and 11.2, and SFRs
ranging from $\log({\rm SFR}/($M$_\sun{\rm yr}^{-1})) = 0.3$ up to 1.7. Two
galaxies fall within the passive region of the rest-frame $UVJ$ colour--colour
diagram. They still are very close to the star-forming sequence, which suggests
that these systems are transitioning to a quenched state.

Our sample covers a wide range of cold molecular gas masses, from
$1.38\times10^9$ up to $3\times10^{10}$ M$_\sun$. The low tail of this gas mass
distribution probes lower values than most other studies of CO at $z>0.05$.
We have compared our results to the PHIBSS2 survey \citep{Freundlich2019}, which is the
best field counterpart to our study at $z\sim0.5$. Looking at the link between galaxy gas
fraction and stellar mass, we find that while 63\% of our galaxies follow
the same trend between \muhh\ and M$_{\rm star}$ as field galaxies, 37\% of our
targets fall below the 1$\sigma$ variance of the relation derived for the
field galaxies. This excess to one side of the field relation deviates
significantly from the expected tail of sources for a Gaussian distribution and reveals
a population that was absent from other surveys. Our cold
molecular gas mass estimates depend on our choice of the CO conversion factor,
$\alpha_\mathrm{CO}$, and the line ratio, $r_{31}$. But our results remain the
same for all reasonable values of these parameters. Nevertheless direct
observation of the CO(1-0) transition should help shed definitive light on this
issue.

Interestingly, the SFRs of the low-\muhh\ galaxies are normal for their
stellar masses. This indicates that their molecular gas reservoir changes, either
in mass or properties, before the galaxy SF activity is impacted. Our sample
displays a much larger scatter in \muhh\ than previously encountered in other studies at
similar redshifts (at least two times larger). This is the case at fixed SF activity and
stellar mass, as represented by the specific SFR normalised to the position of
the galaxies on the mass sequence.

Although our galaxies have been selected in the vicinity of a galaxy
cluster, we have not identified any correspondence between the low gas fraction
of the galaxies and the local density of their environment.


\begin{acknowledgements}
  
We thank the anonymous referee whose detailed comments helped to improve the
presentation of the paper. This paper makes use of the following ALMA data:
ADS/JAO.ALMA\#2015.1.01324.S and ADS/JAO.ALMA\#2017.1.00257.S.  ALMA is a
partnership of ESO (representing its member states), NSF (USA) and NINS (Japan),
together with NRC (Canada), MOST and ASIAA (Taiwan), and KASI (Republic of
Korea), in cooperation with the Republic of Chile.  The Joint ALMA Observatory
is operated by ESO, AUI/NRAO and NAOJ.  The authors are indebted to the
\textit{International Space Science Institute} (ISSI), Bern, Switzerland, for
supporting and funding the international team ``The Effect of Dense Environments
on Gas in Galaxies over 10 Billion Years of Cosmic Time''.  We are grateful to
the \texttt{Numpy} \citep{numpy:2006, numpy:2011}, \texttt{SciPy}
\citep{scipy:2020}, \texttt{Matplotlib} \citep{matplotlib:2007},
\texttt{IPython} \citep{ipython:2007} and \texttt{Astropy} \citep{astropy:2013,
  astropy:2018} teams for providing the scientific community with essential
Python tools.

\end{acknowledgements}


\bibliographystyle{aa}
\bibliography{library}


\begin{appendix}


\section{ALMA maps and spectra of our galaxies \label{section-maps}}

In this appendix we present the $i$-band images, the ALMA intensity maps, and the spectra of all of our targets.
The low-\muhh\ targets are indicated as such by a label at the bottom left of the $i$-band image.

\begin{figure*}[htbp]
    \centering	
	\includegraphics[scale=0.3]{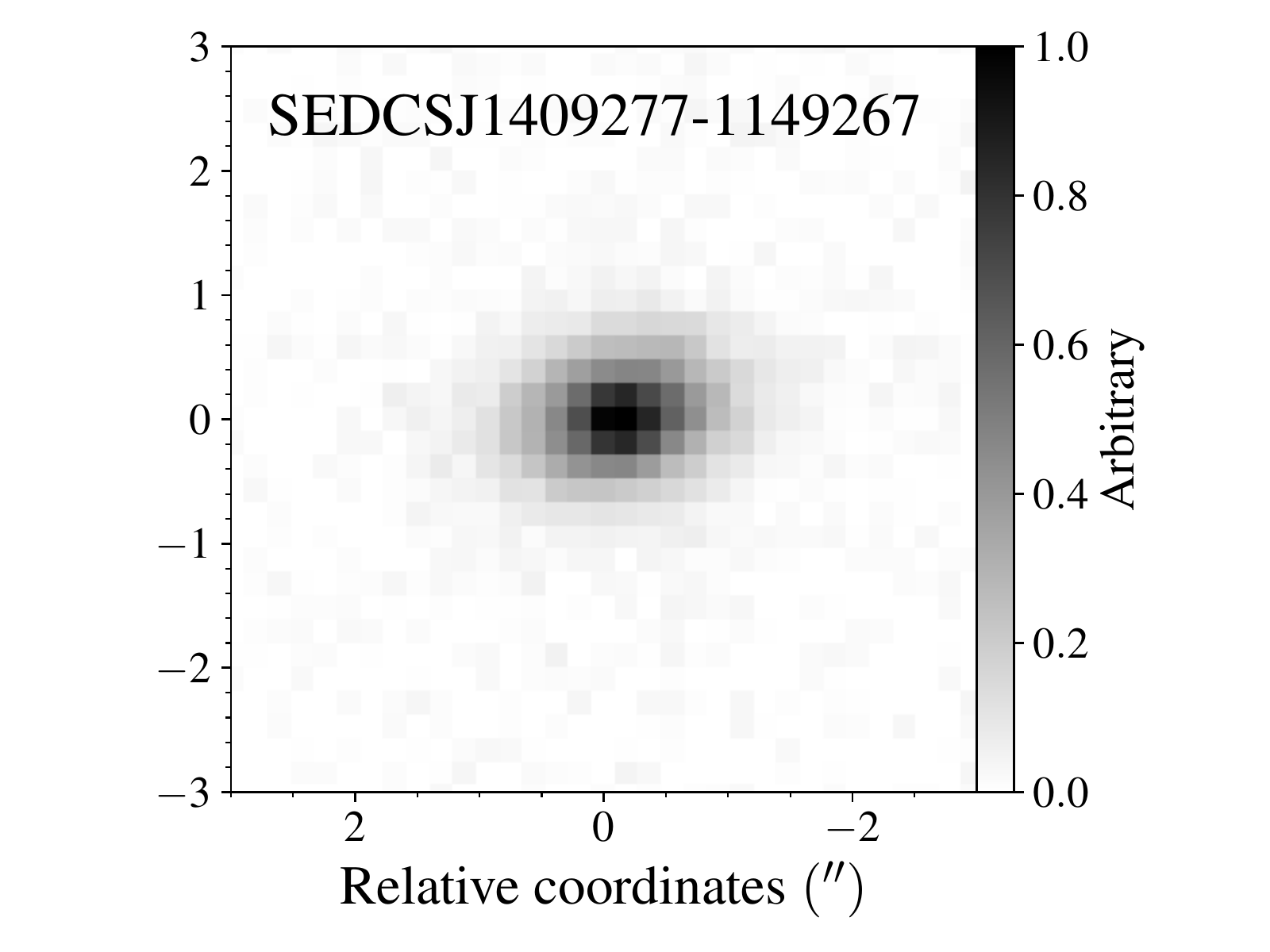}
	\includegraphics[scale=0.3]{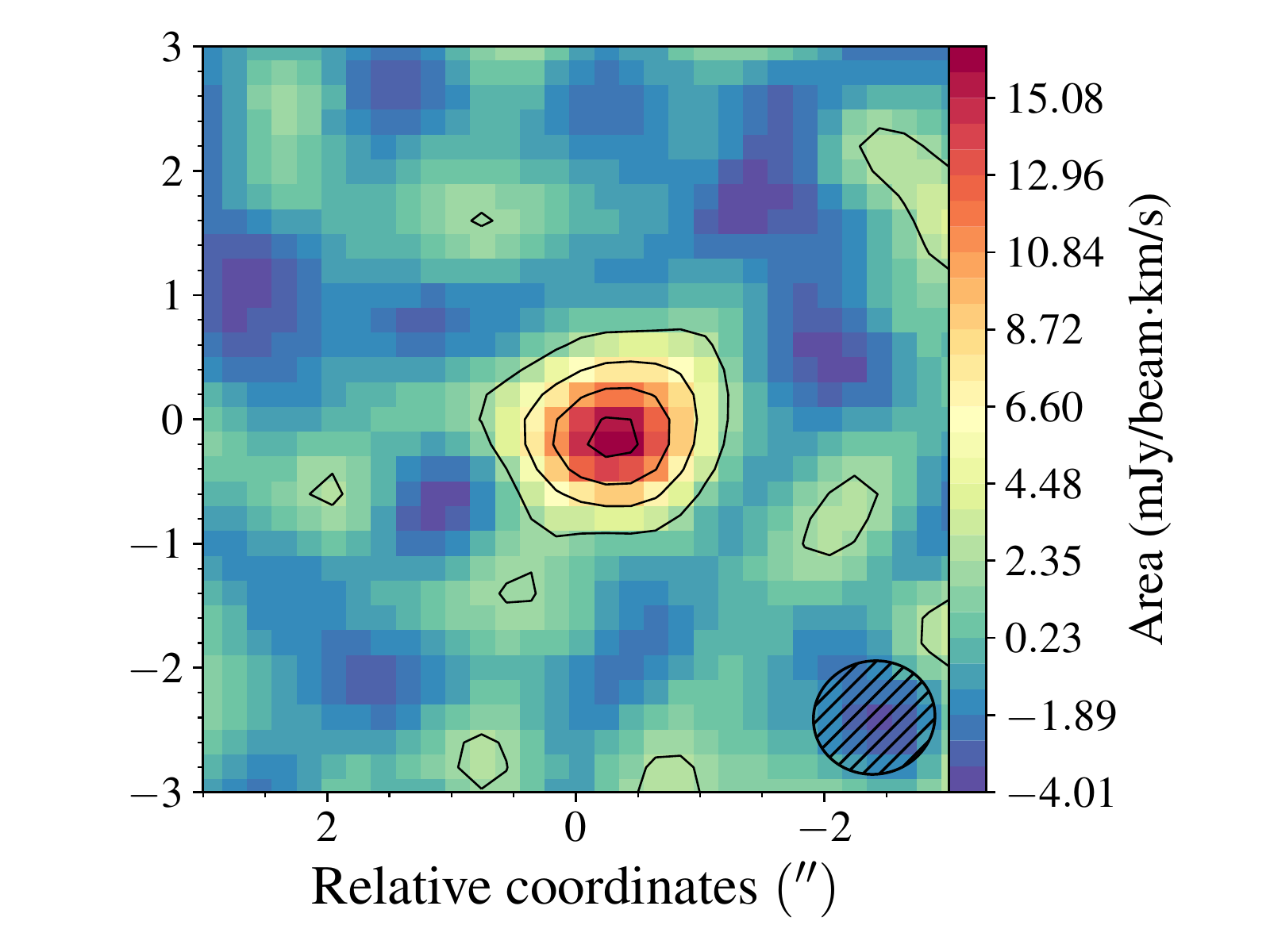}
	\includegraphics[scale=0.3]{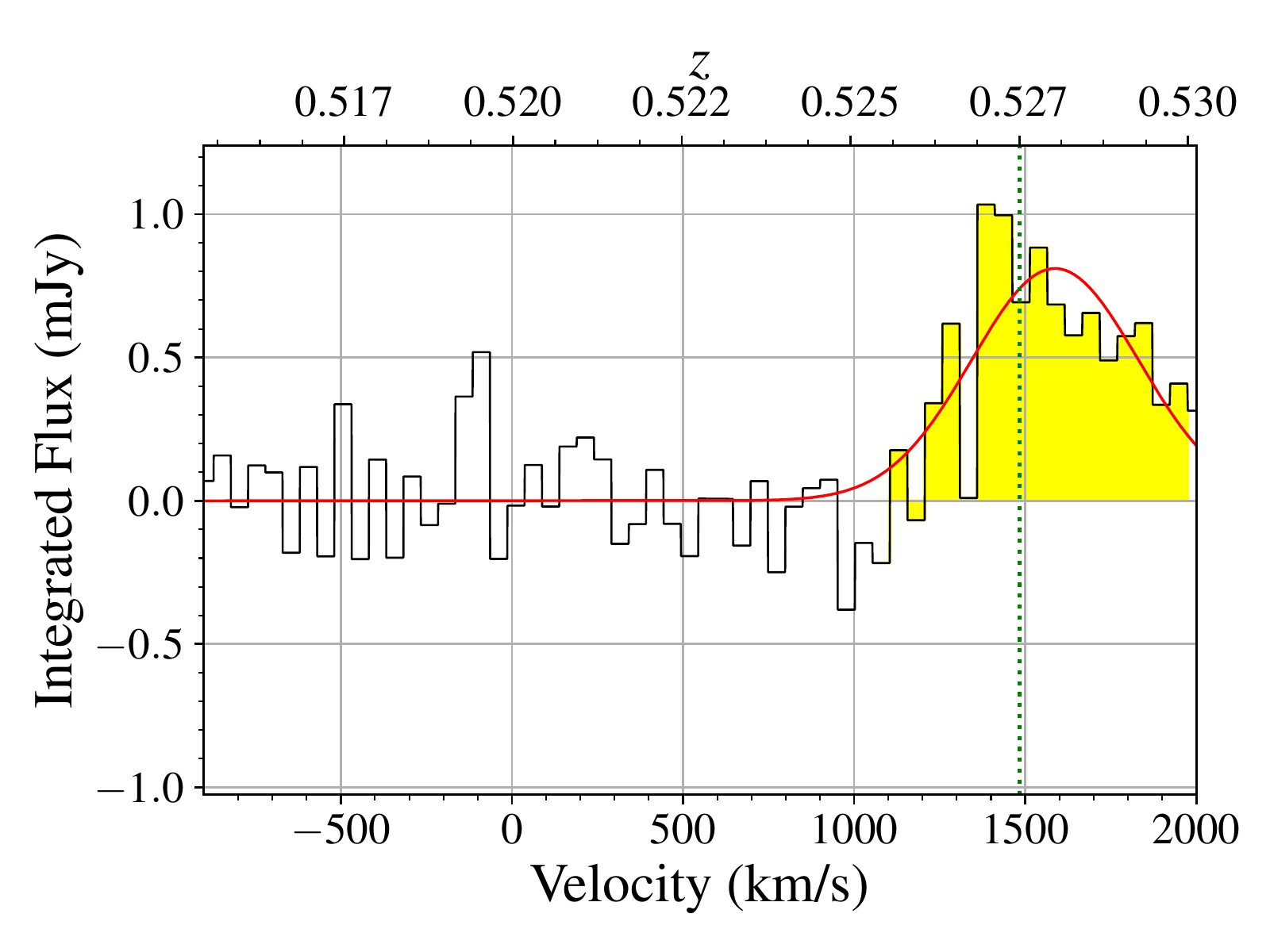}

	\includegraphics[scale=0.3]{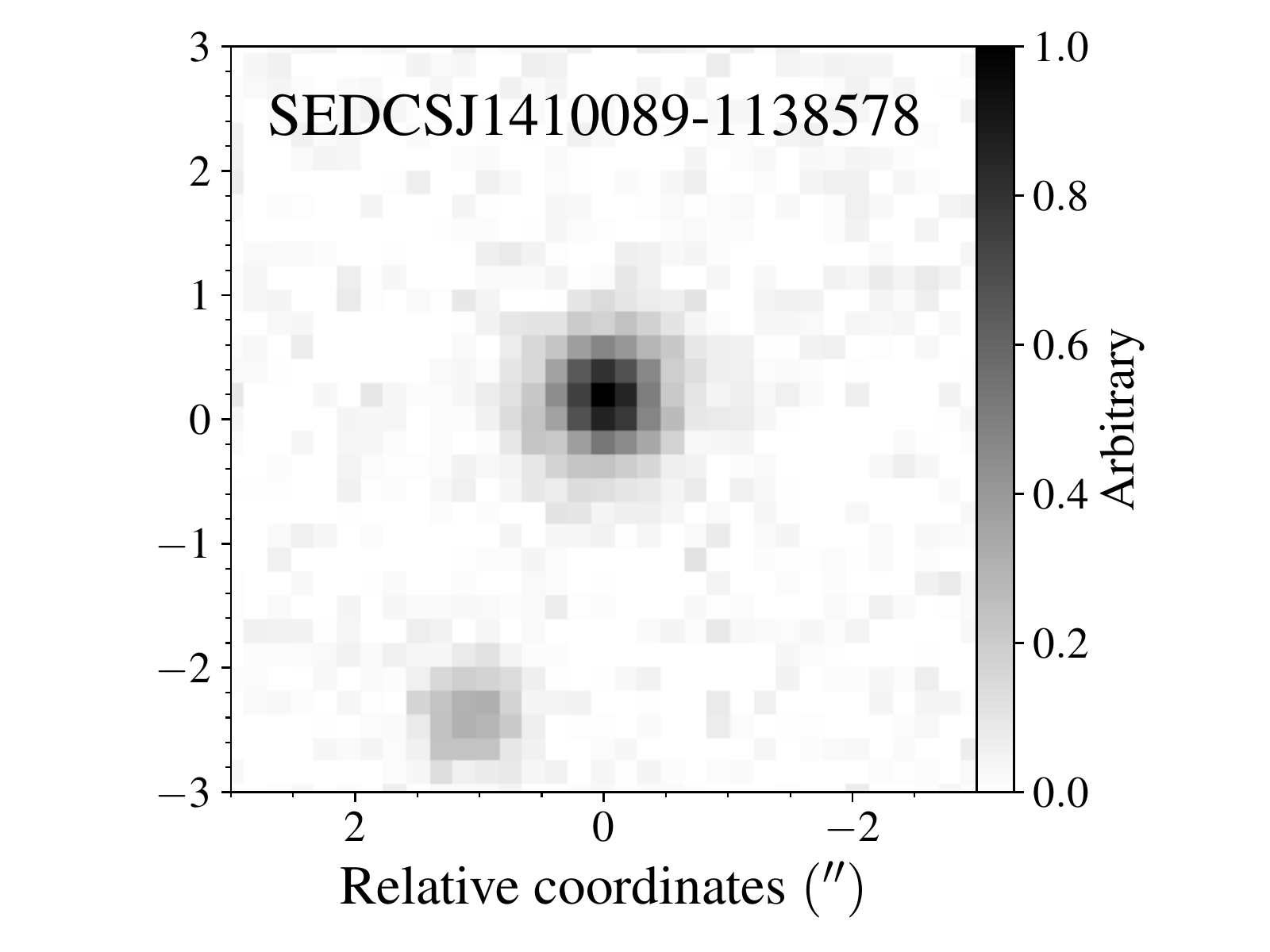}
	\includegraphics[scale=0.3]{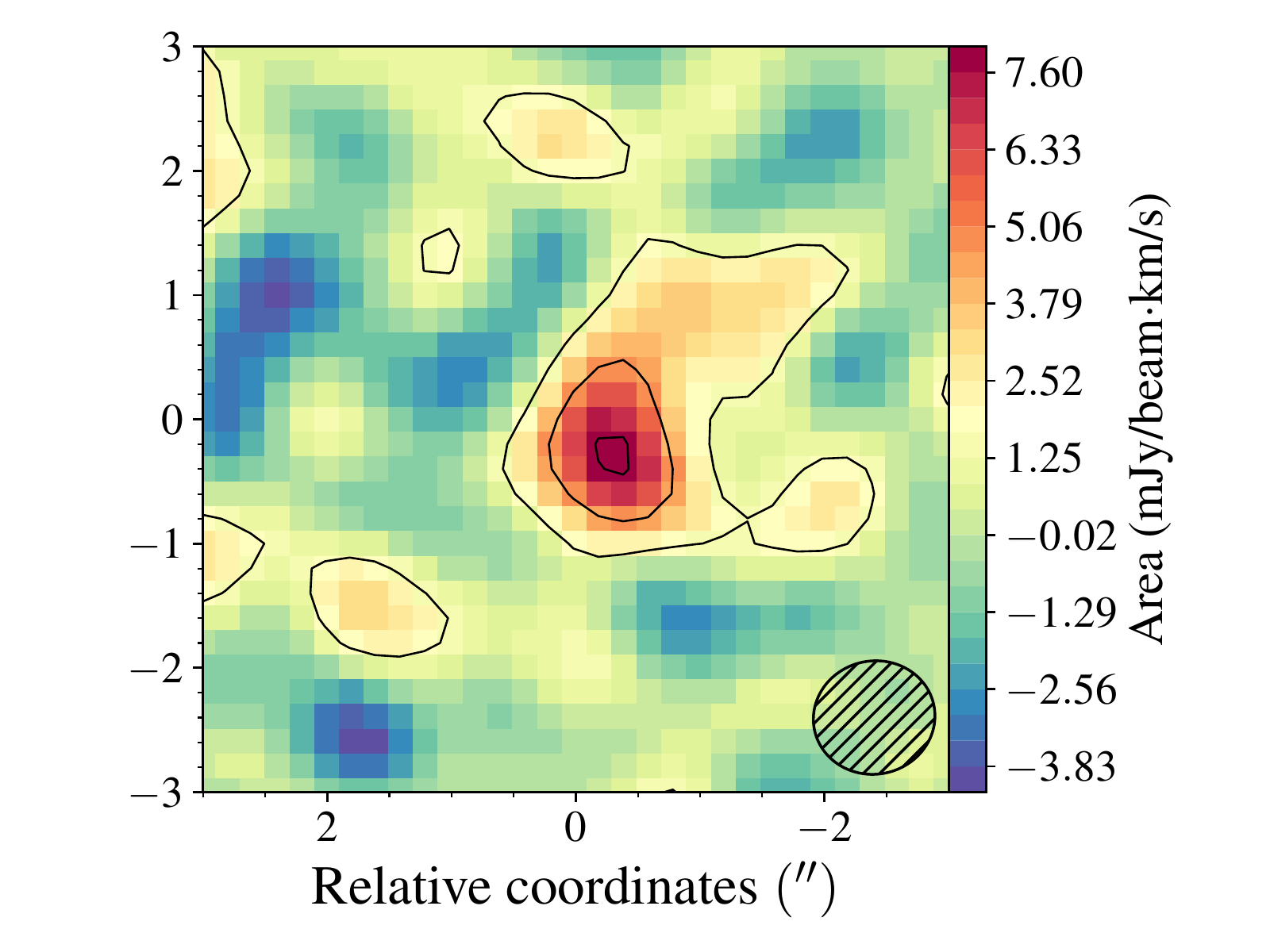}
	\includegraphics[scale=0.3]{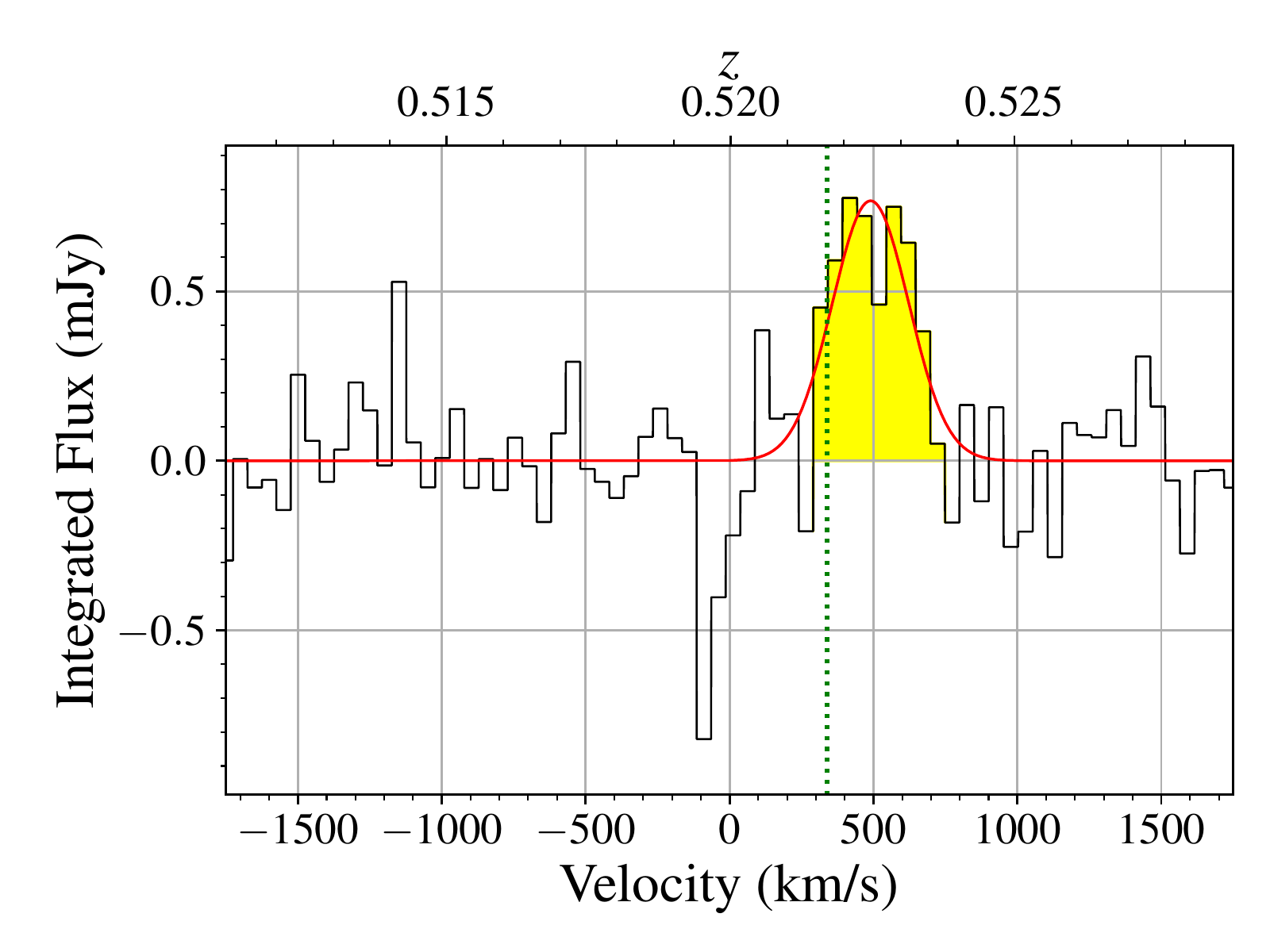}

	\includegraphics[scale=0.3]{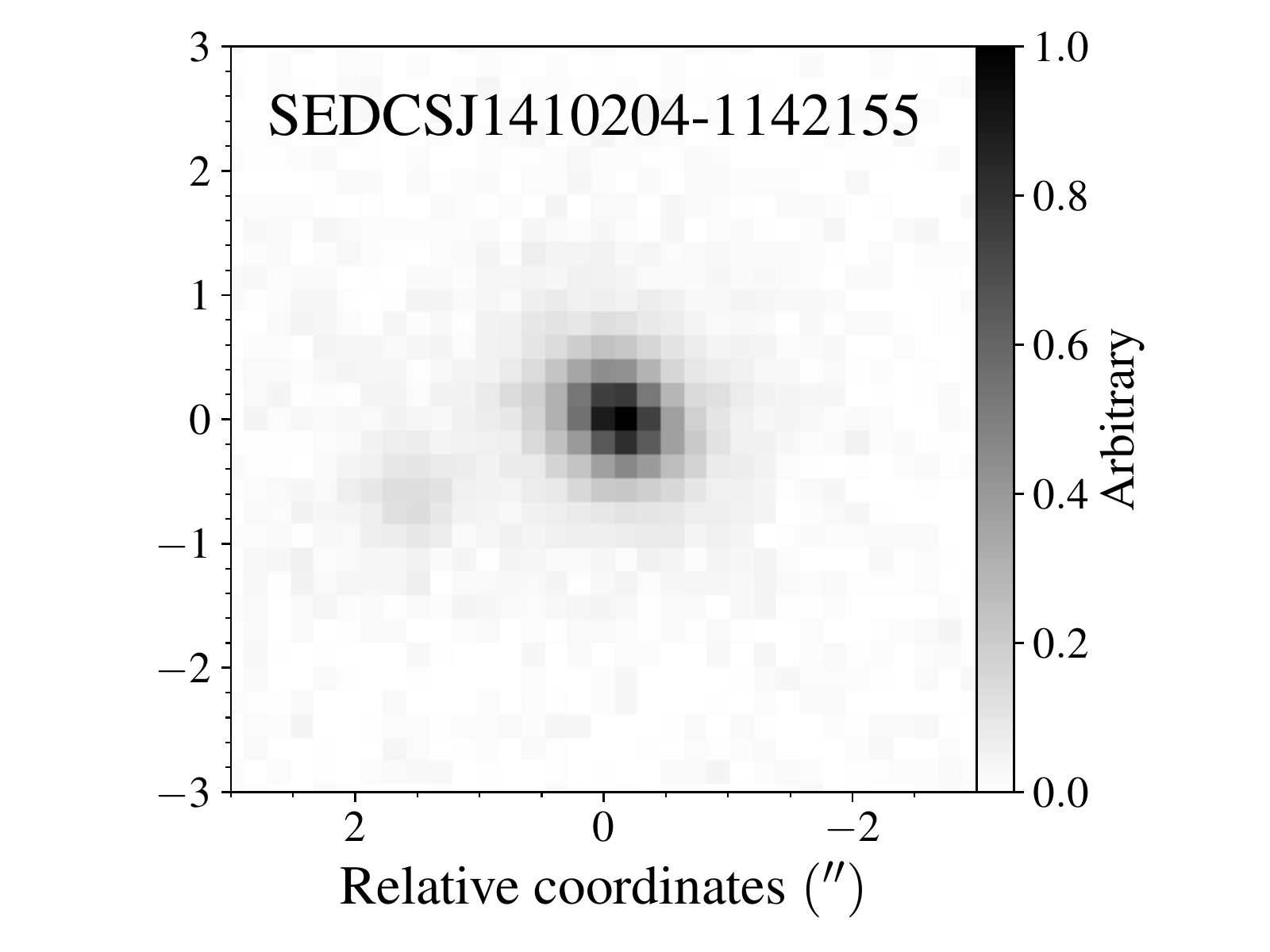}
	\includegraphics[scale=0.3]{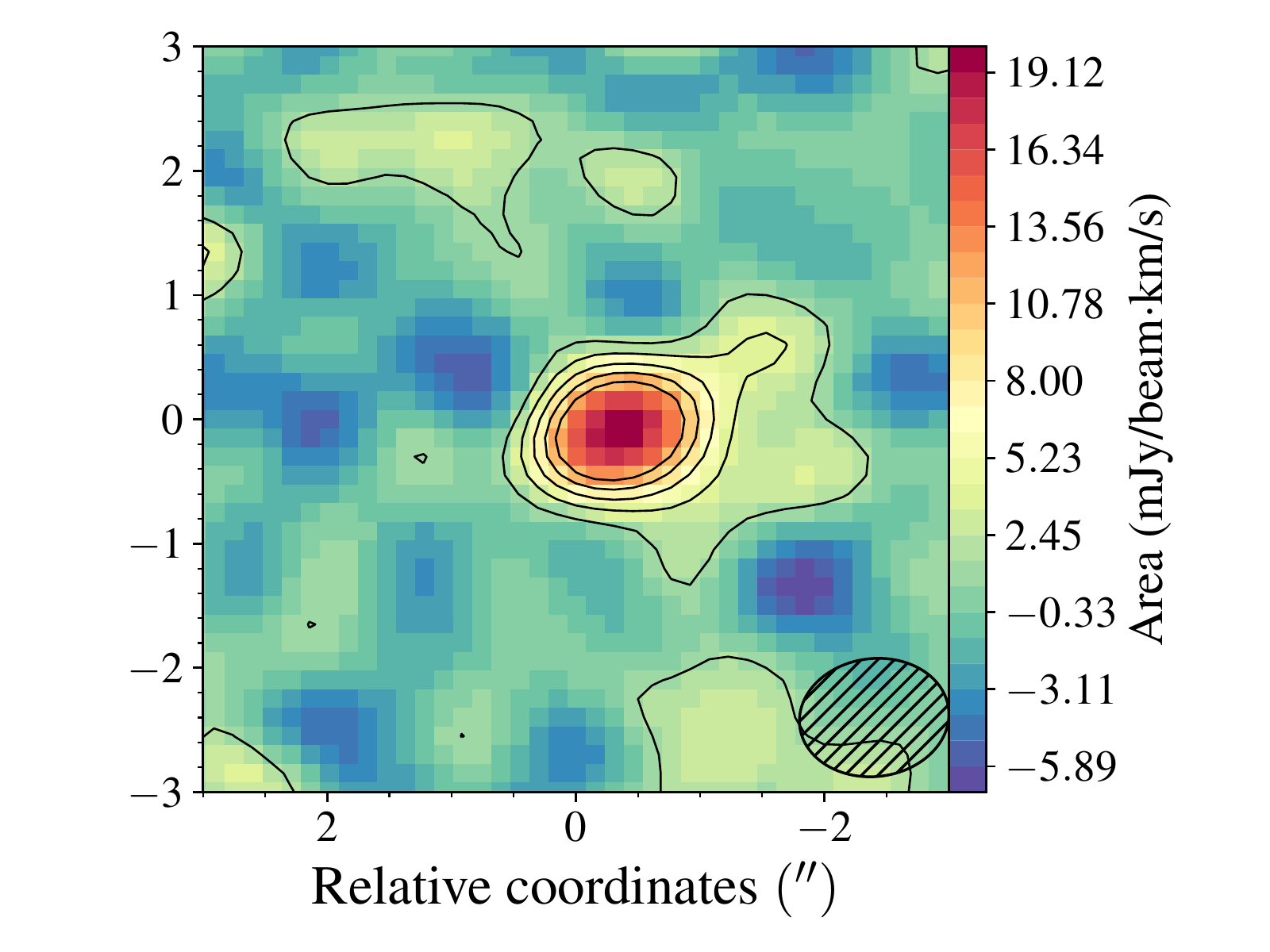}
	\includegraphics[scale=0.3]{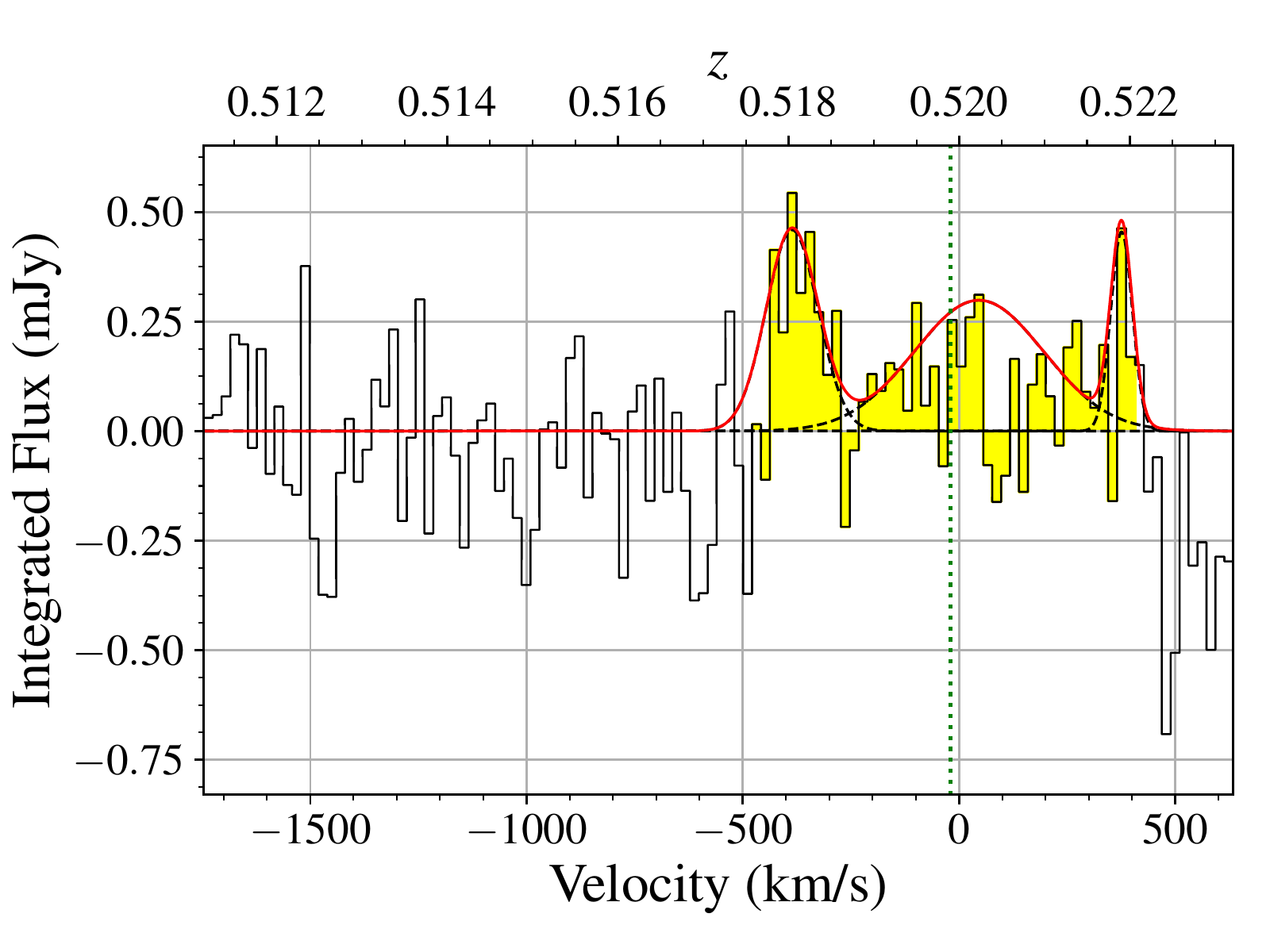}

	\includegraphics[scale=0.3]{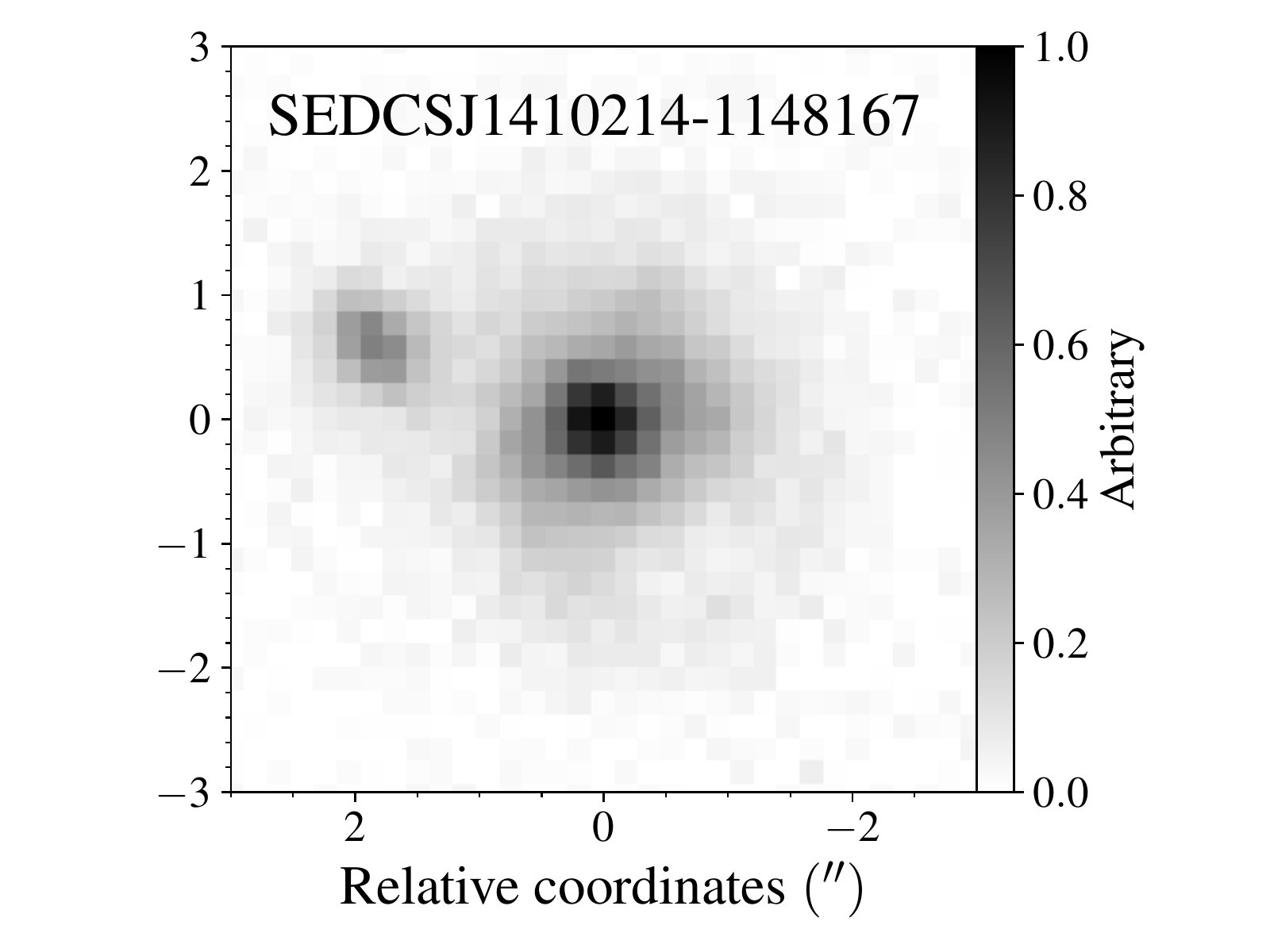}
	\includegraphics[scale=0.3]{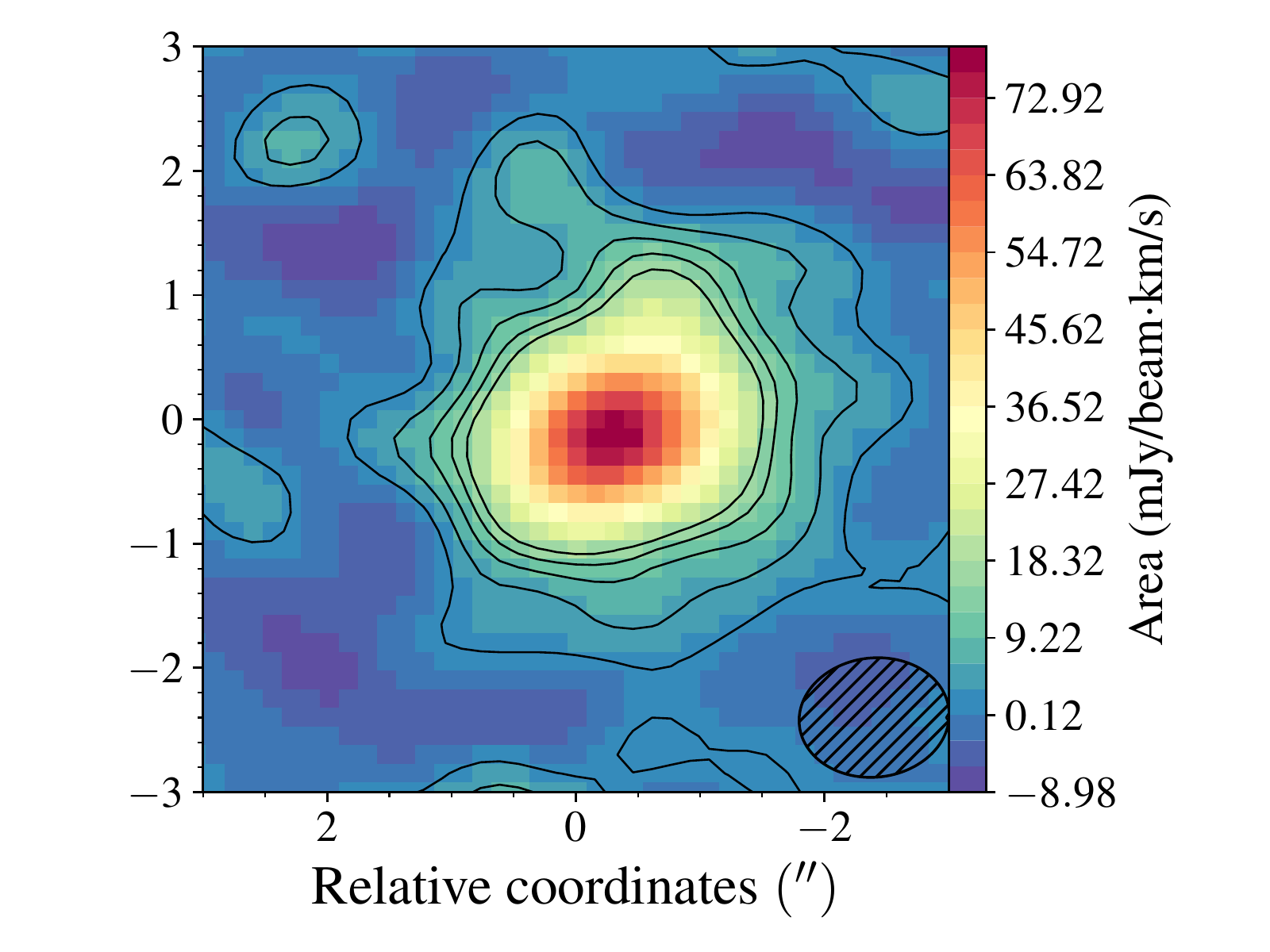}
	\includegraphics[scale=0.3]{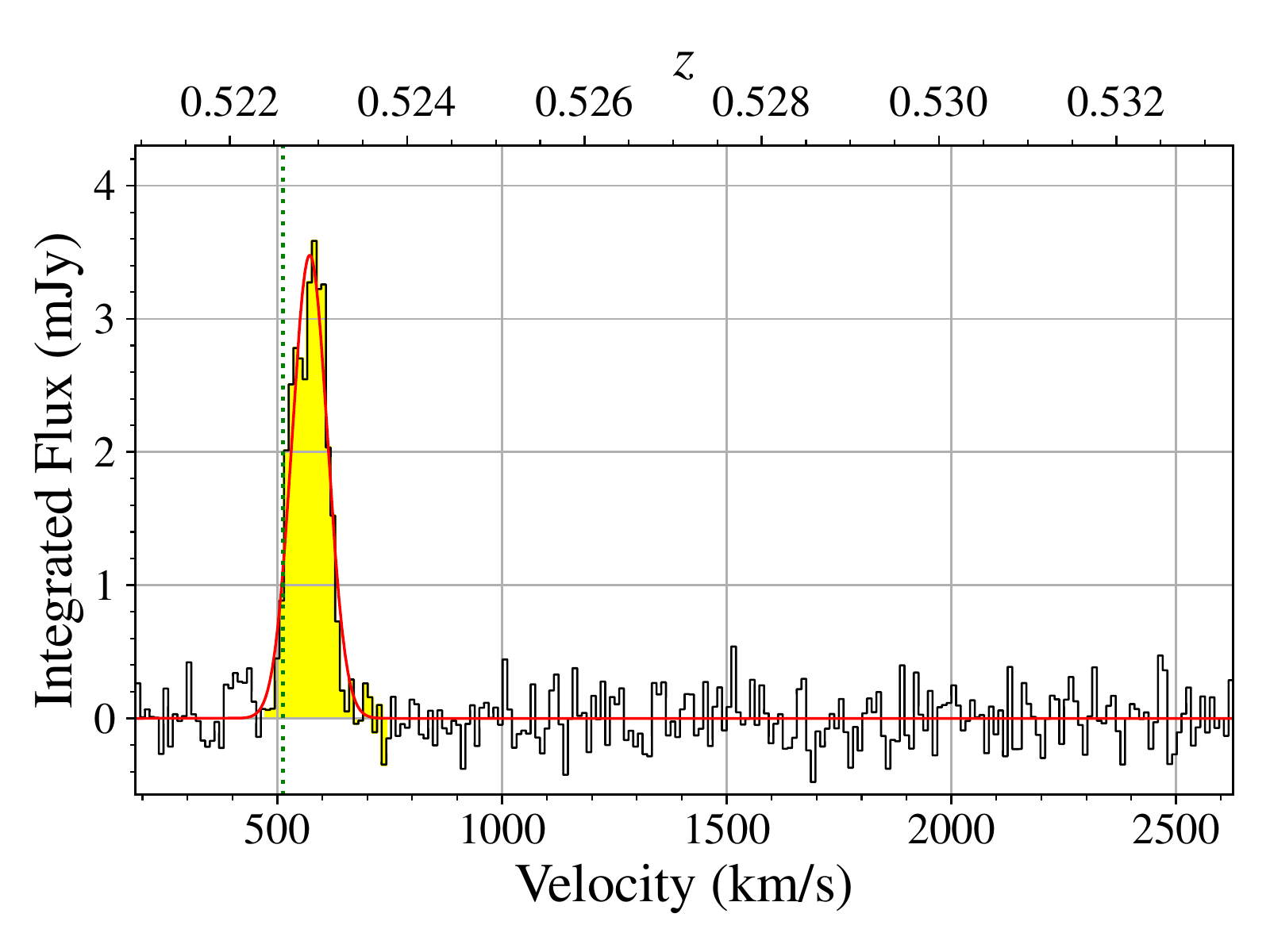}

	\includegraphics[scale=0.3]{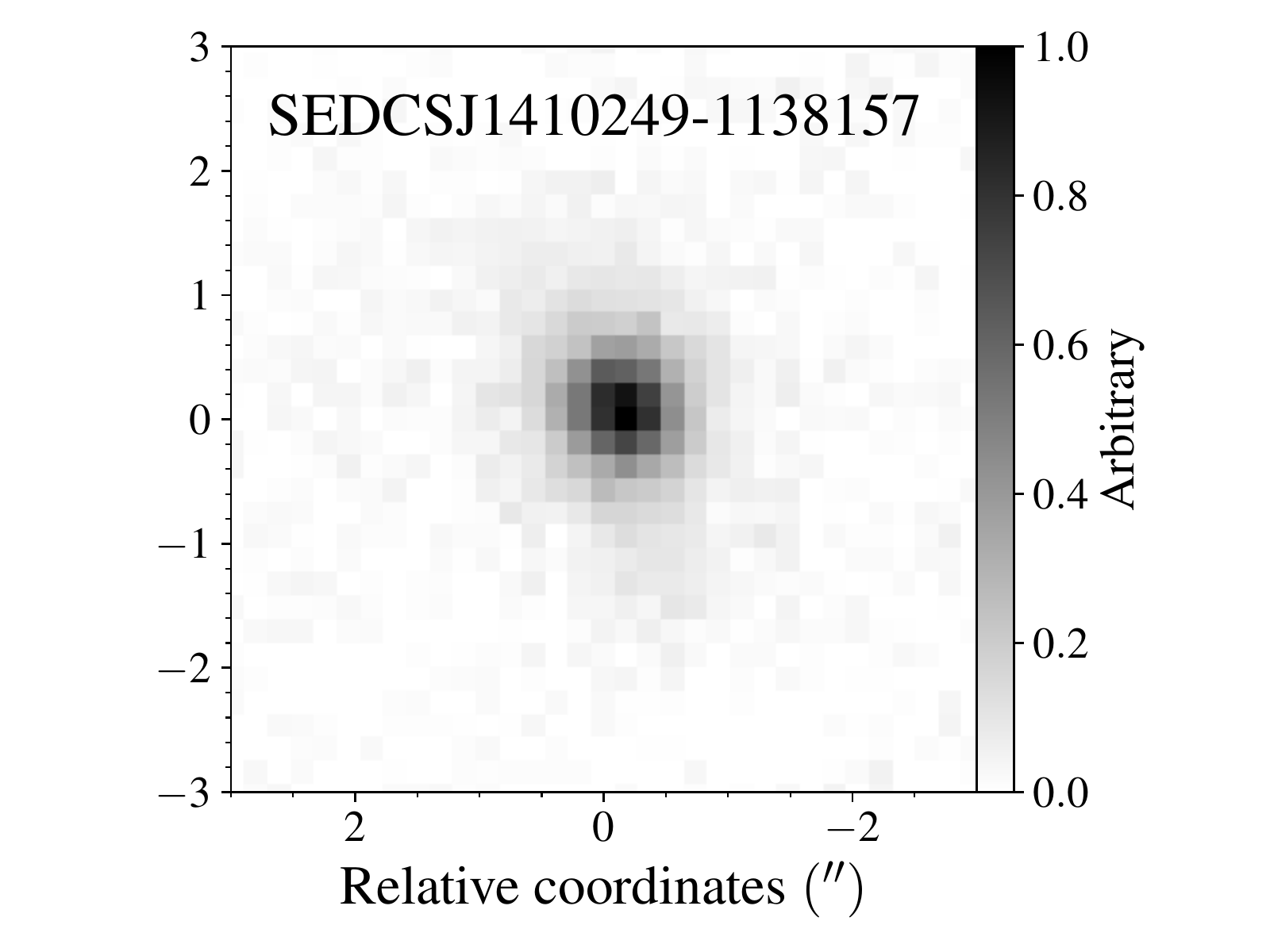}
	\includegraphics[scale=0.3]{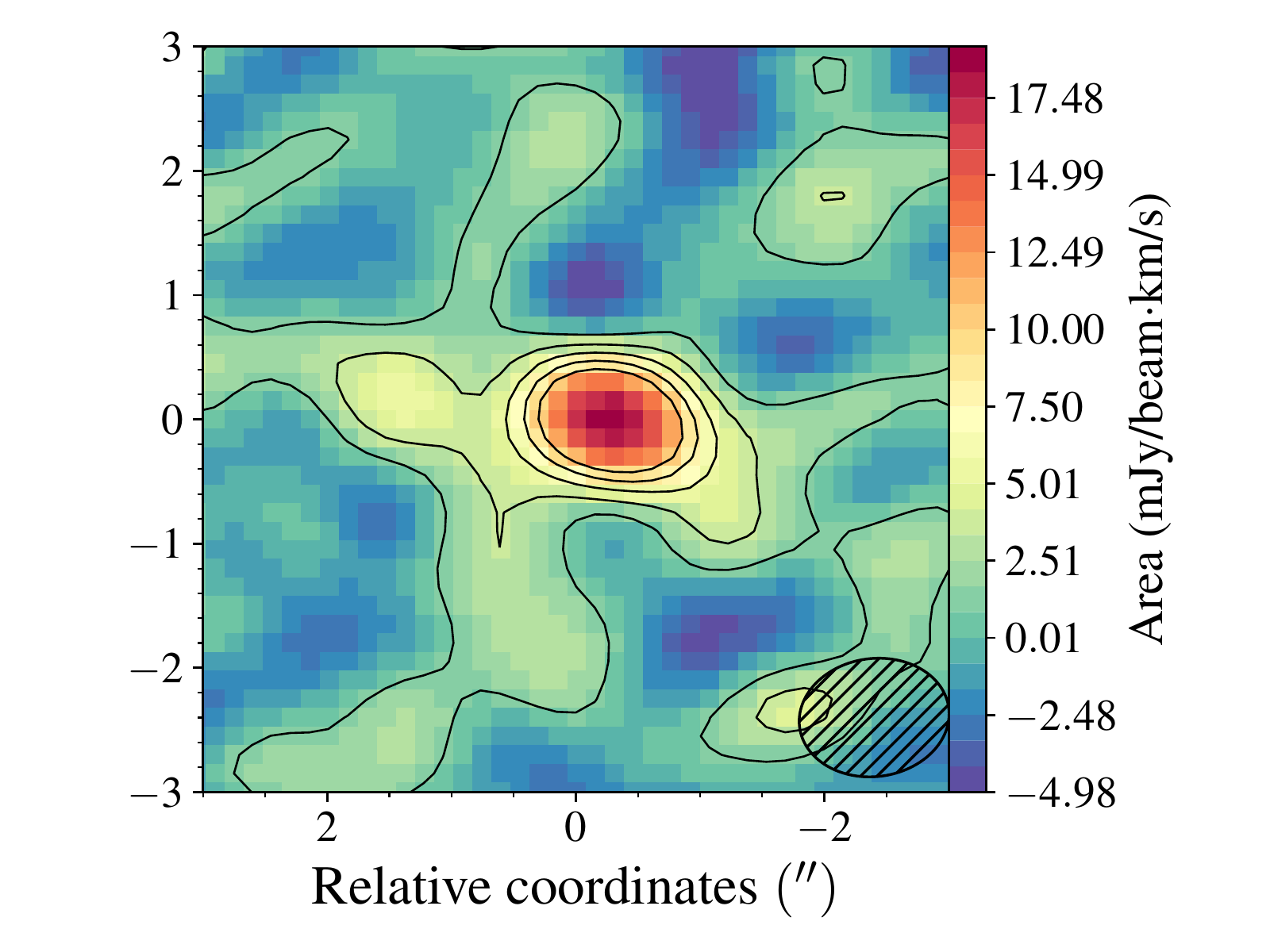}
	\includegraphics[scale=0.3]{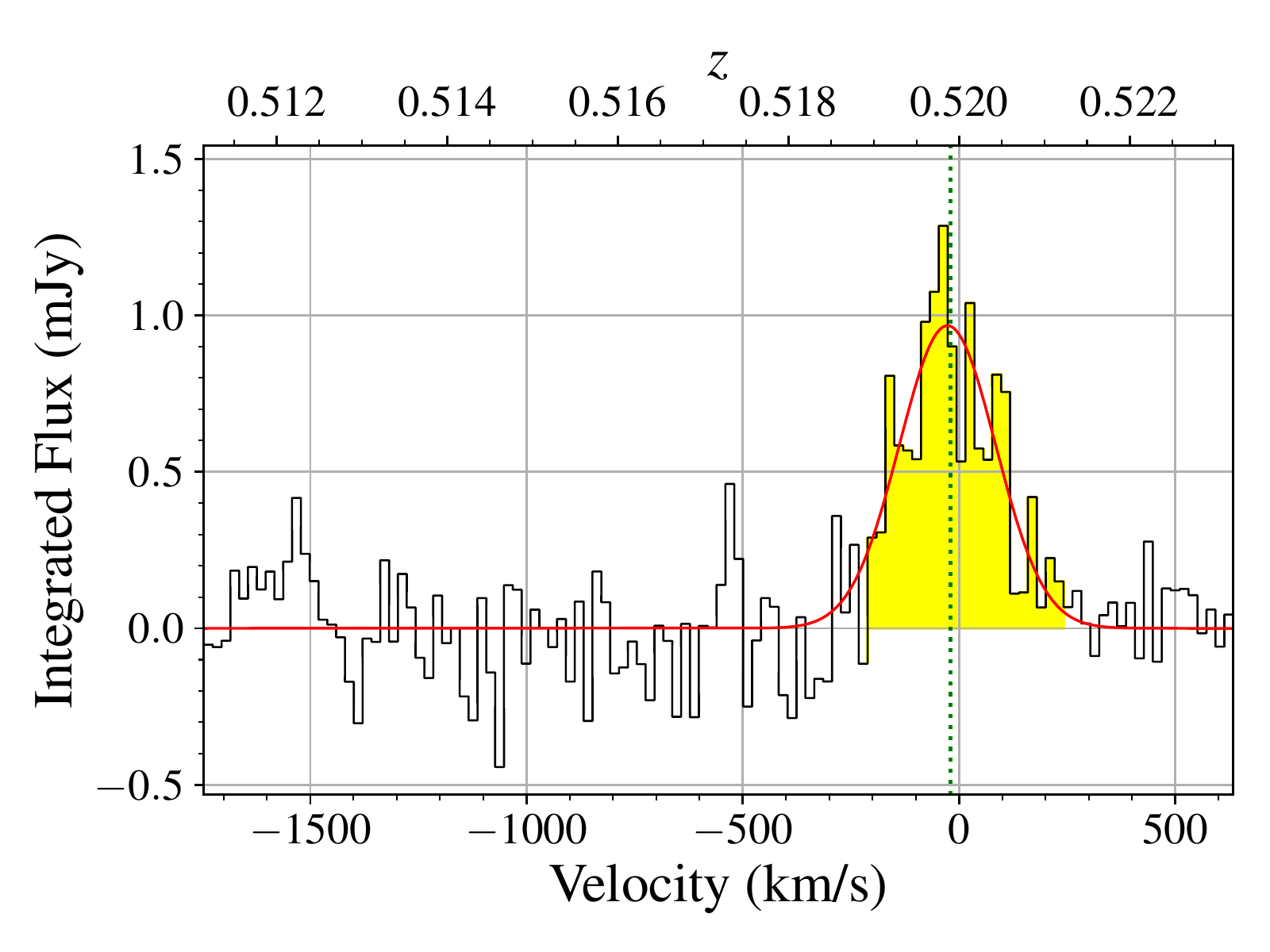}

    \caption{\textit{Left}: CFHT/MEGACAM \textit{i}-band images of our galaxies, unless stated otherwise,
    in a $6\arcsec\times6\arcsec$ snapshot, centred on the galaxies 
    coordinates. \textit{Middle}: ALMA map of the CO(3-2) emission around our 
    galaxies. Spatial scale is the same as in the left panel.
    The colour wedge of the intensity map is in mJy$/$beam\,km\,s$^{-1}$. The 
    contours are defined such that they are spaced by 2 times the rms starting at 1 time above
    the rms. In the bottom right corner is the beam size. \textit{Right}: The 
    spectrum shows the flux, $S_\mathrm{CO}$, spatially integrated as indicated in
    Sect. \ref{alma_obs}, of the source in mJy in function of the velocity in km\,s$^{-1}$, 
    with respect to a fixed frequency. 
    The Gaussian profiles are fits of the emission lines from which we derived our FWHMs. 
    The yellow filled zones correspond to the spectral extent of the emissions.
    The green vertical line corresponds to the spectroscopic redshift.}
 \label{maps}
\end{figure*}    

\begin{figure*}[htbp]\ContinuedFloat
	\centering	
	\includegraphics[scale=0.3]{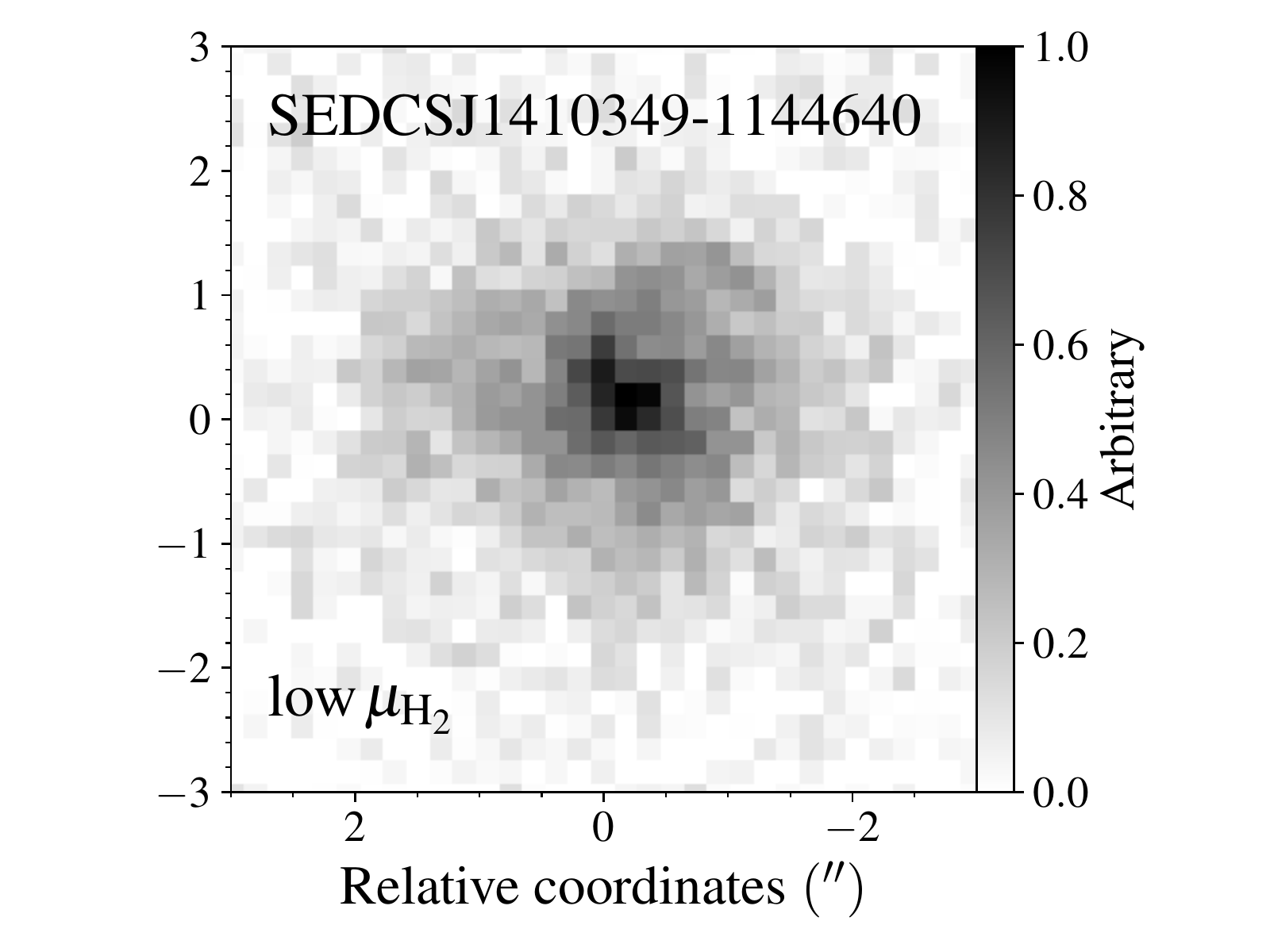}
	\includegraphics[scale=0.3]{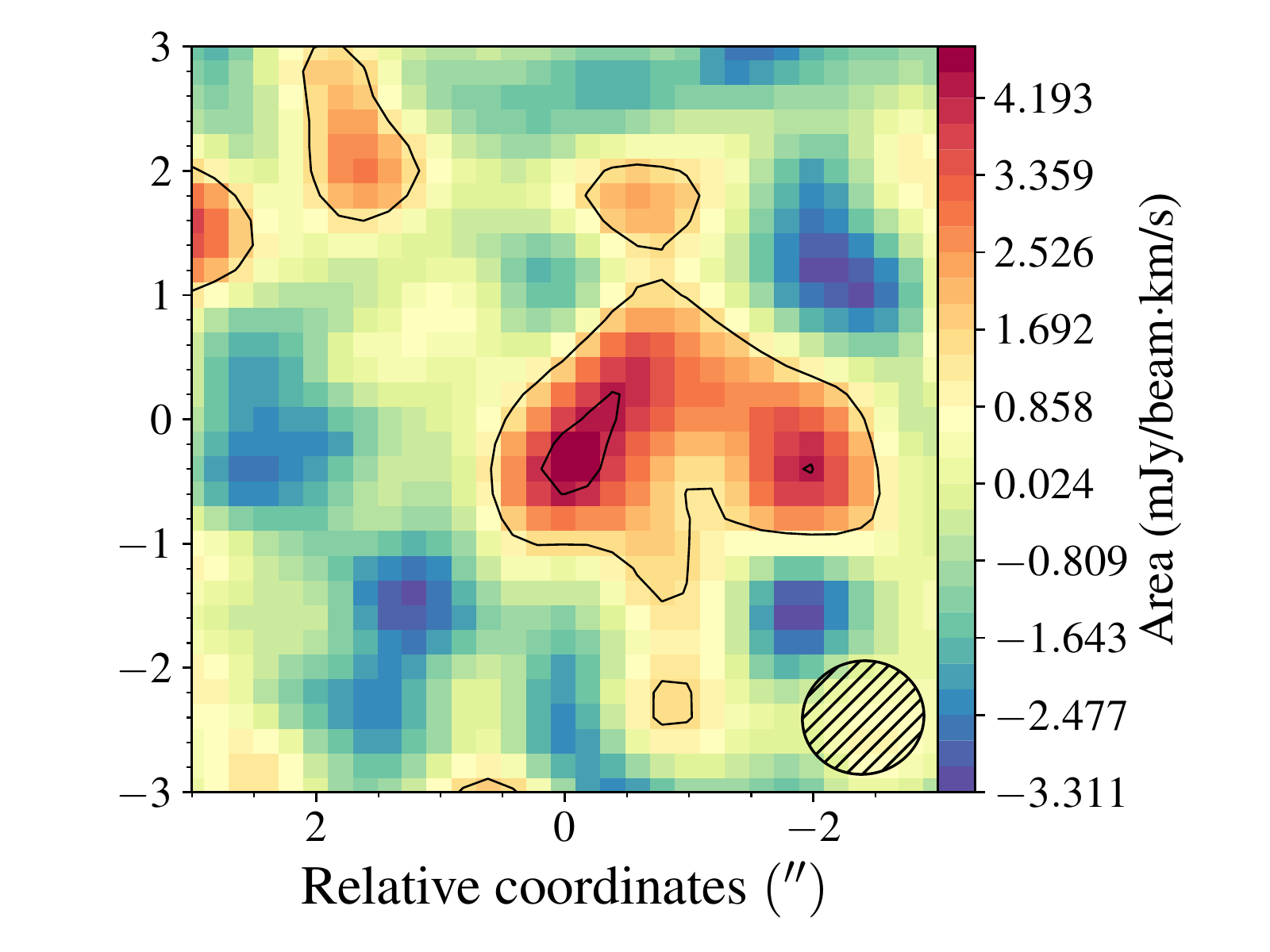}
	\includegraphics[scale=0.3]{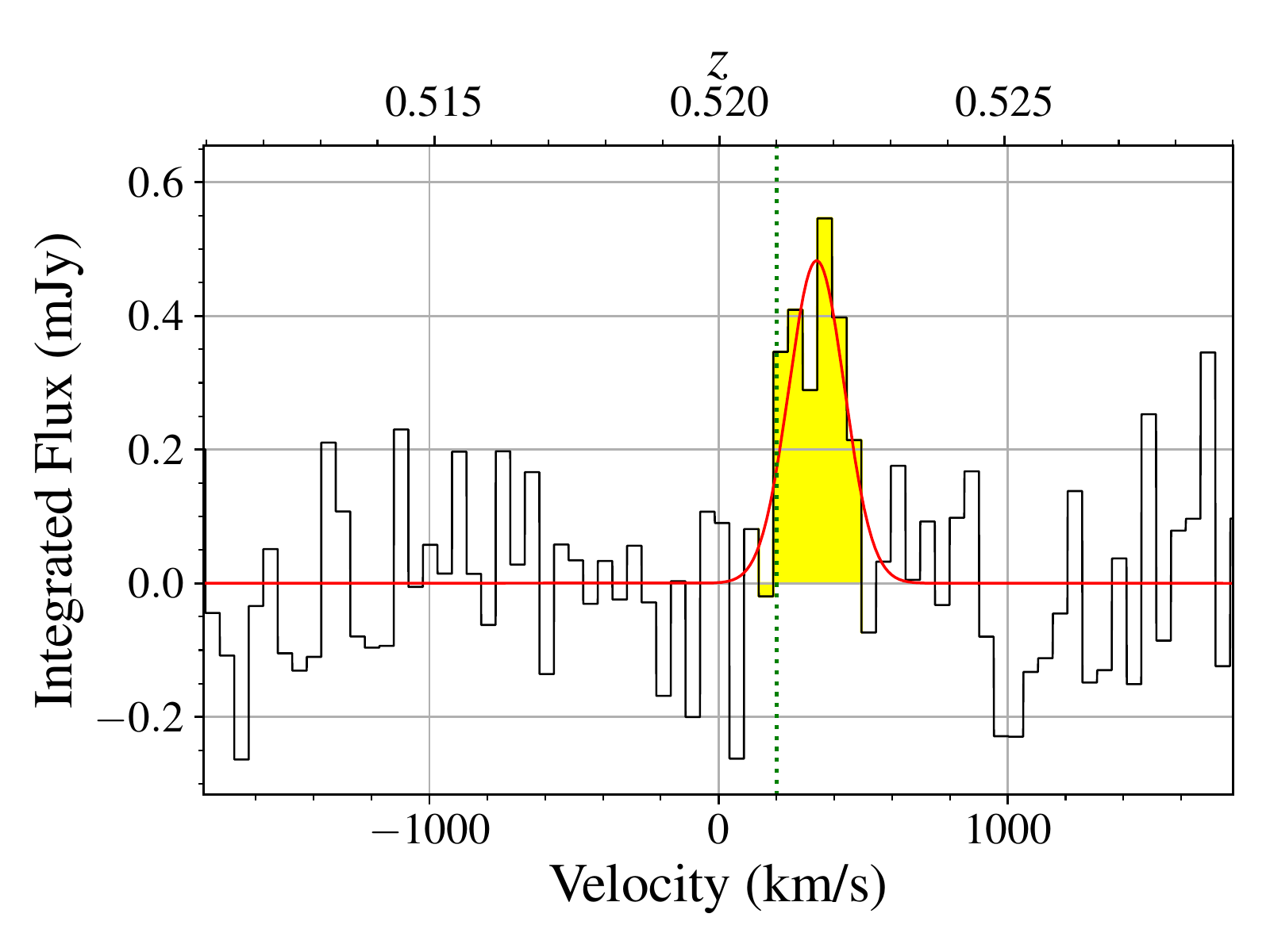}
			
	\includegraphics[scale=0.3]{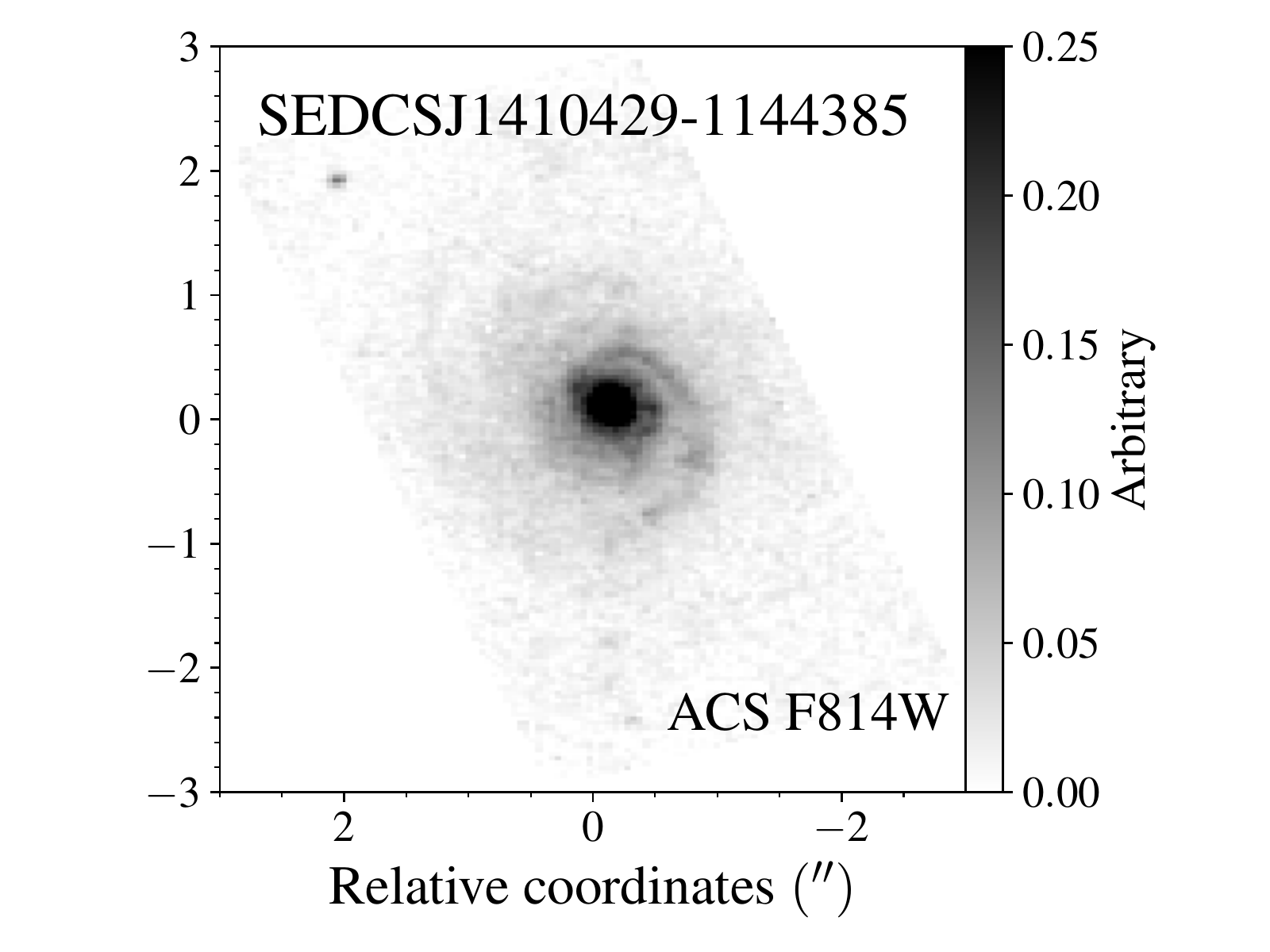}
	\includegraphics[scale=0.3]{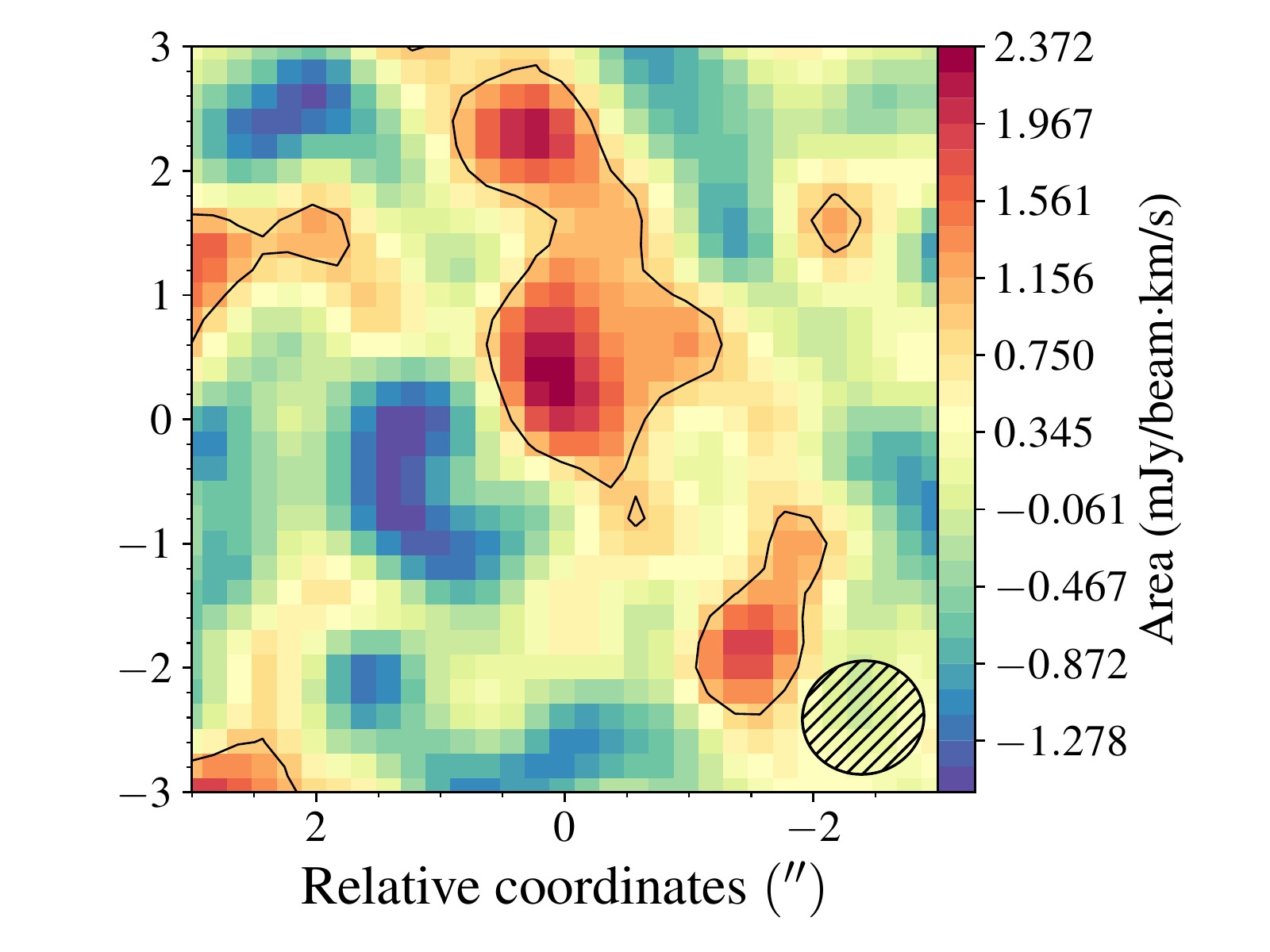}
	\includegraphics[scale=0.3]{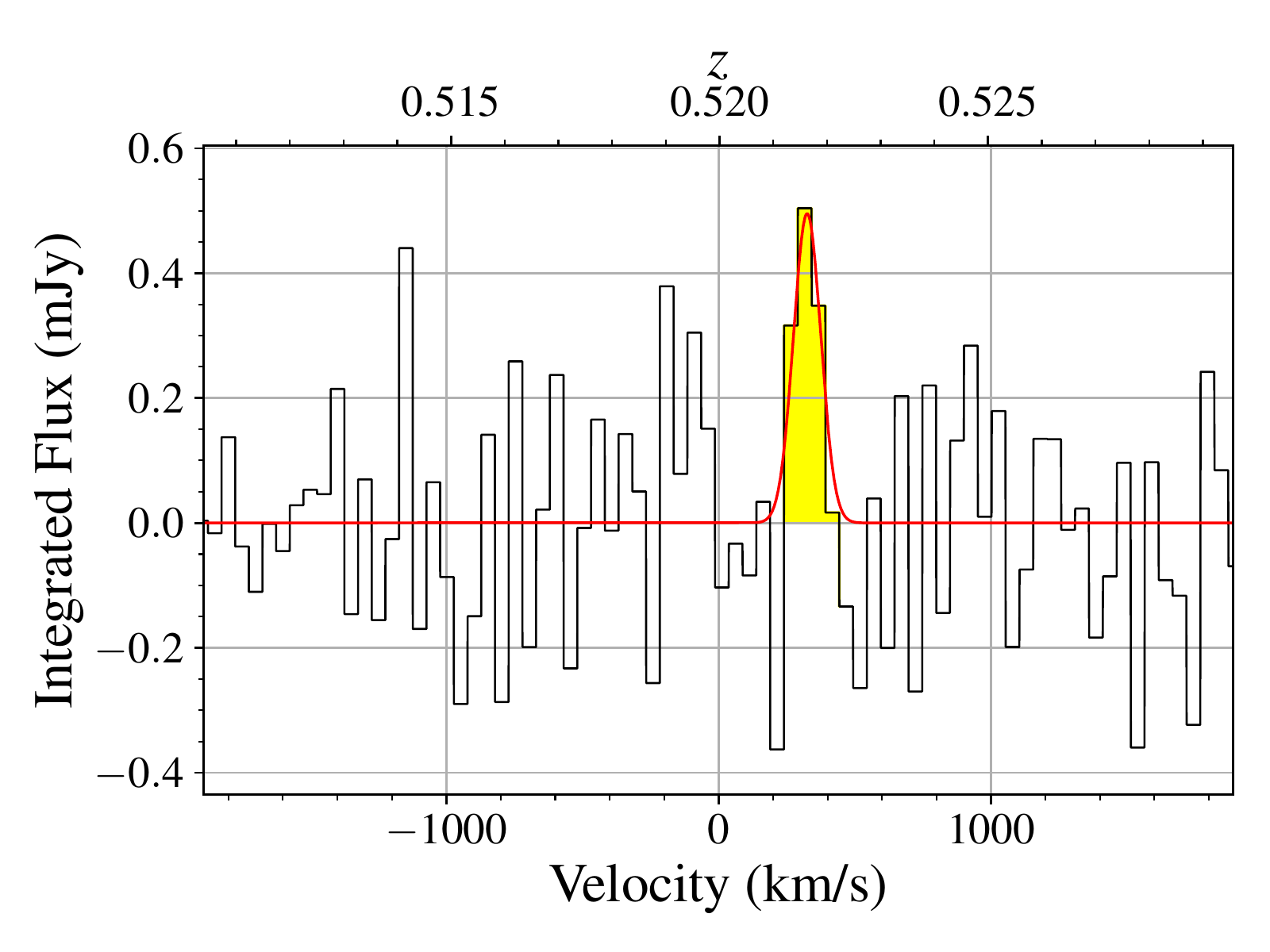}

	\includegraphics[scale=0.3]{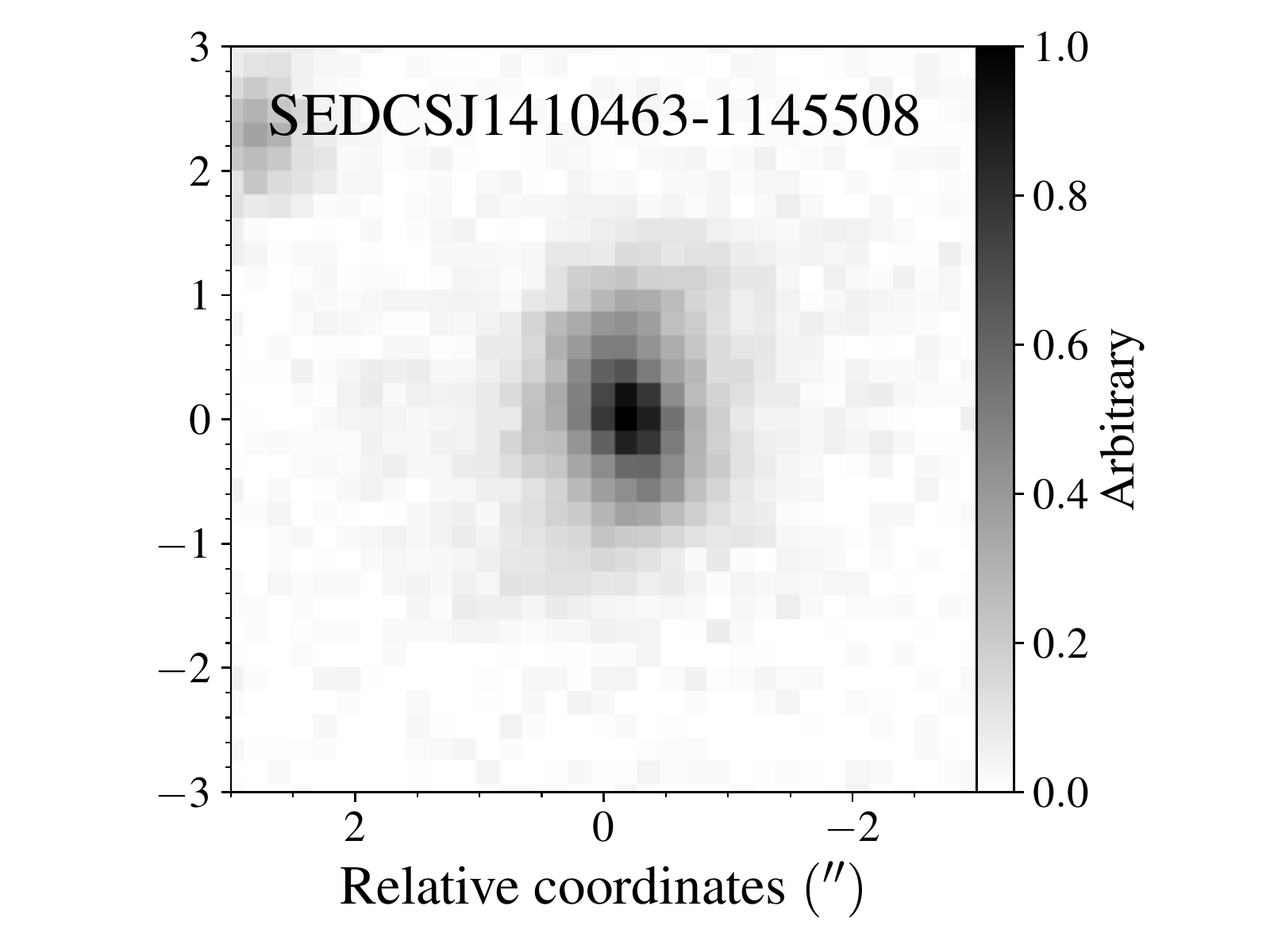}
	\includegraphics[scale=0.3]{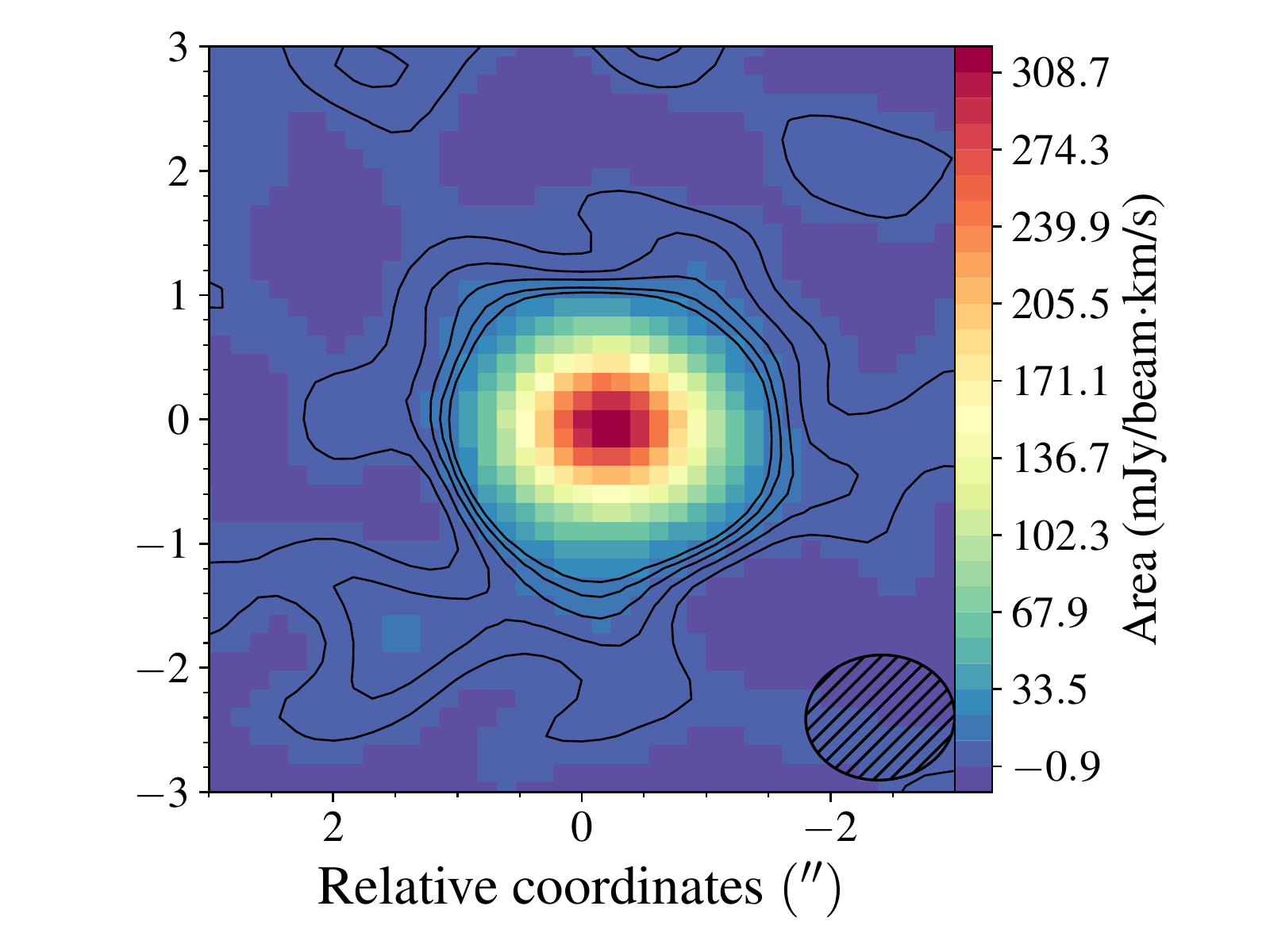}
	\includegraphics[scale=0.3]{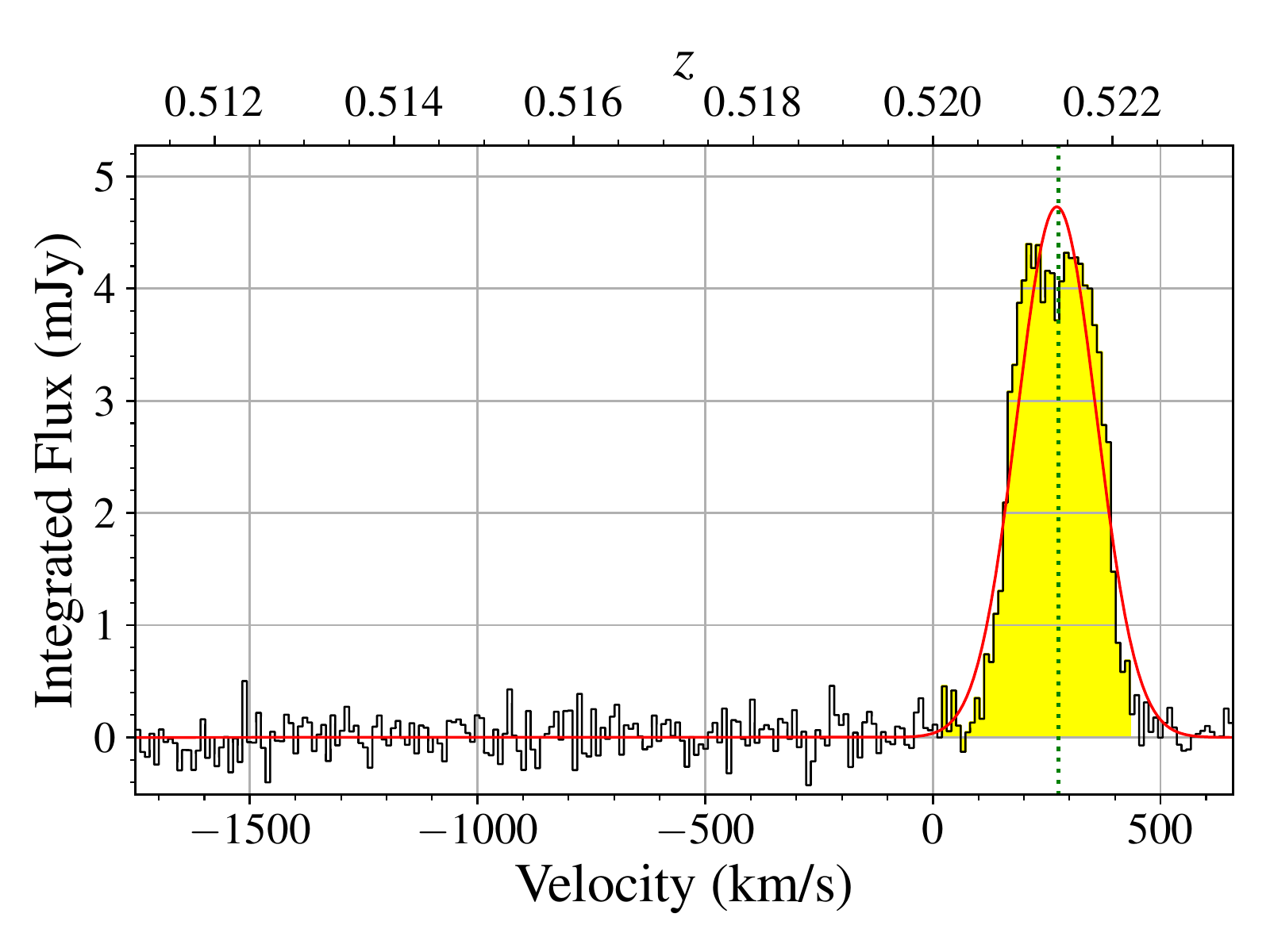}

	\includegraphics[scale=0.3]{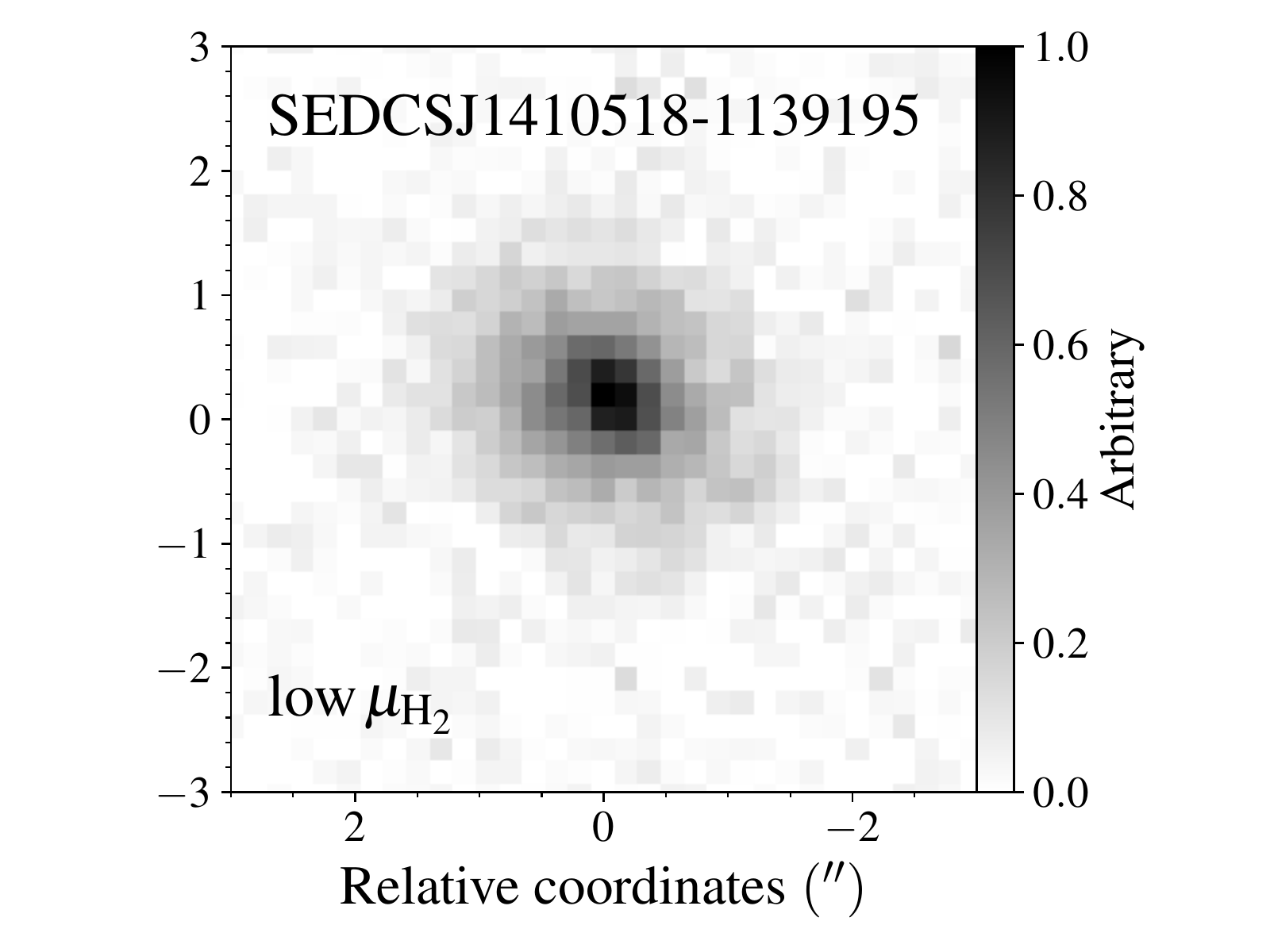}
	\includegraphics[scale=0.3]{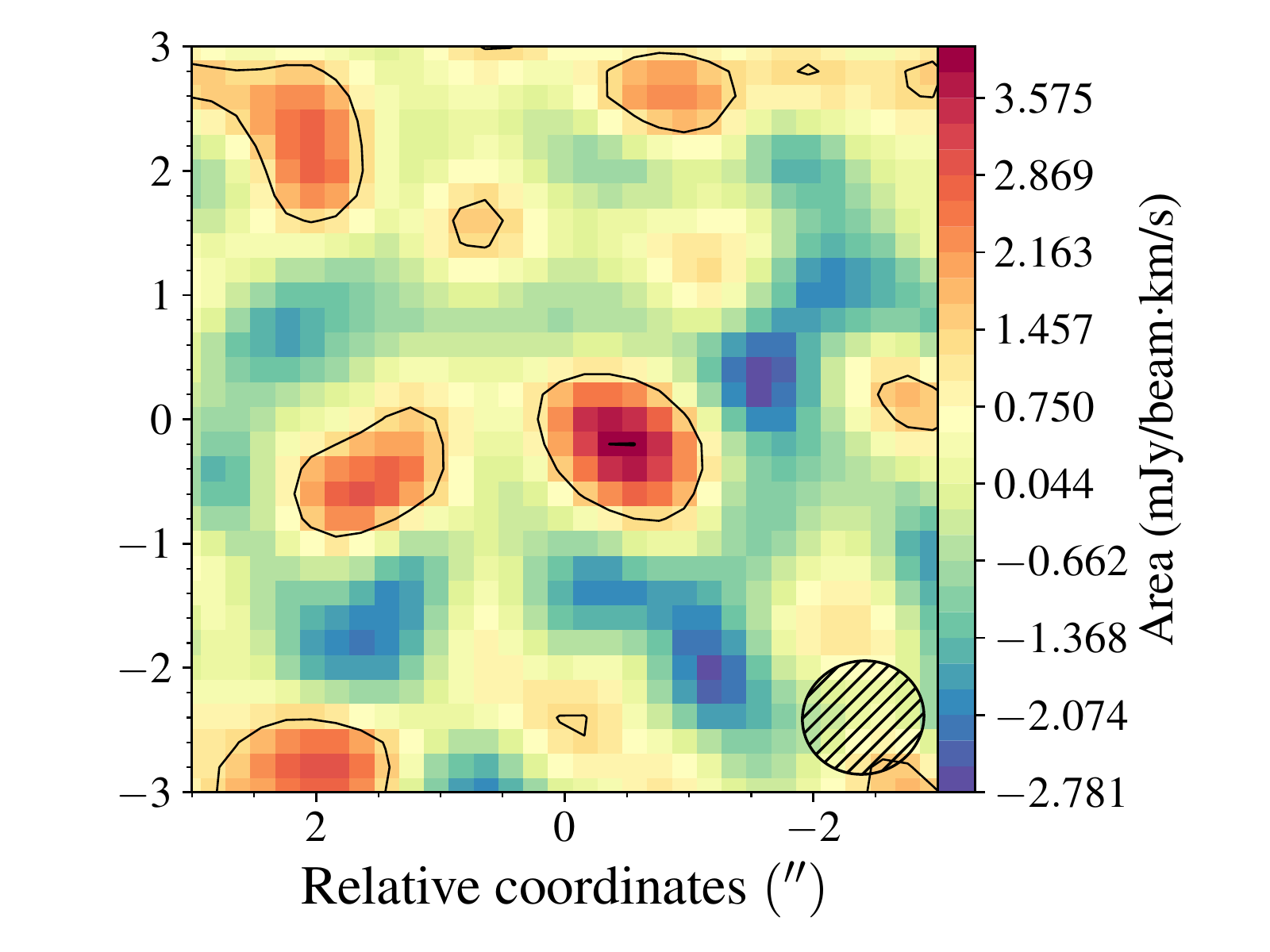}
	\includegraphics[scale=0.3]{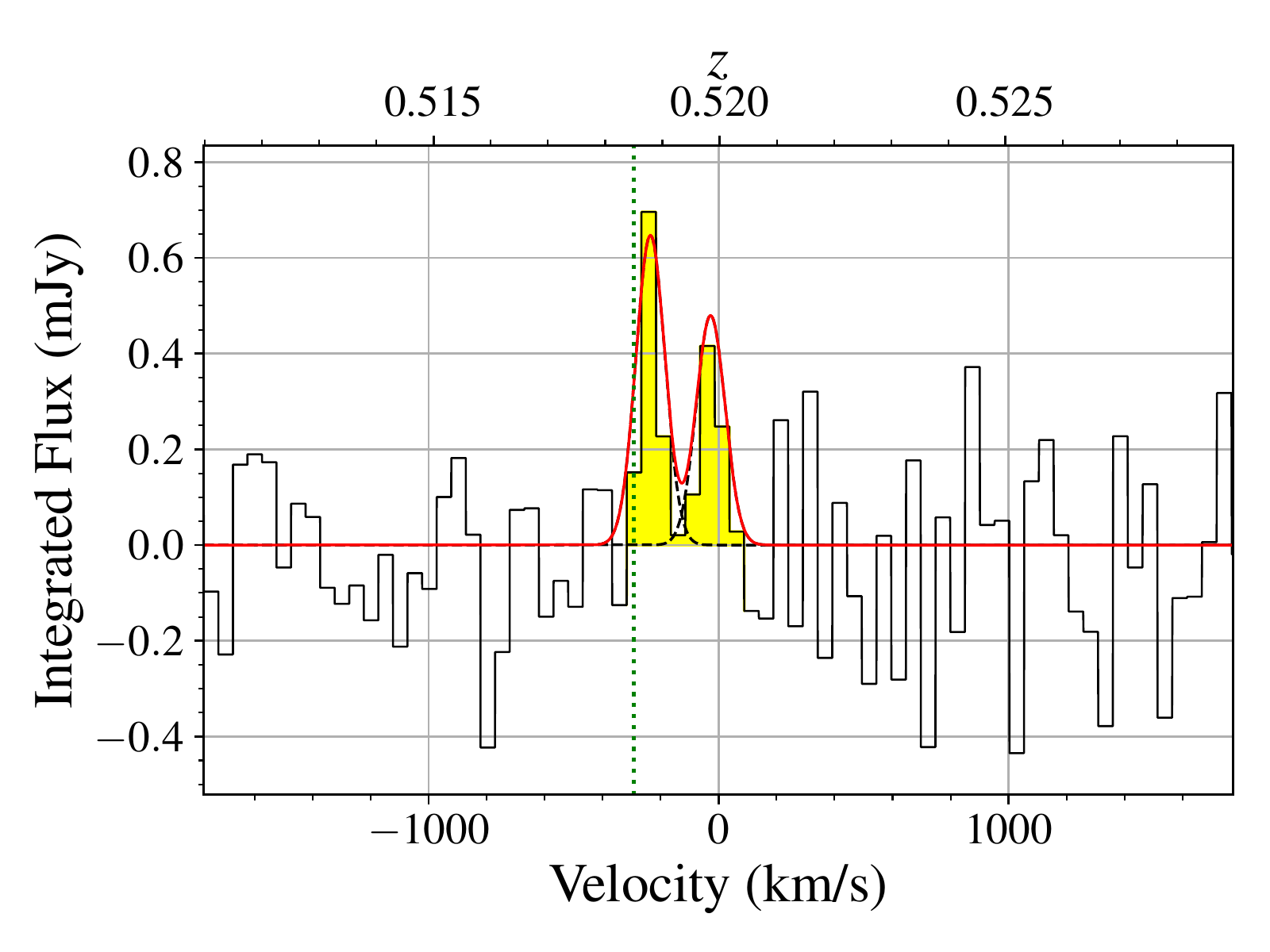}

	\includegraphics[scale=0.3]{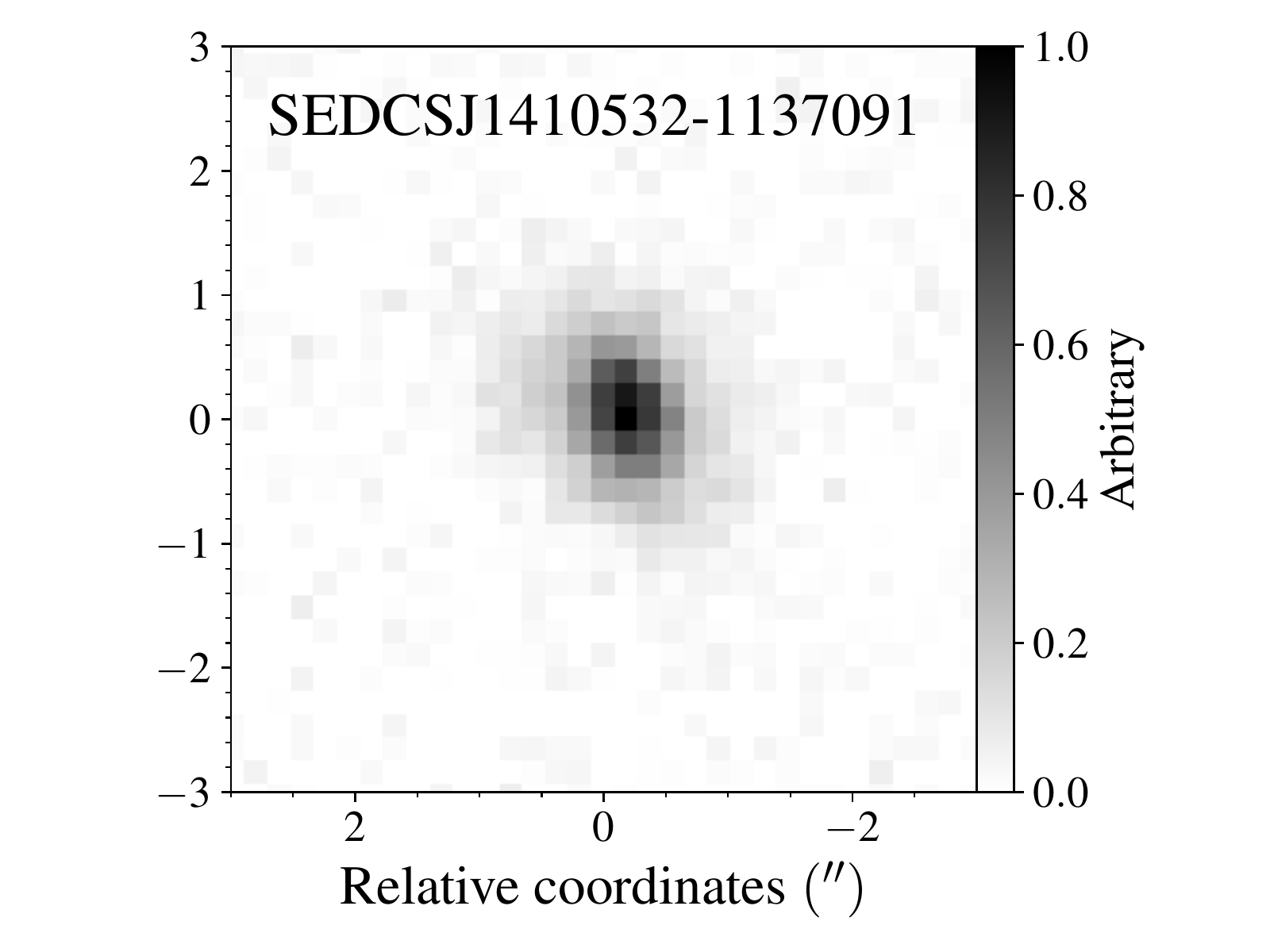}
	\includegraphics[scale=0.3]{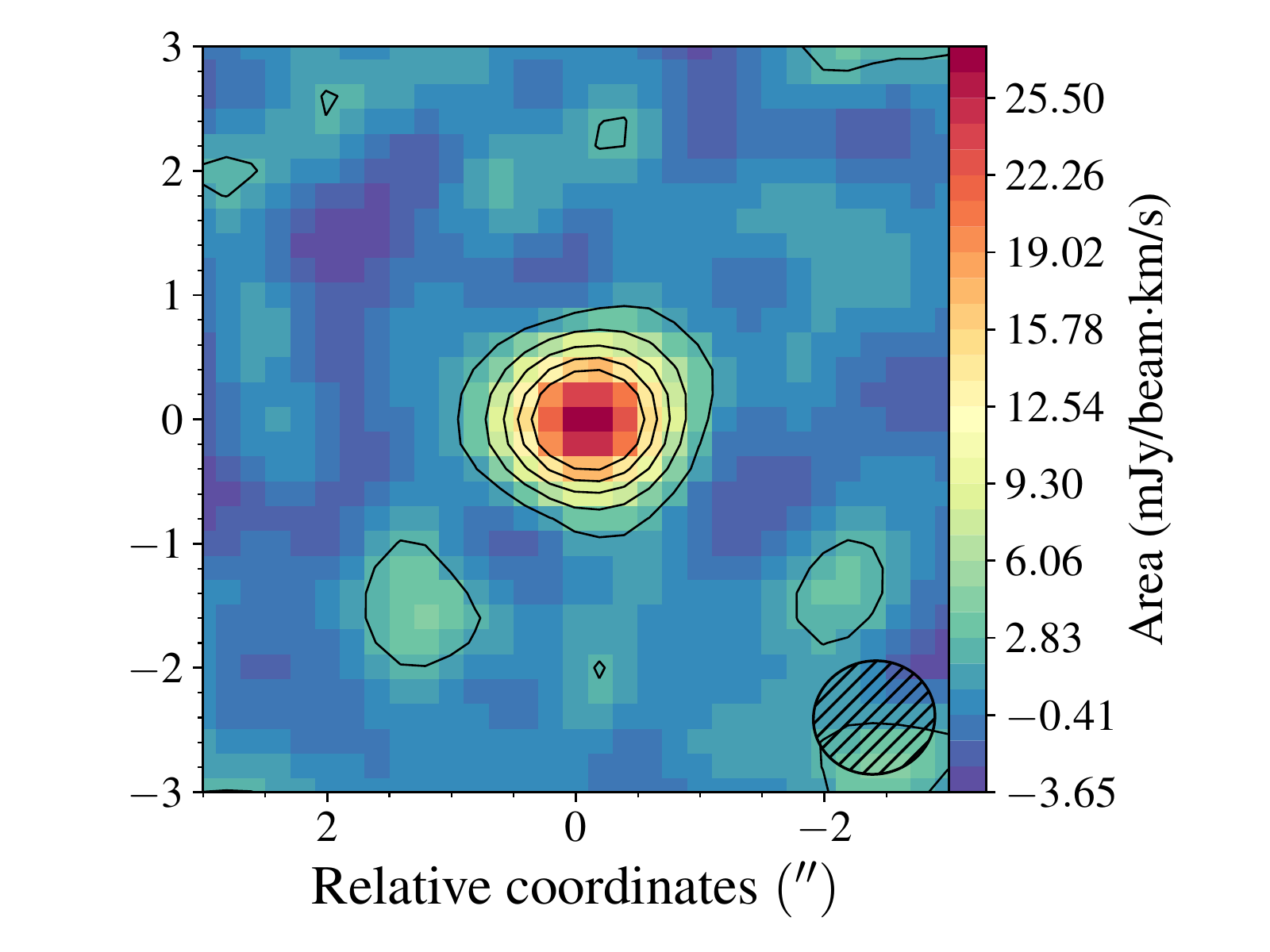}
	\includegraphics[scale=0.3]{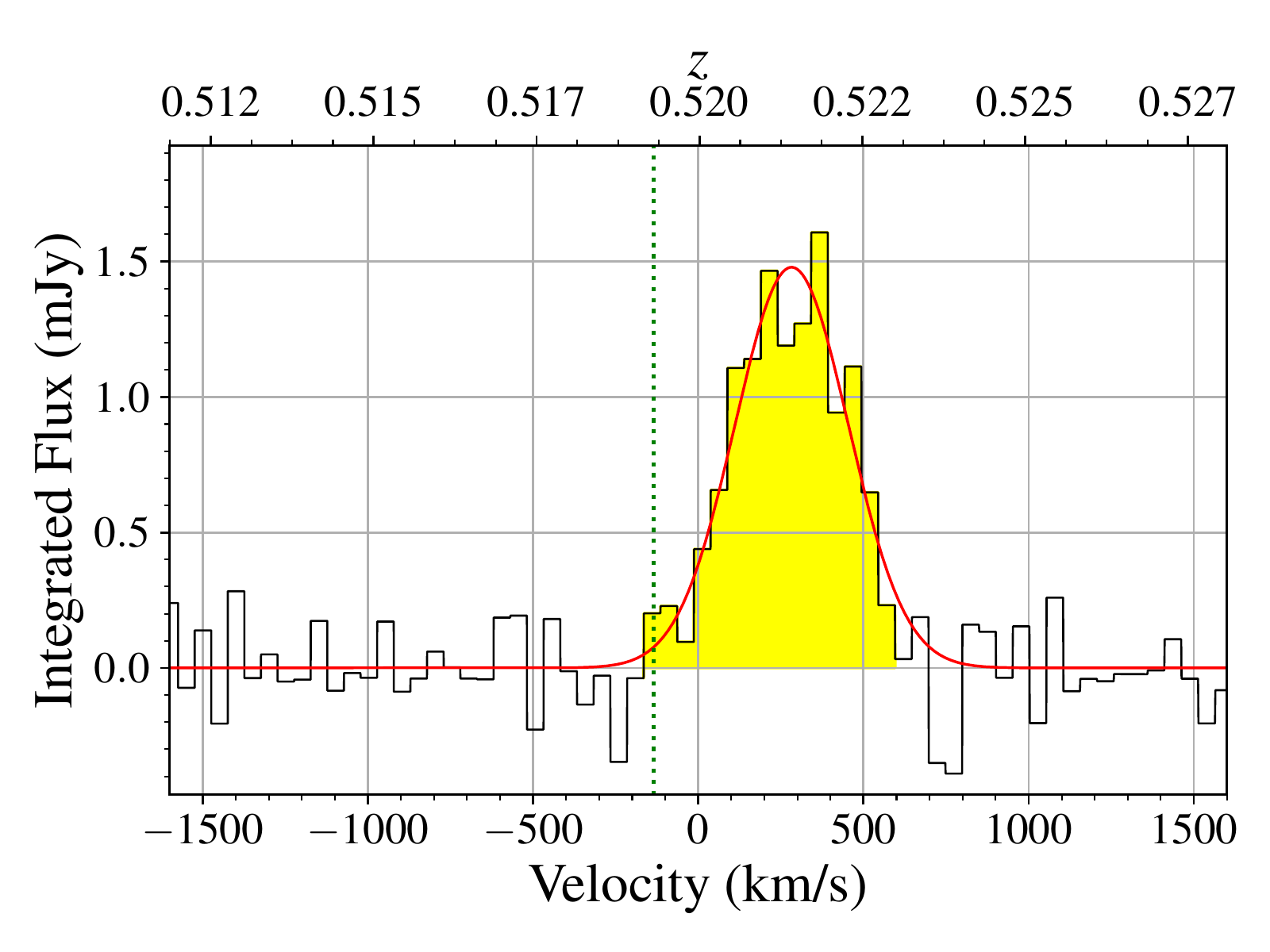}

	\includegraphics[scale=0.3]{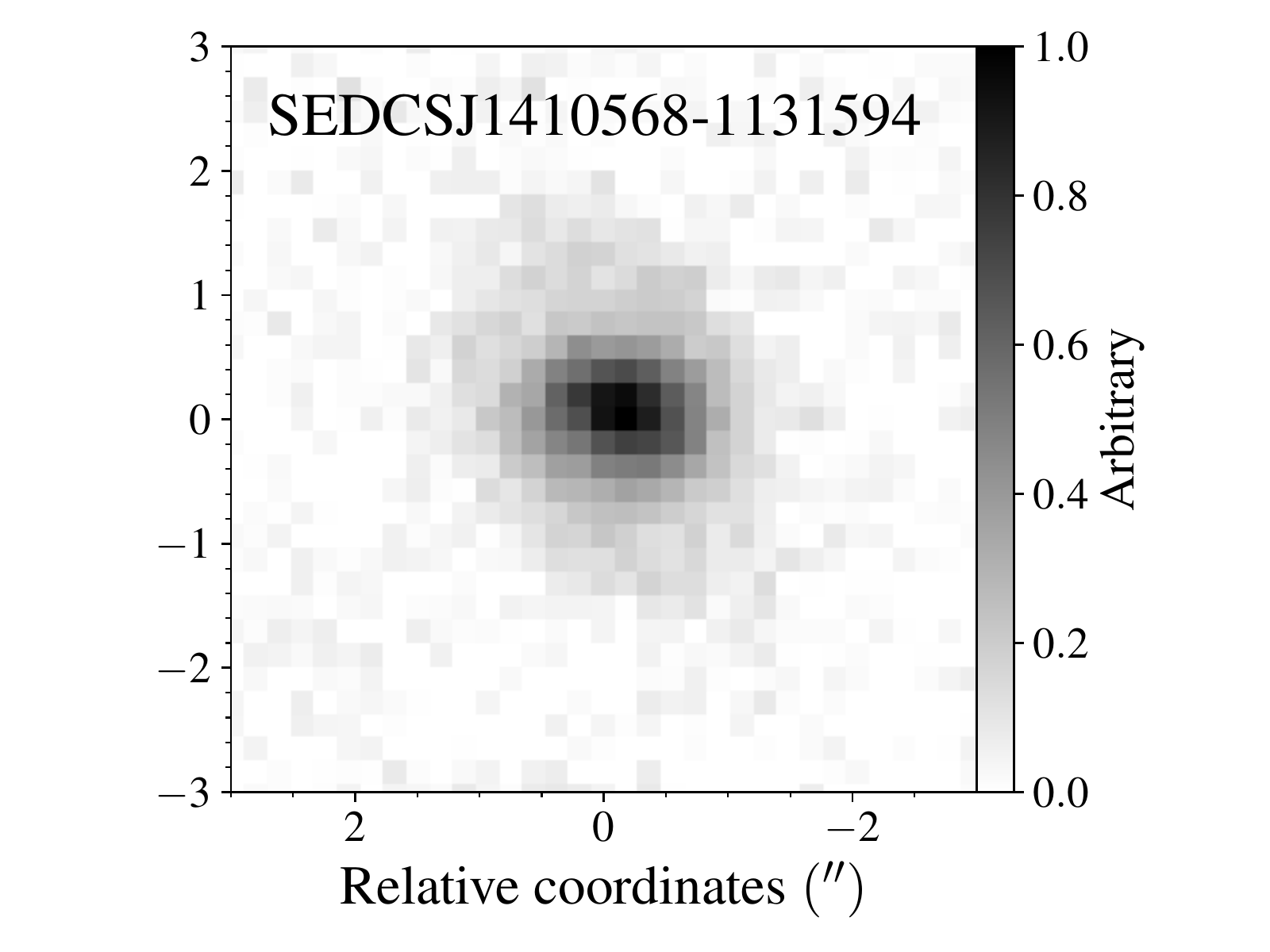}
	\includegraphics[scale=0.3]{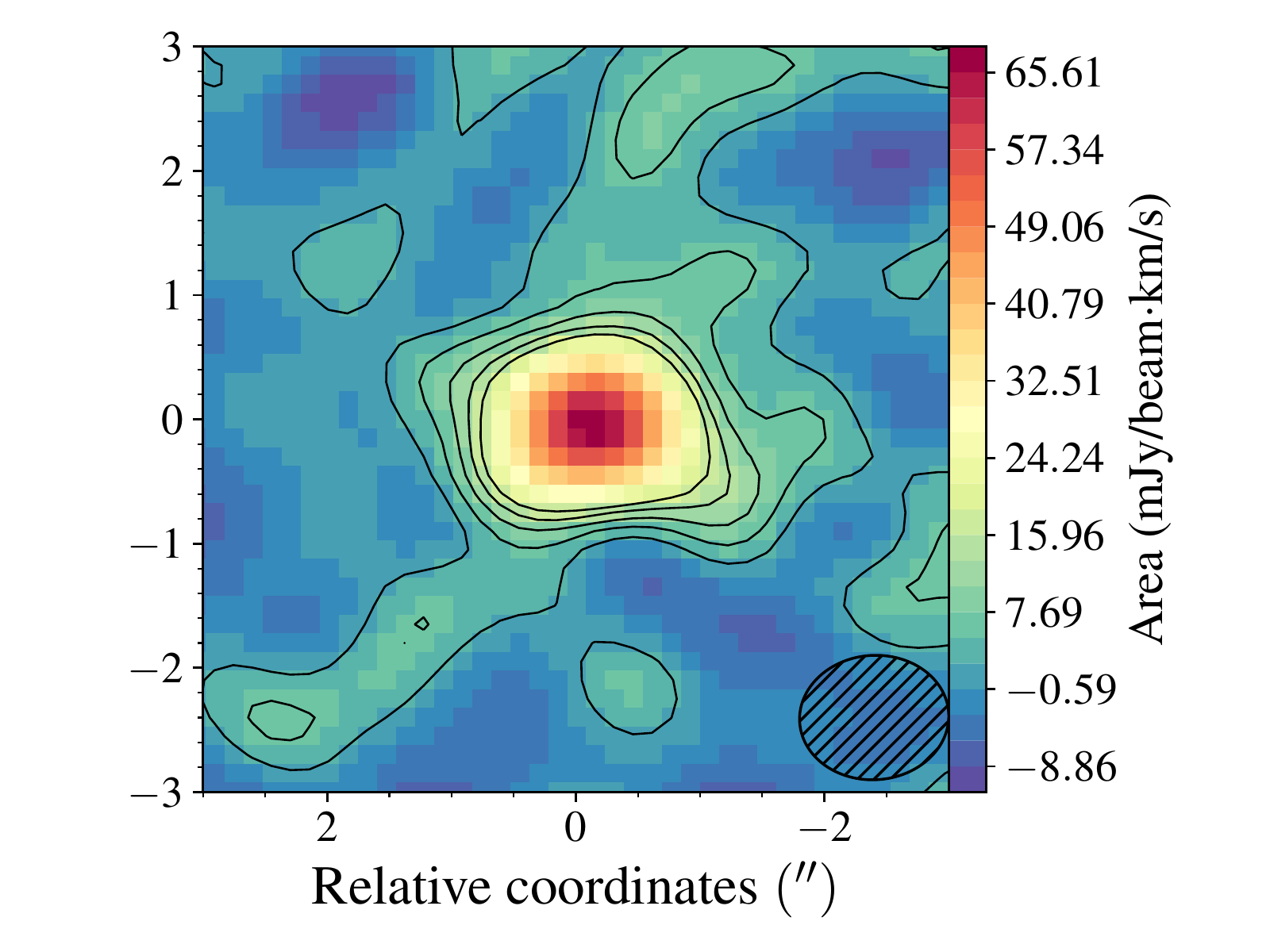}
	\includegraphics[scale=0.3]{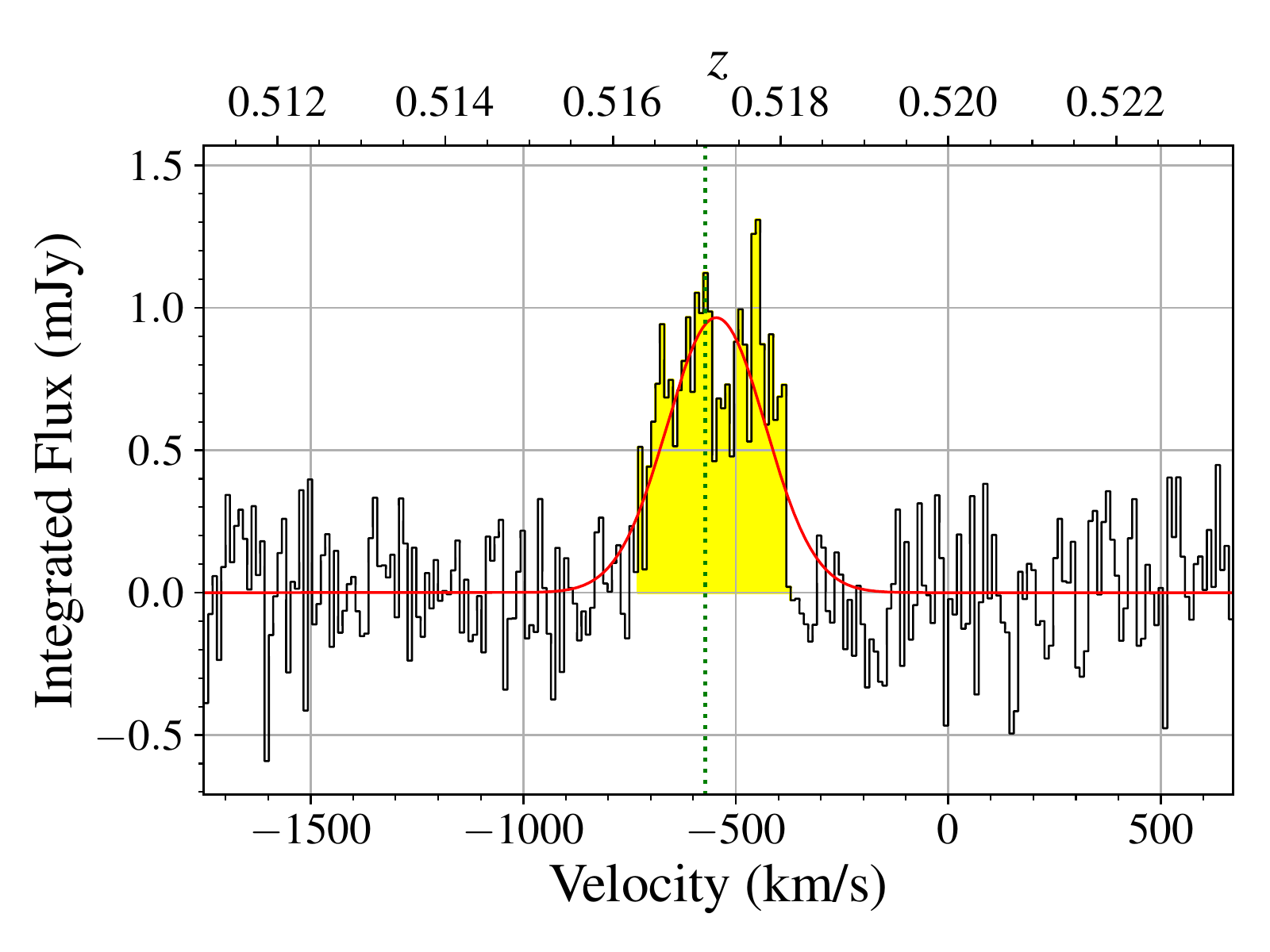}

    \caption{Continued.}
\end{figure*}

\begin{figure*}[htbp]\ContinuedFloat
\centering
	\includegraphics[scale=0.3]{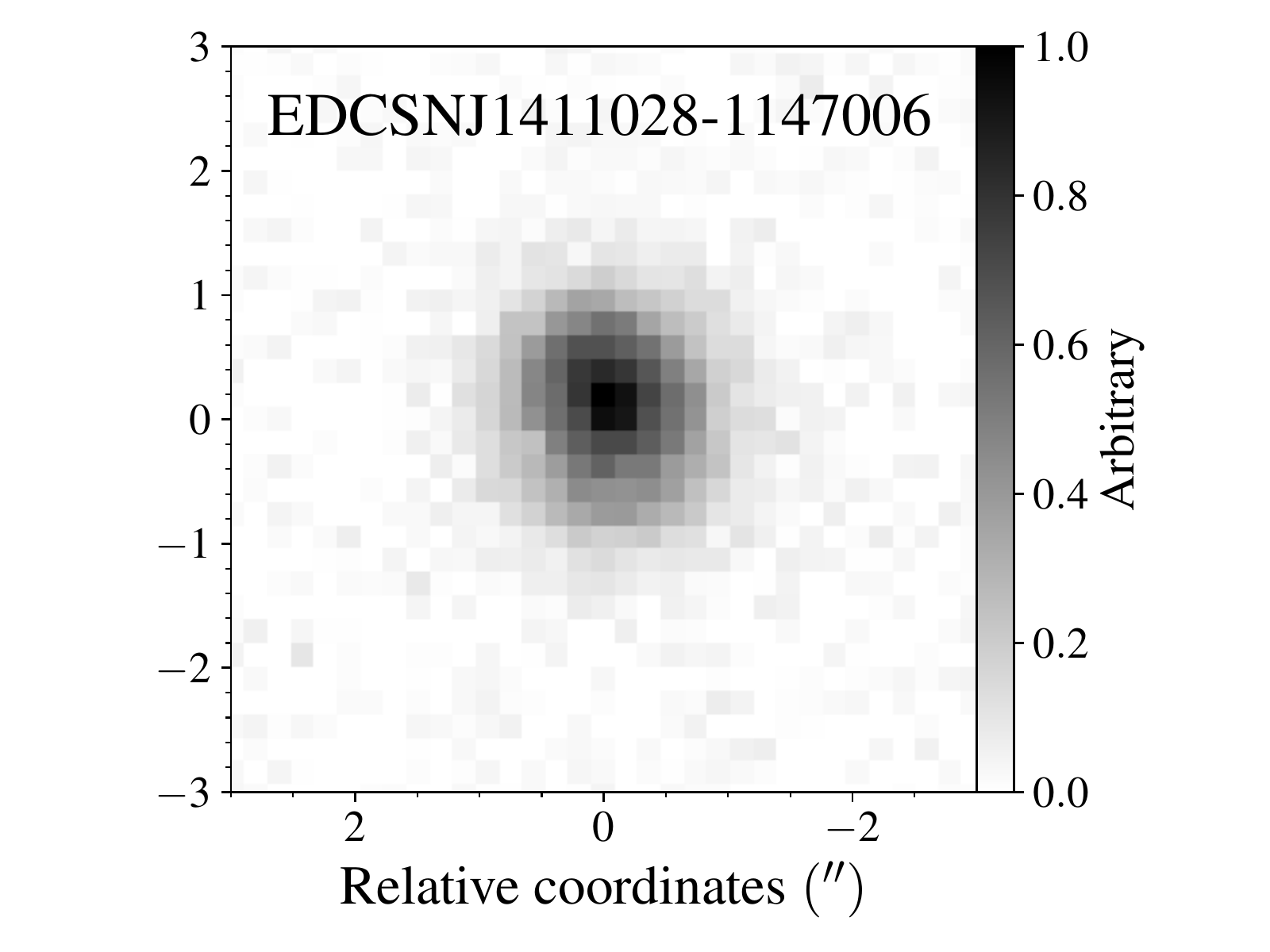}
	\includegraphics[scale=0.3]{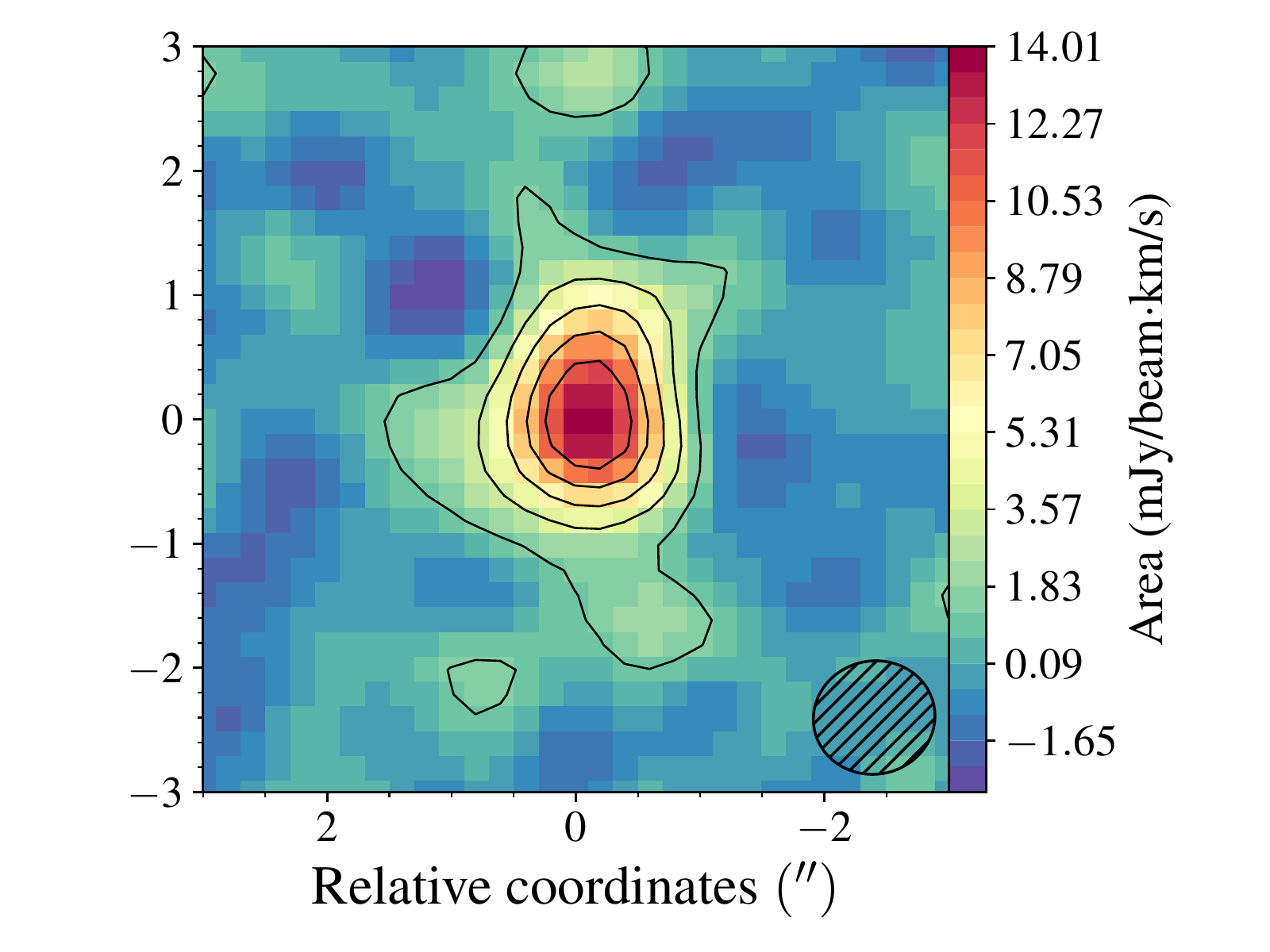}
	\includegraphics[scale=0.3]{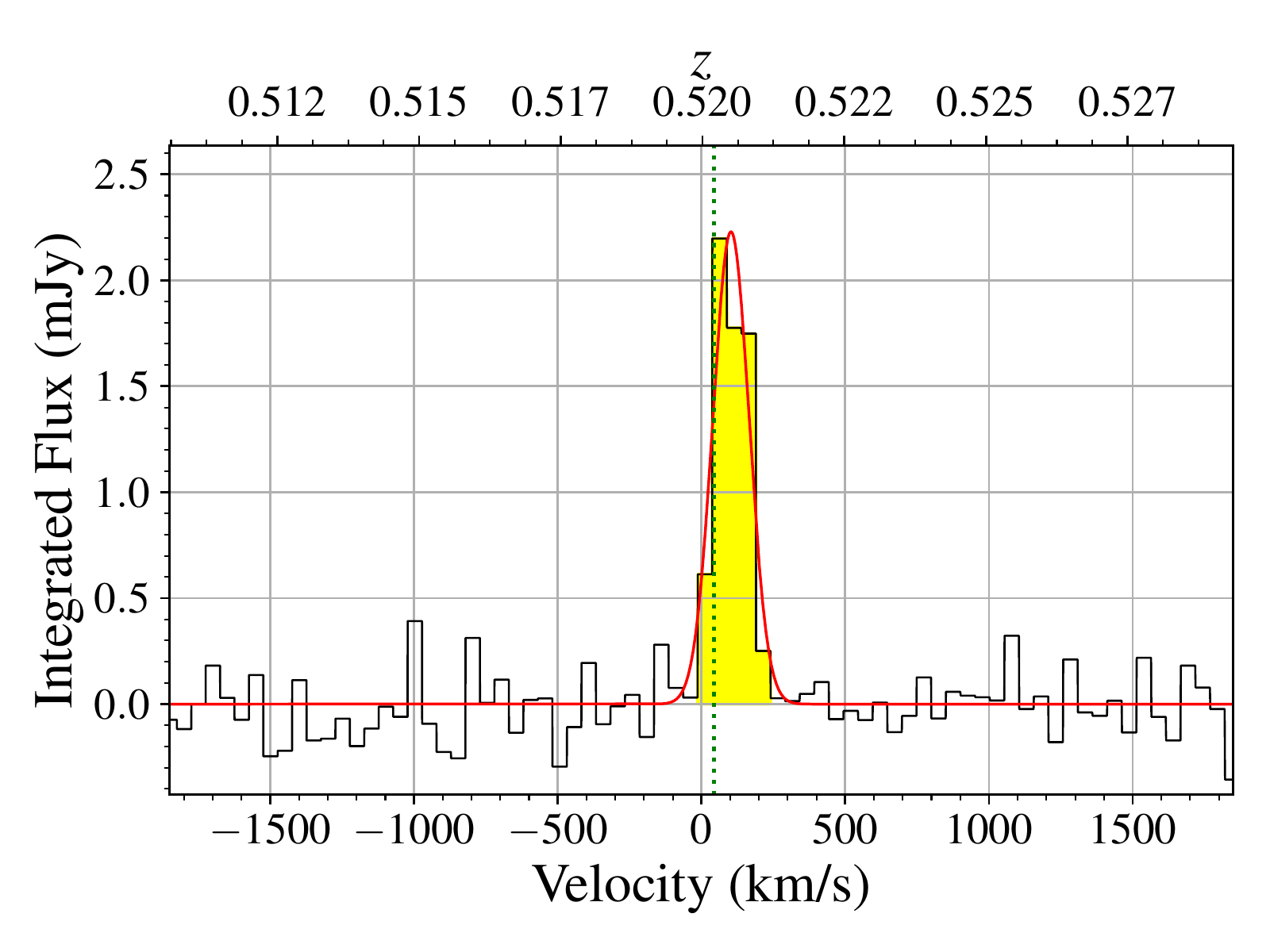}

	\includegraphics[scale=0.3]{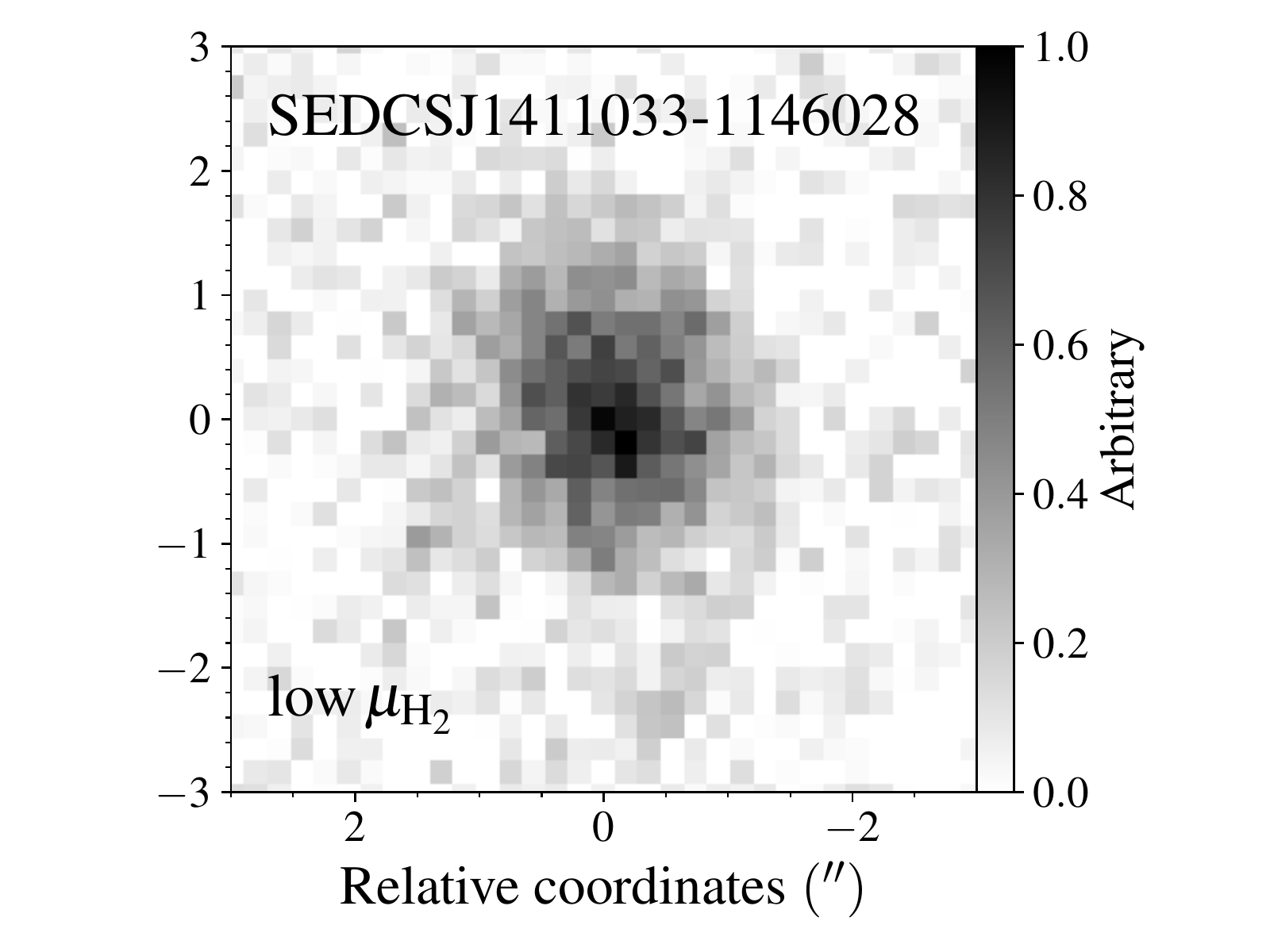}
	\includegraphics[scale=0.3]{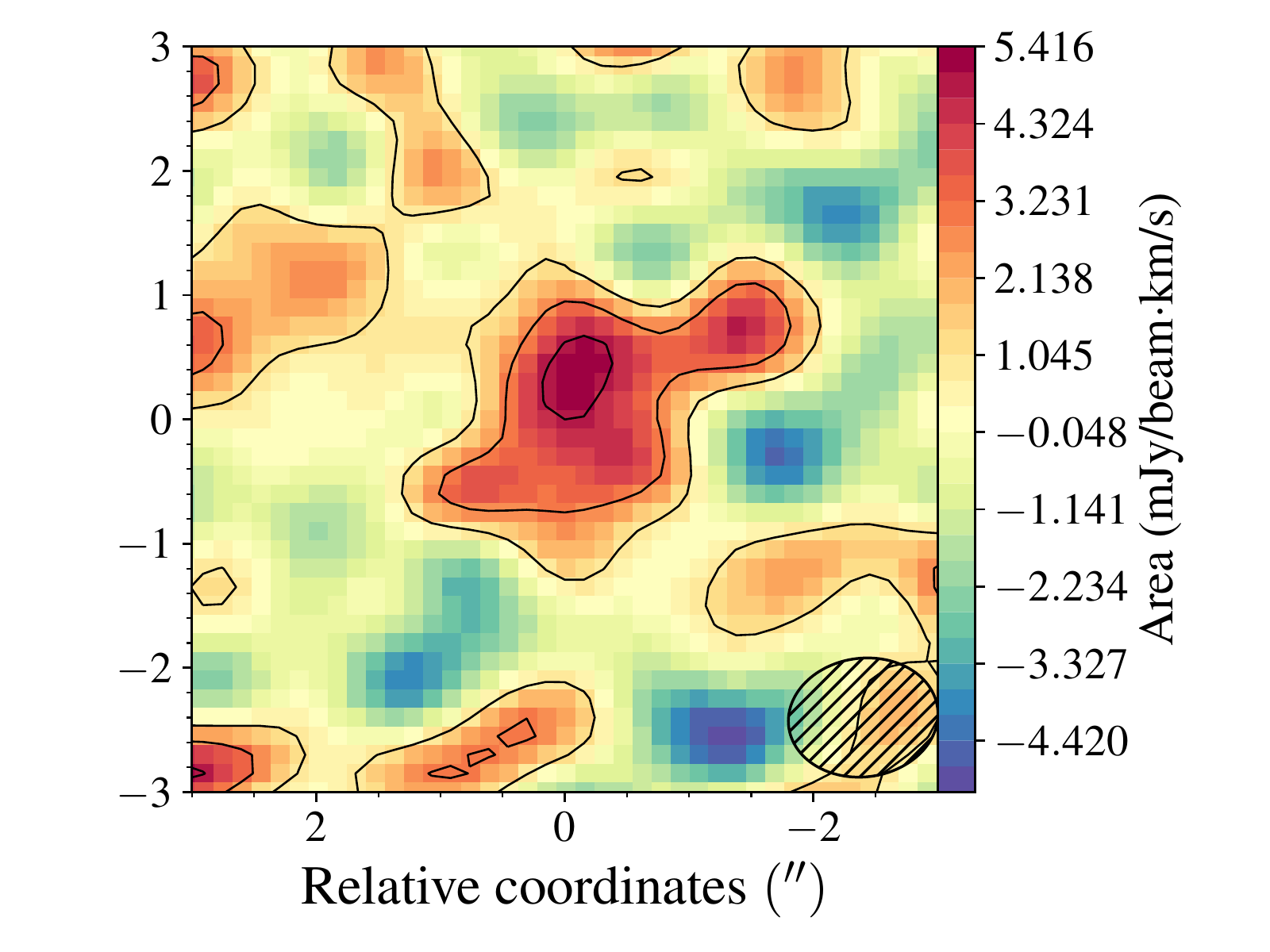}
	\includegraphics[scale=0.3]{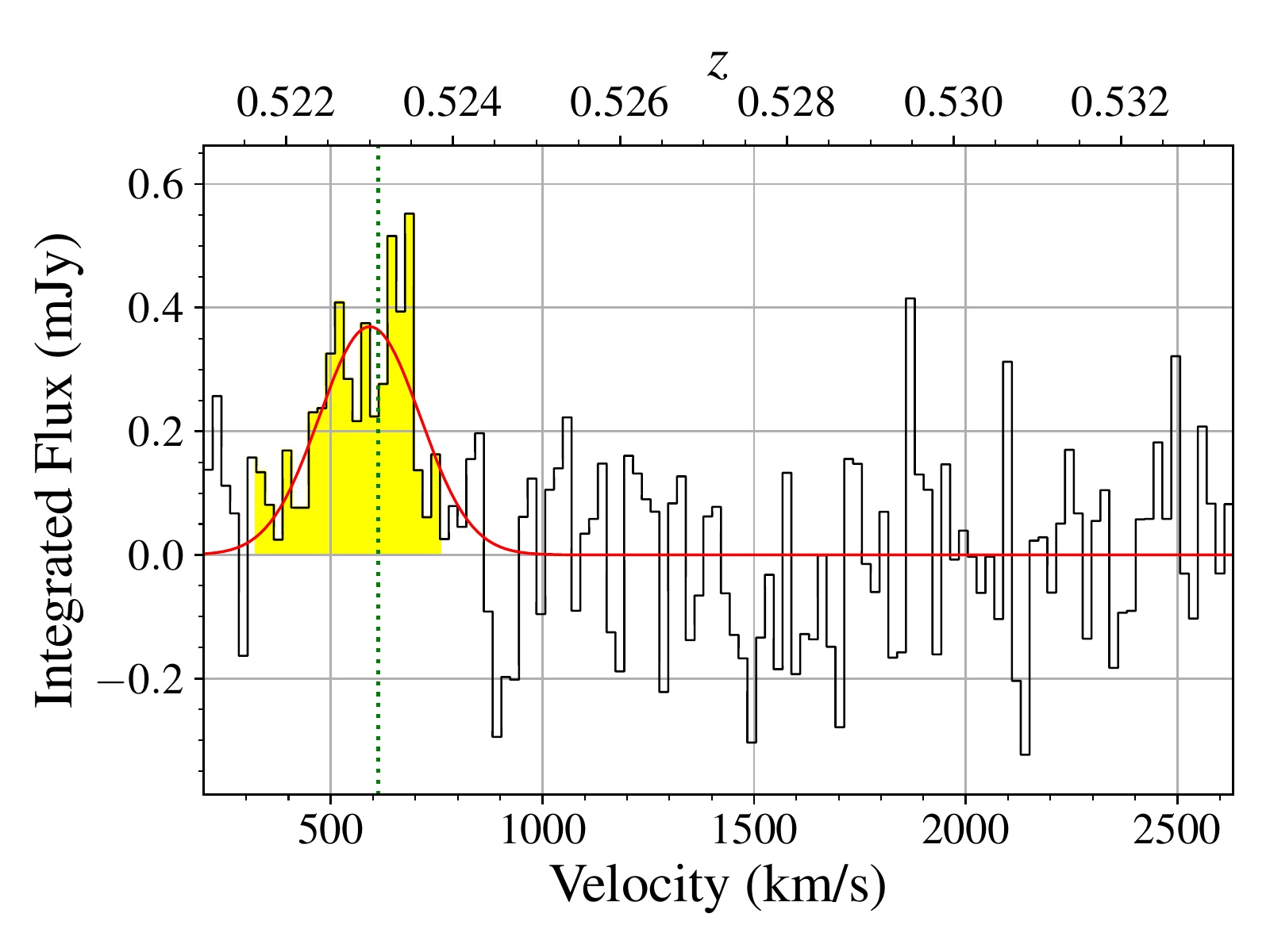}

	\includegraphics[scale=0.3]{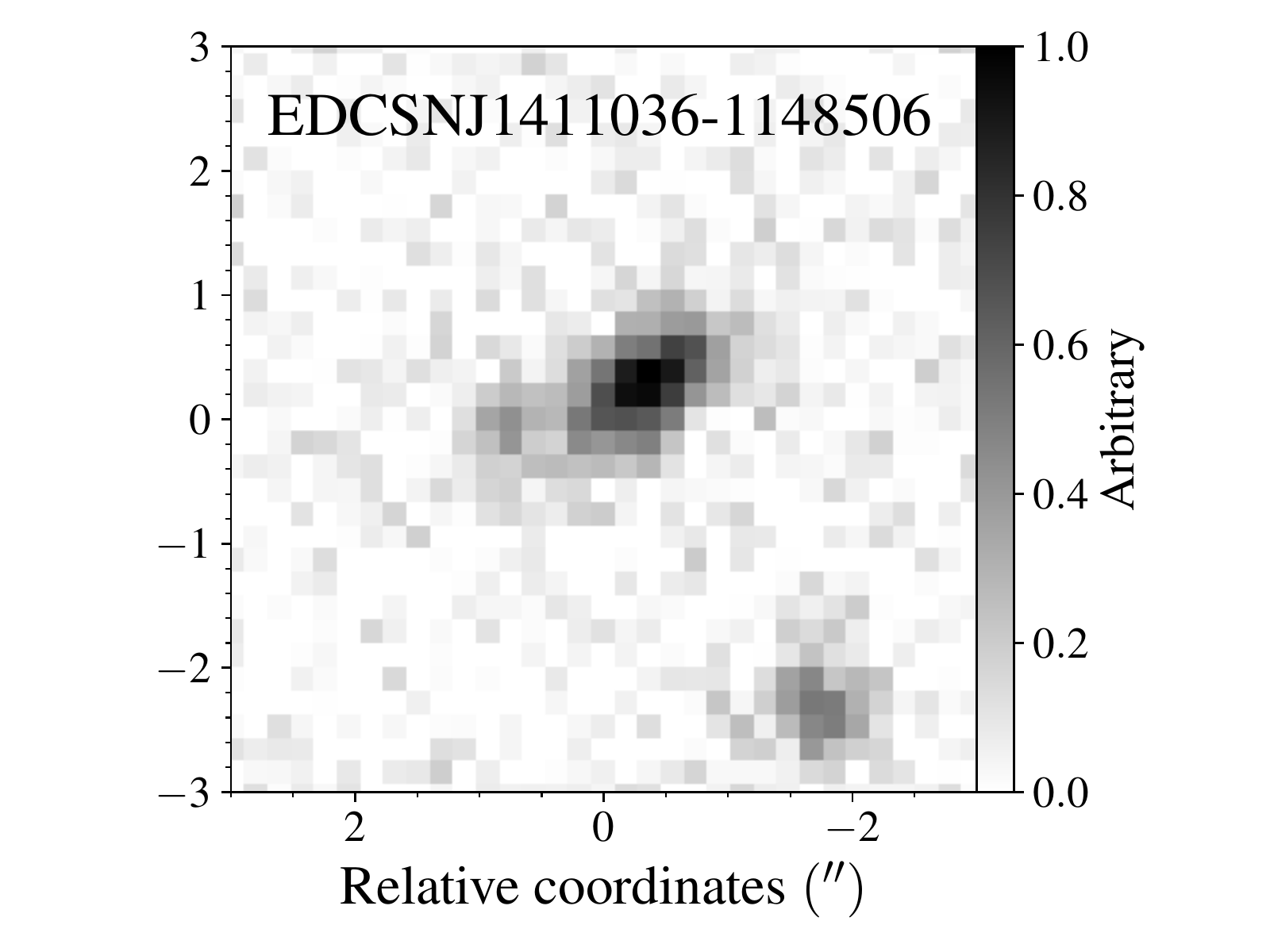}
	\includegraphics[scale=0.3]{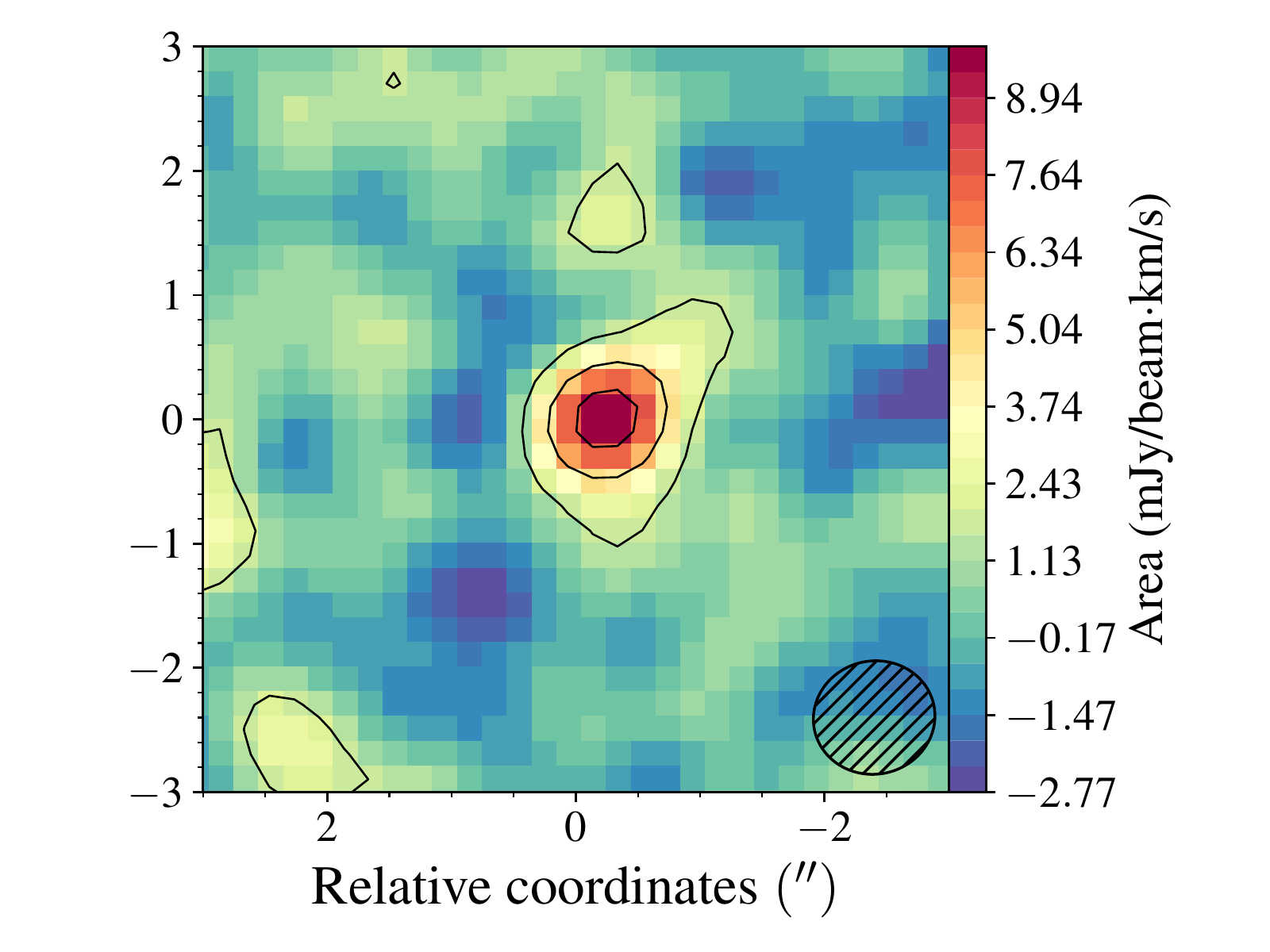}
	\includegraphics[scale=0.3]{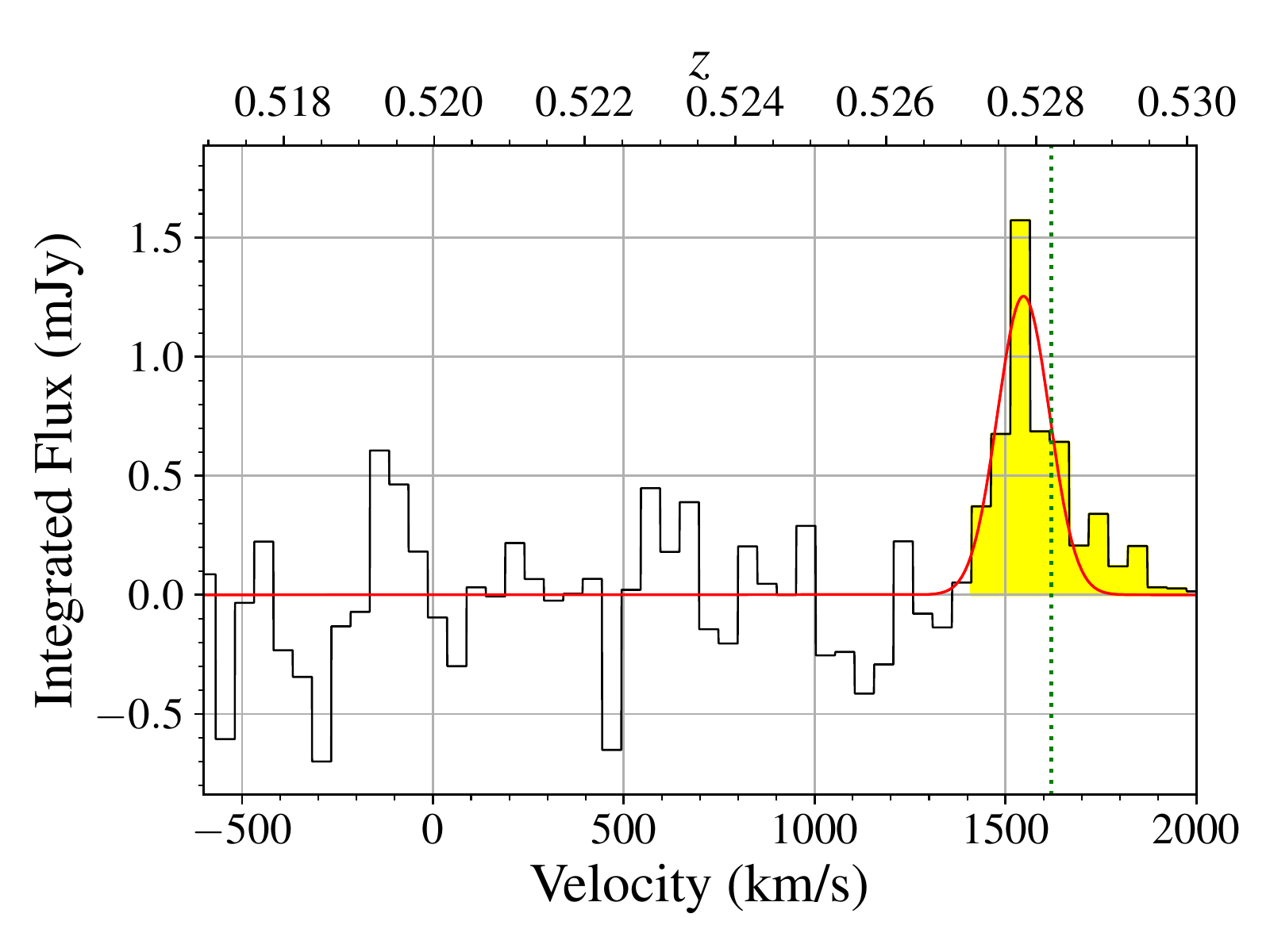}

	\includegraphics[scale=0.3]{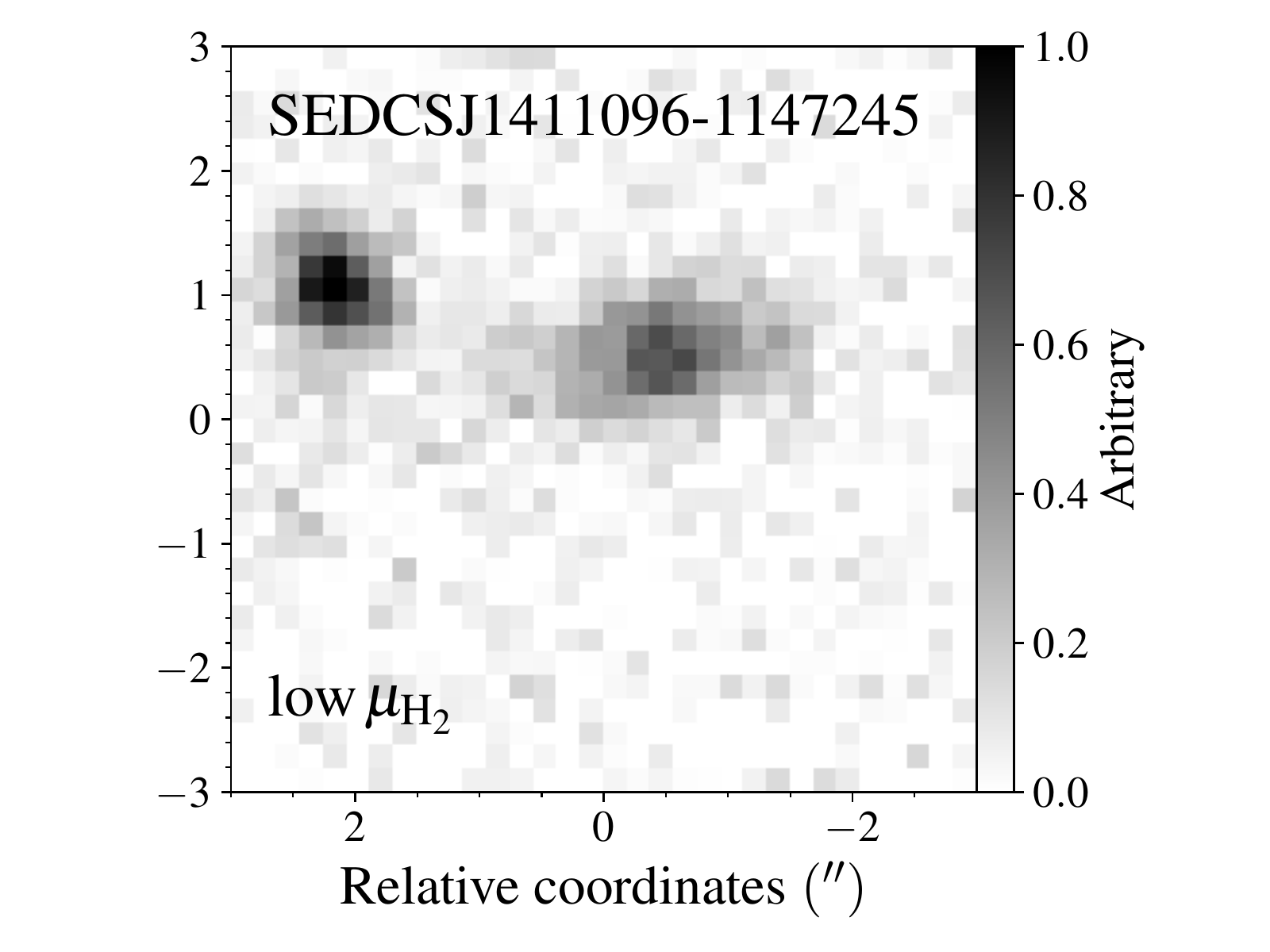}
	\includegraphics[scale=0.3]{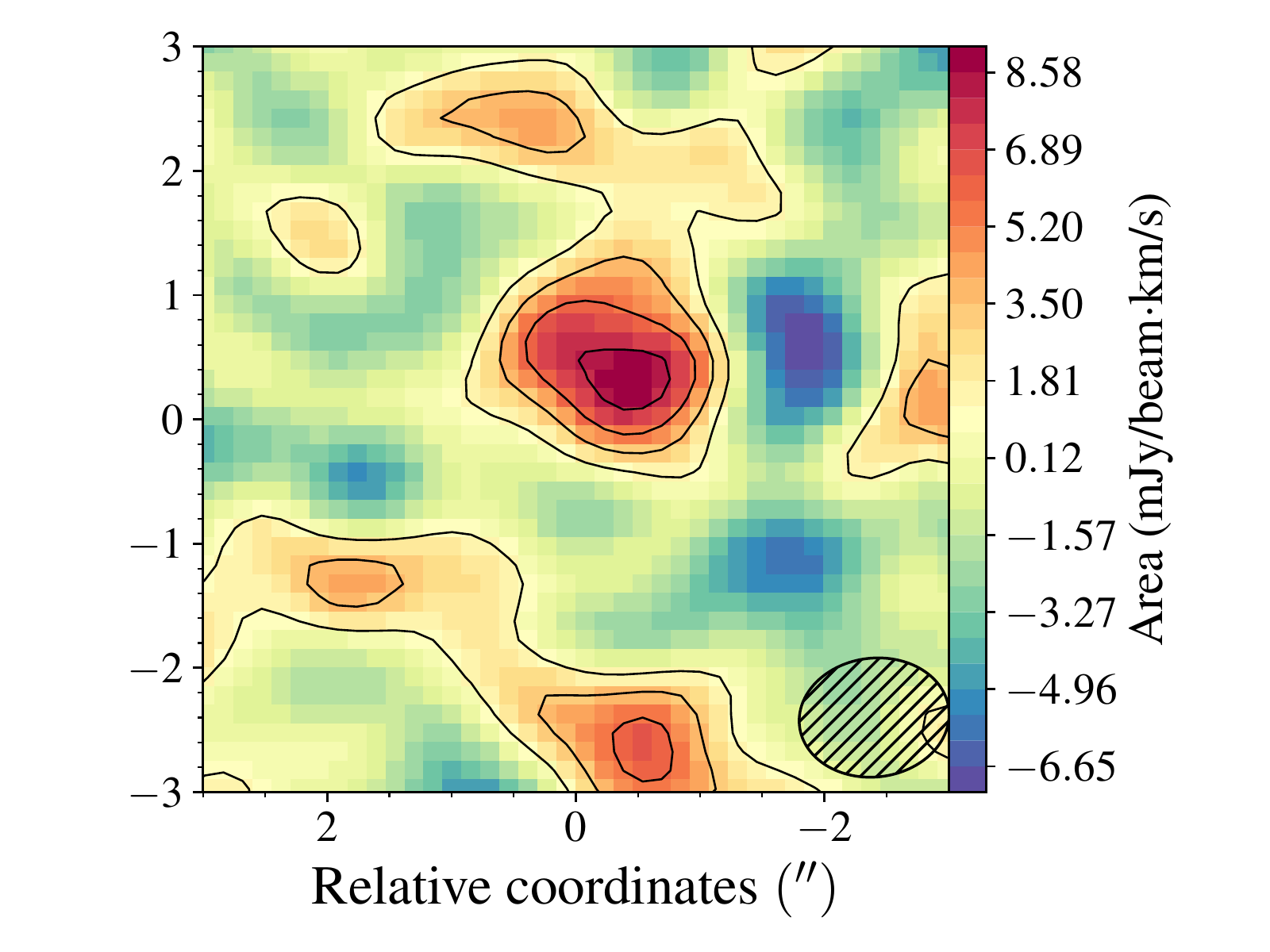}
	\includegraphics[scale=0.3]{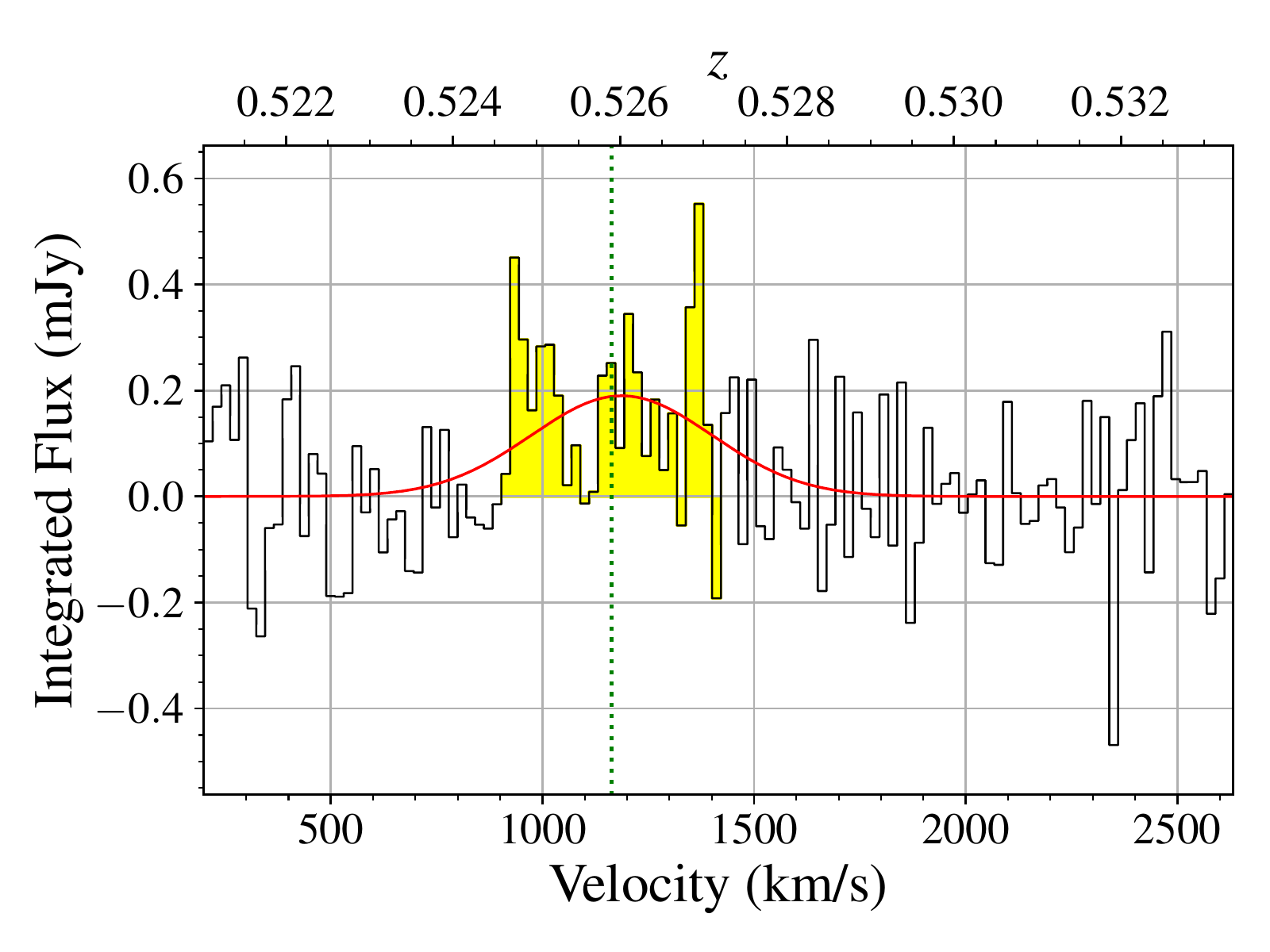}

	\includegraphics[scale=0.3]{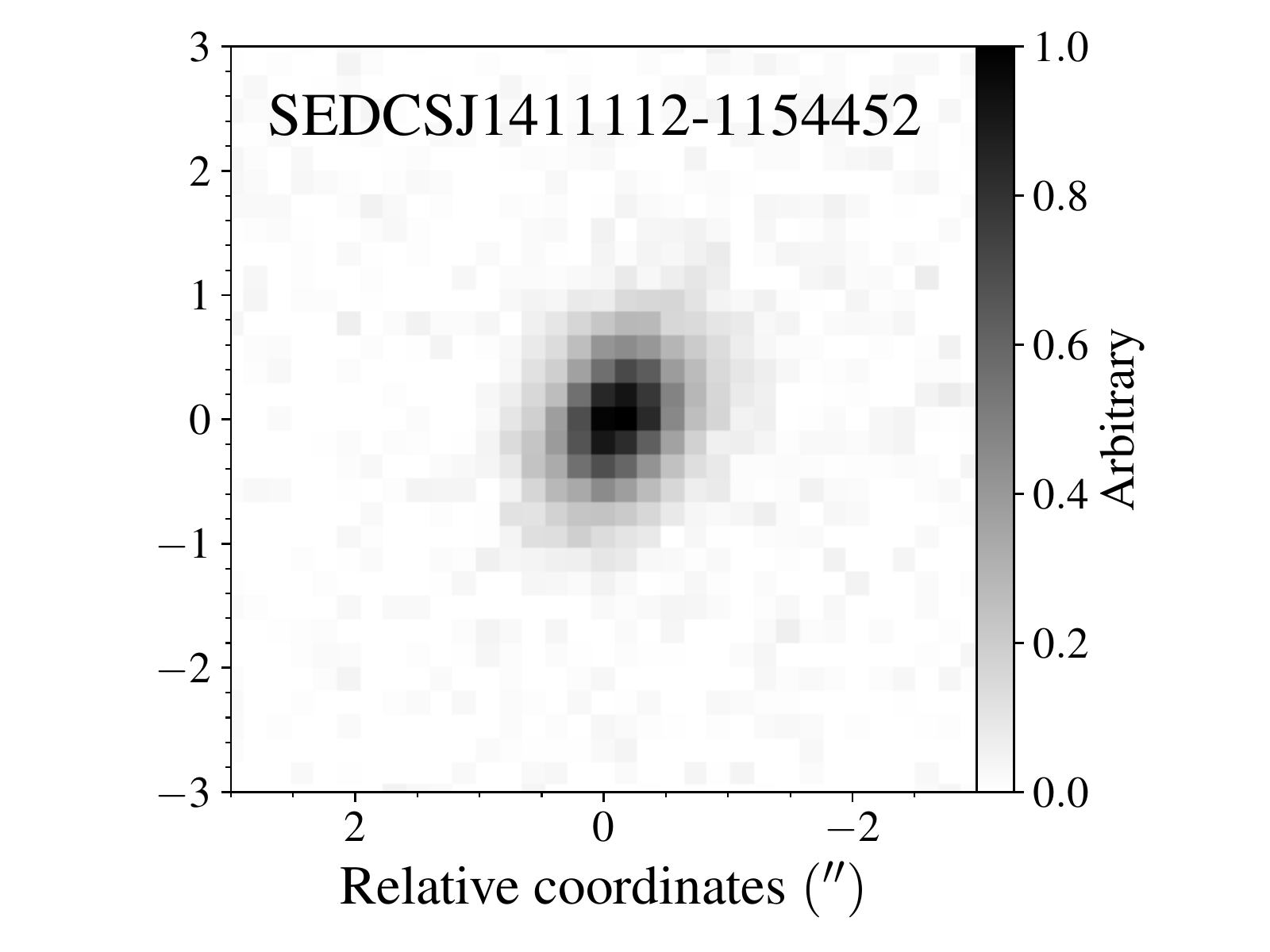}
	\includegraphics[scale=0.3]{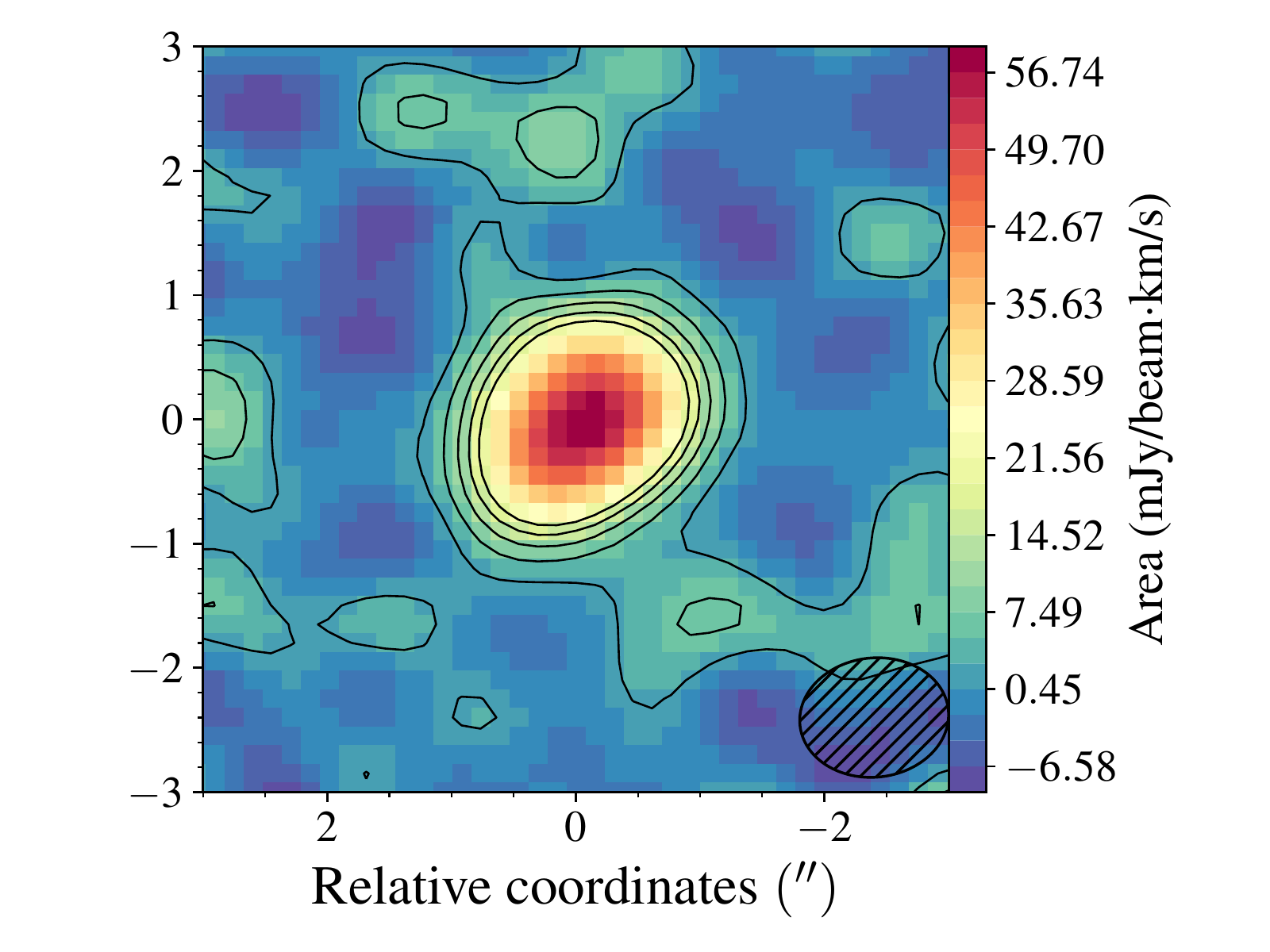}
	\includegraphics[scale=0.3]{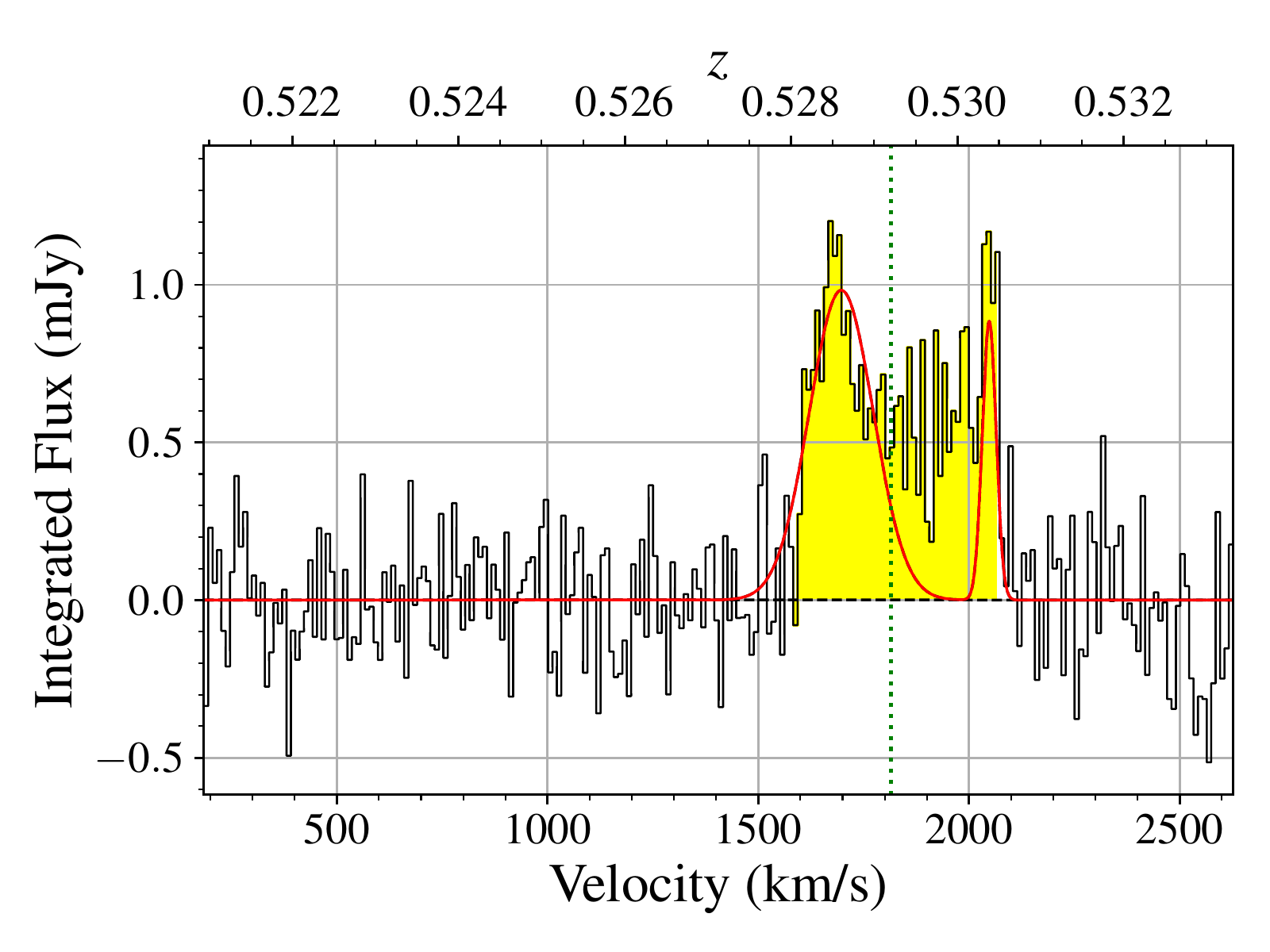}

	\includegraphics[scale=0.3]{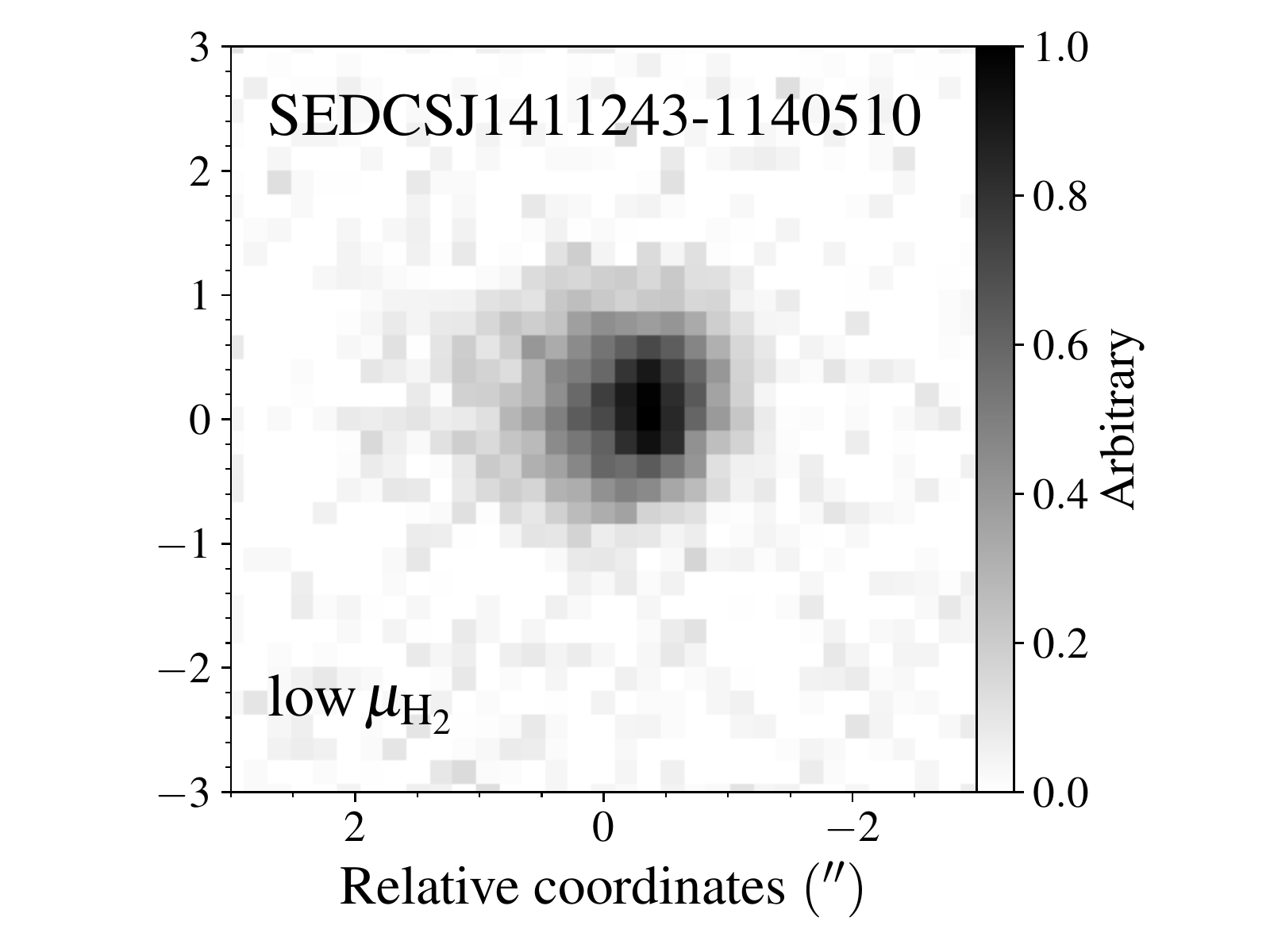}
	\includegraphics[scale=0.3]{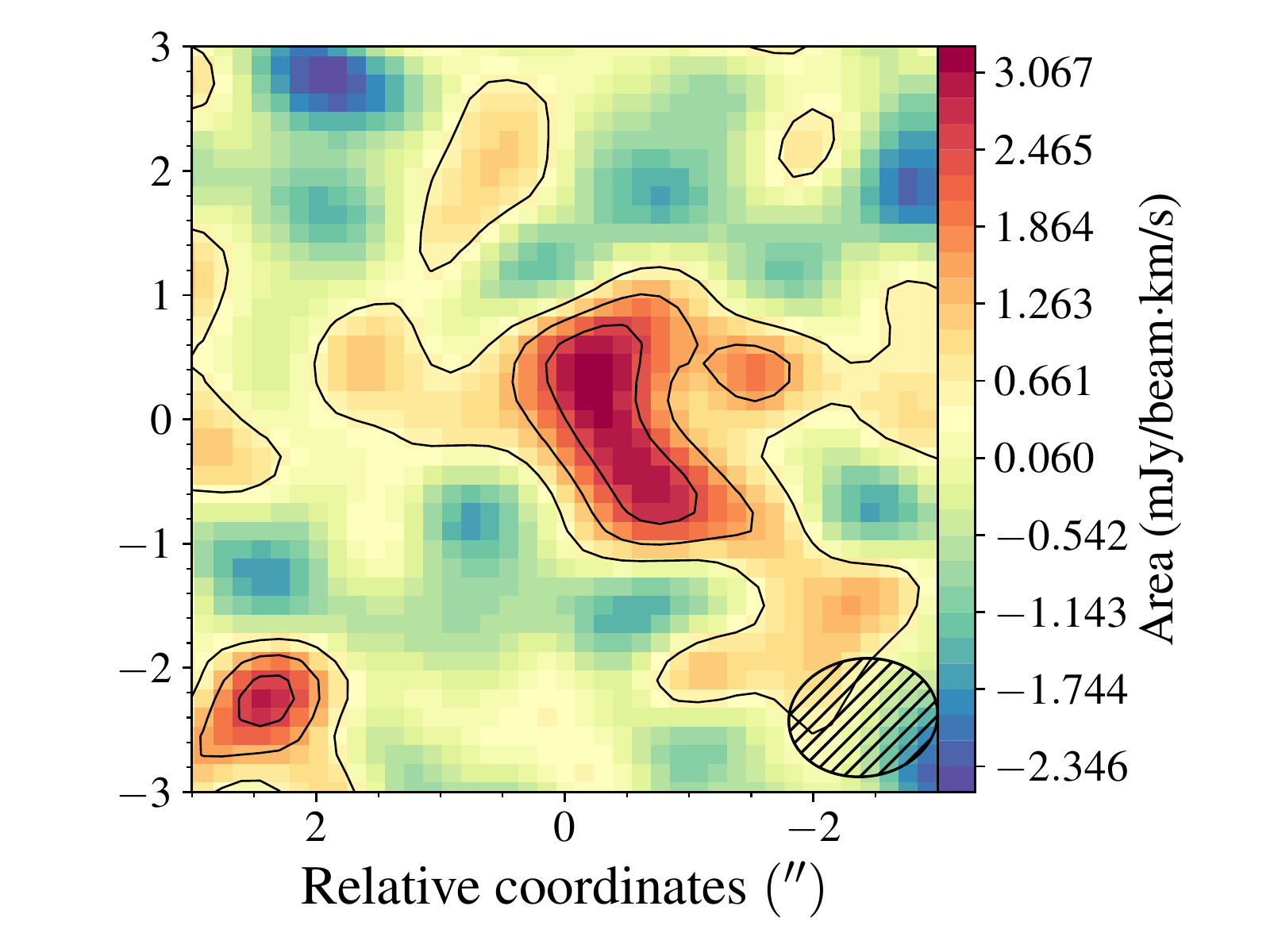}
	\includegraphics[scale=0.3]{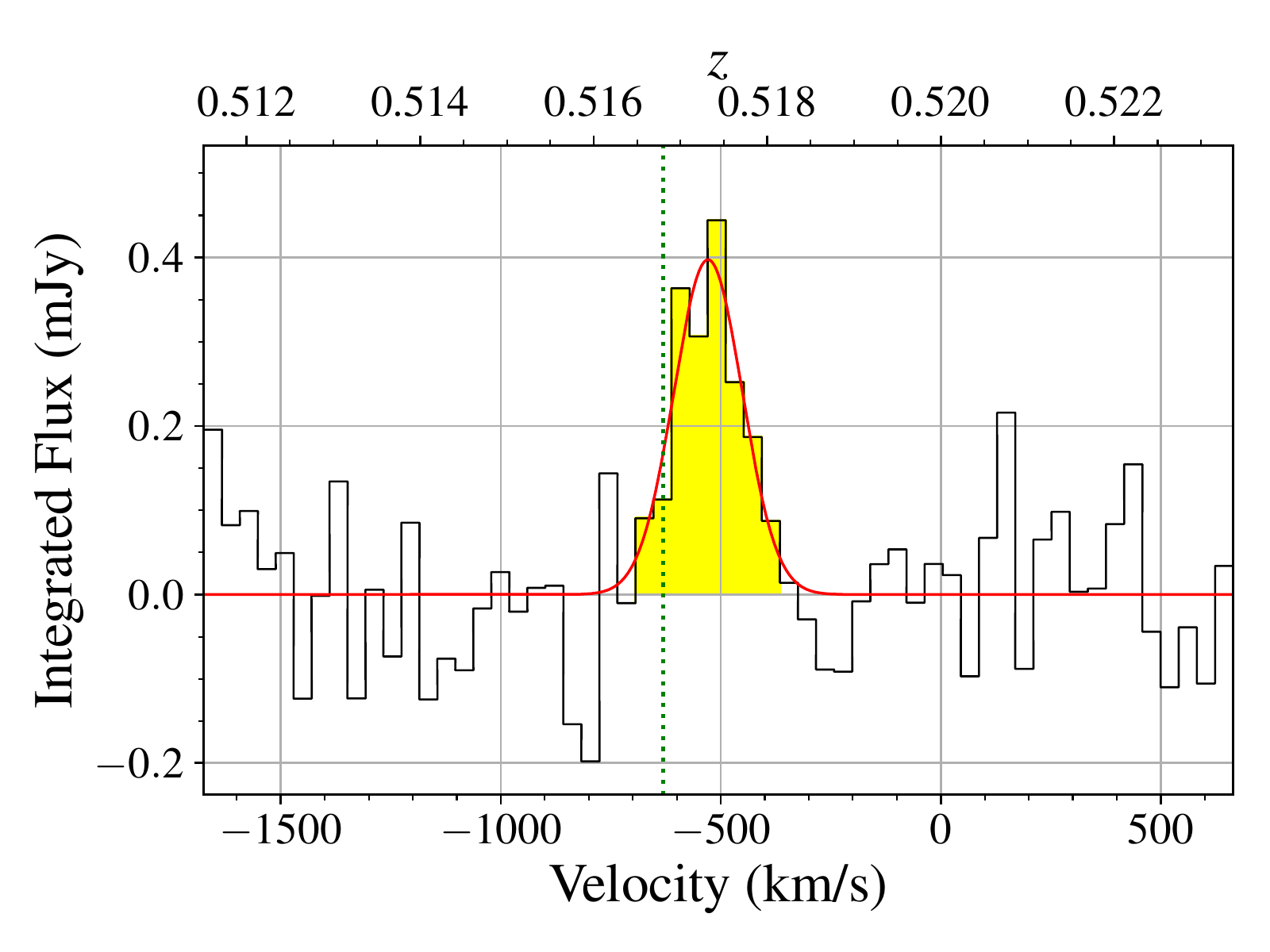}

    \caption{Continued.}
\end{figure*}

\begin{figure*}[htbp]\ContinuedFloat
\centering
	\includegraphics[scale=0.3]{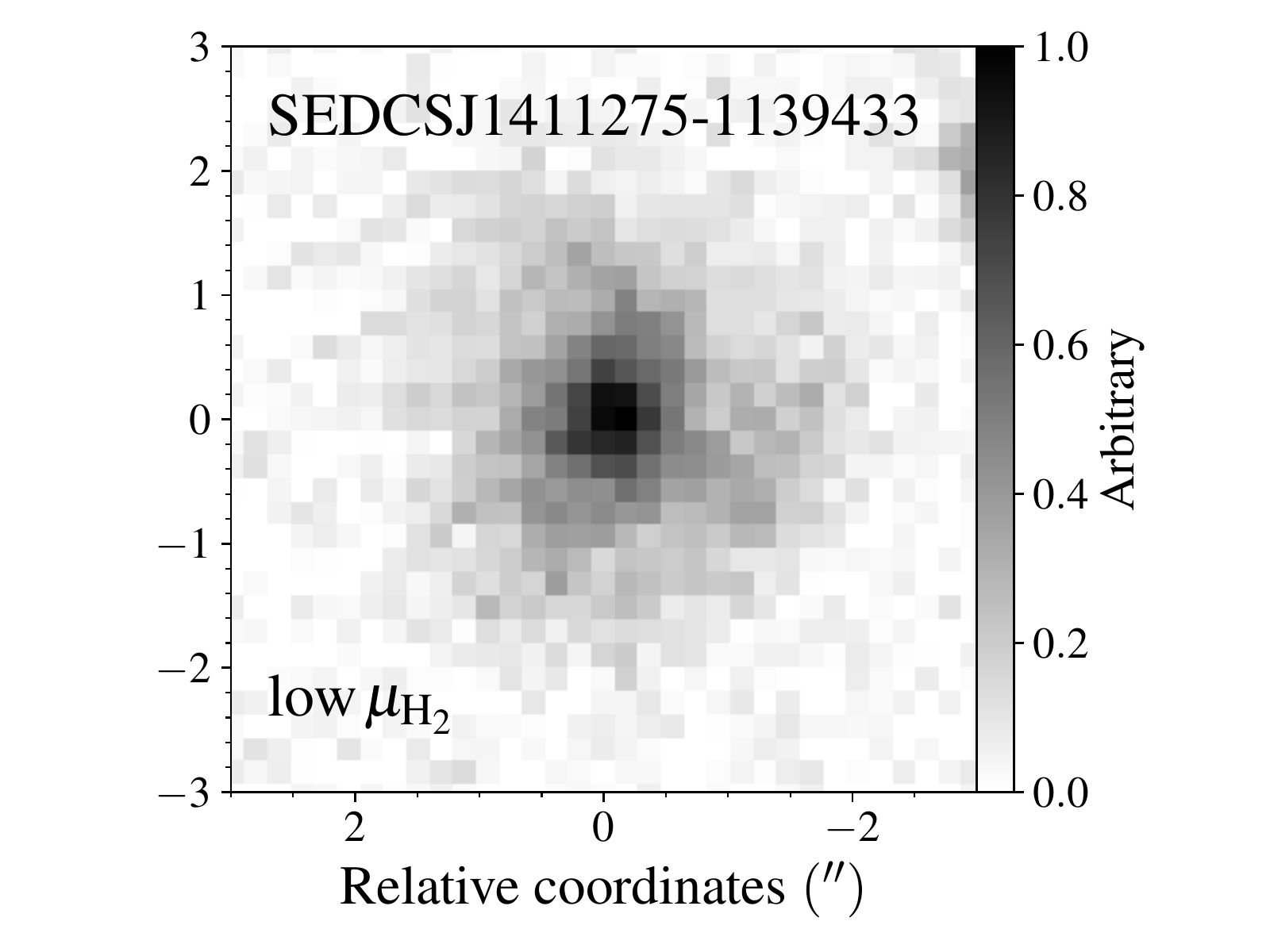}
	\includegraphics[scale=0.3]{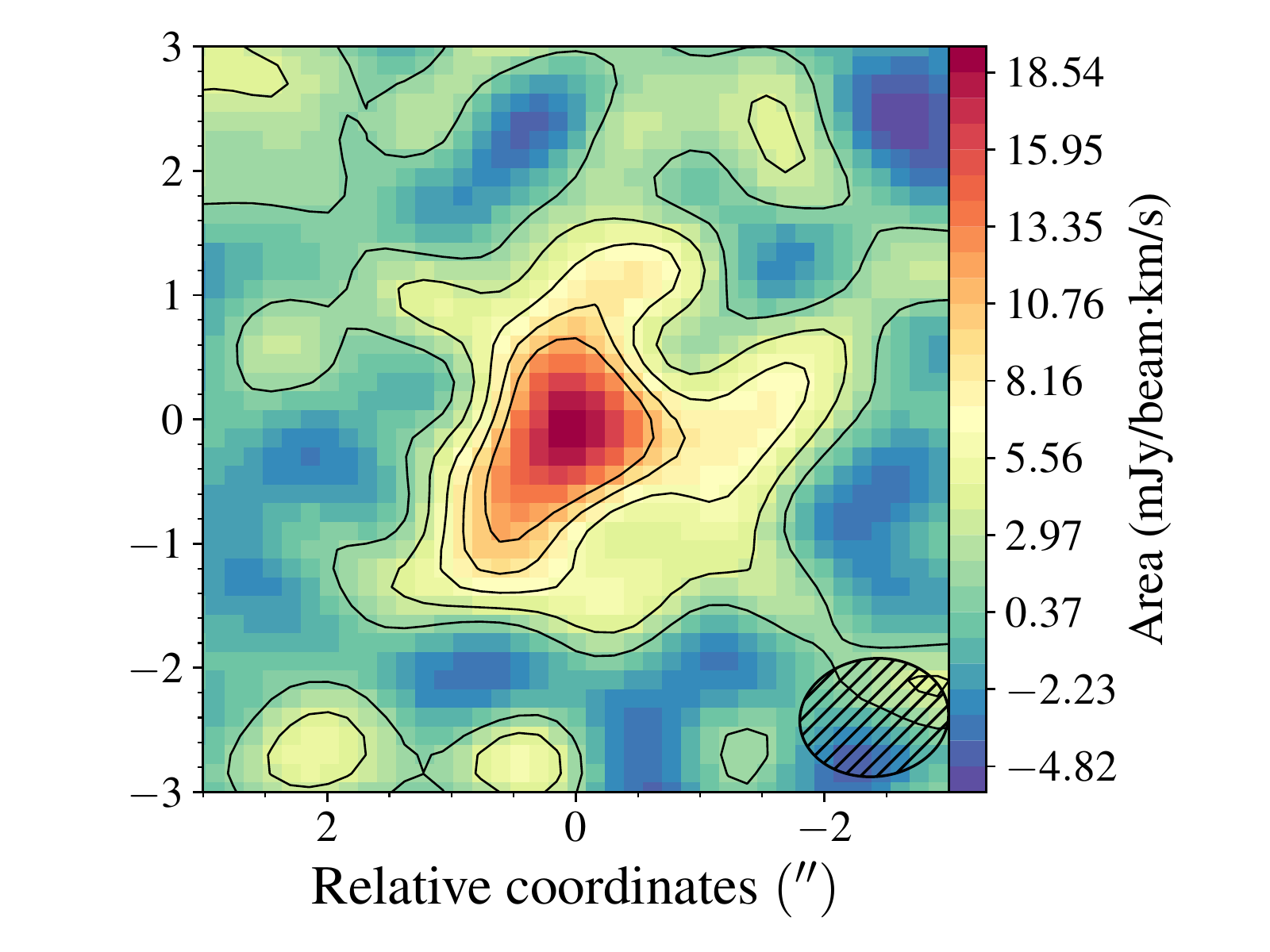}
	\includegraphics[scale=0.3]{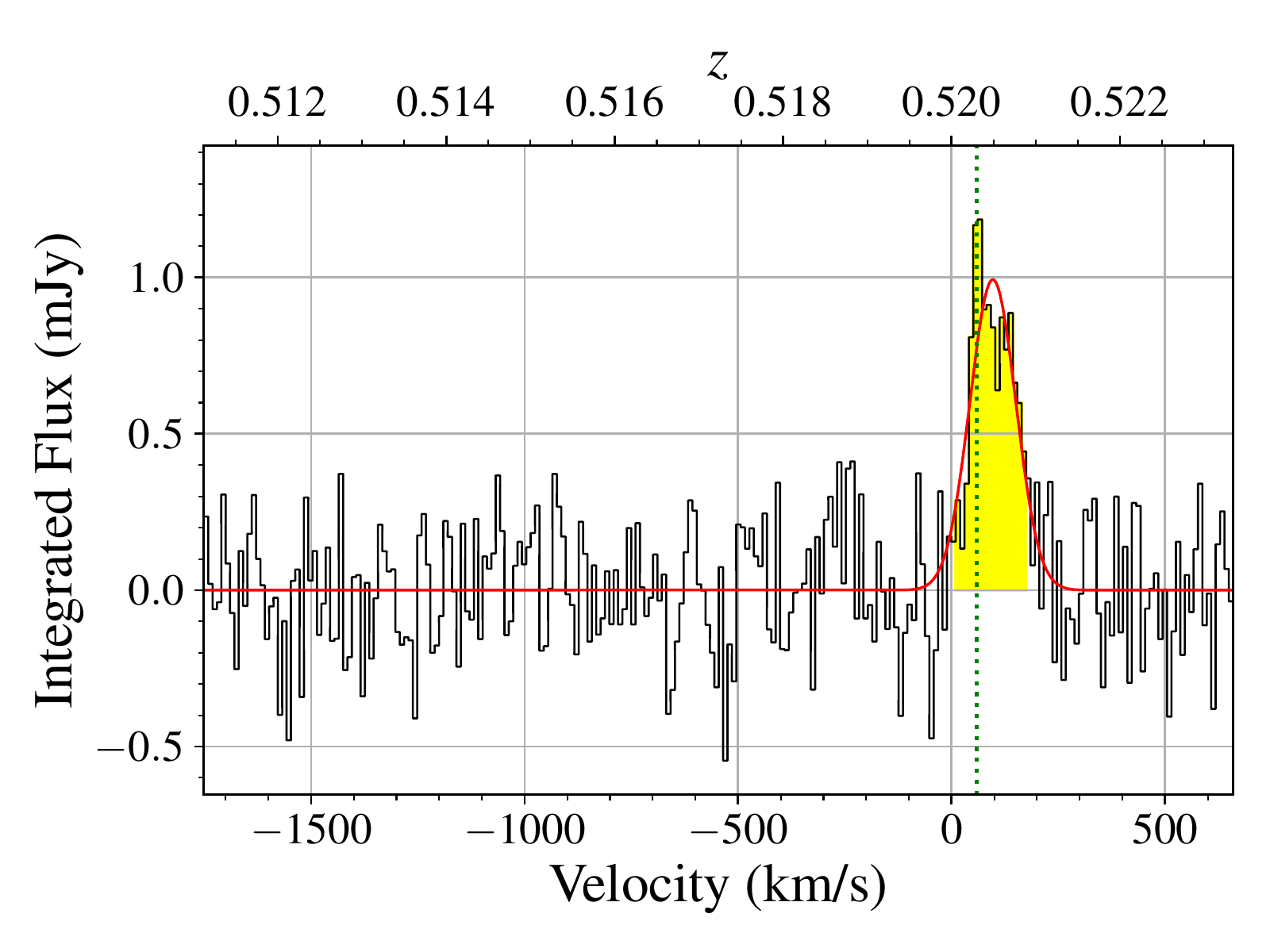}

	\includegraphics[scale=0.3]{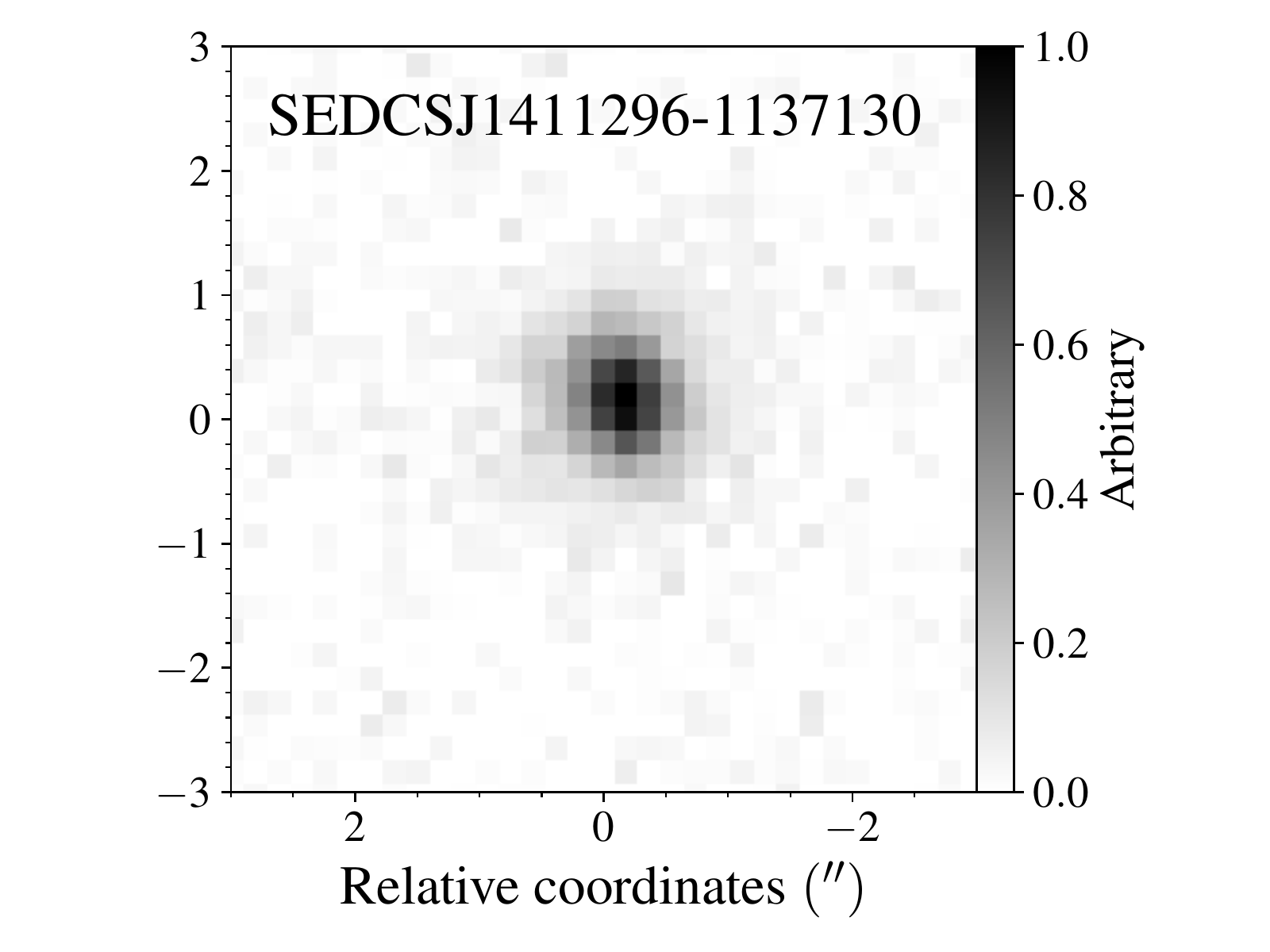}
	\includegraphics[scale=0.3]{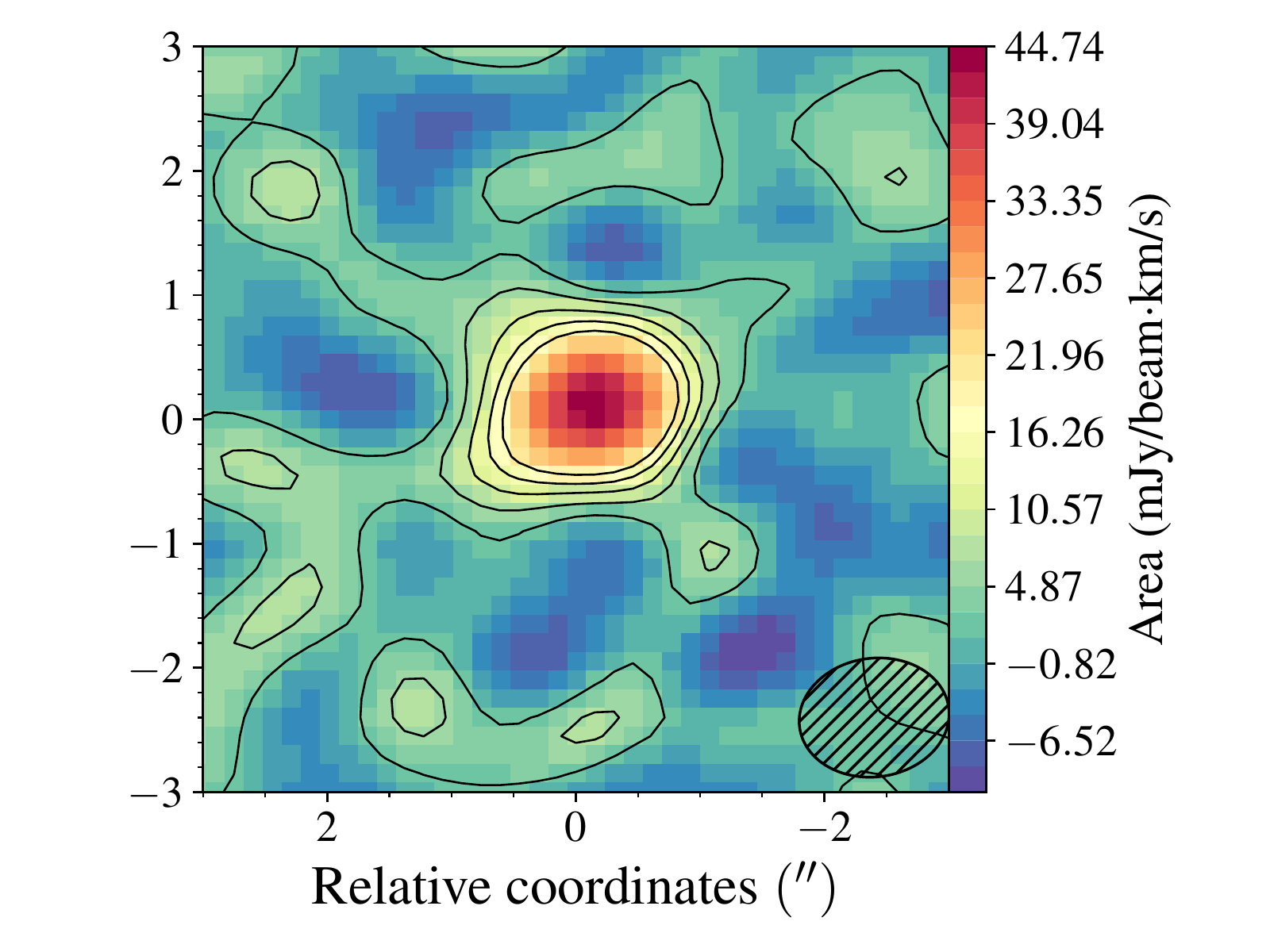}
	\includegraphics[scale=0.3]{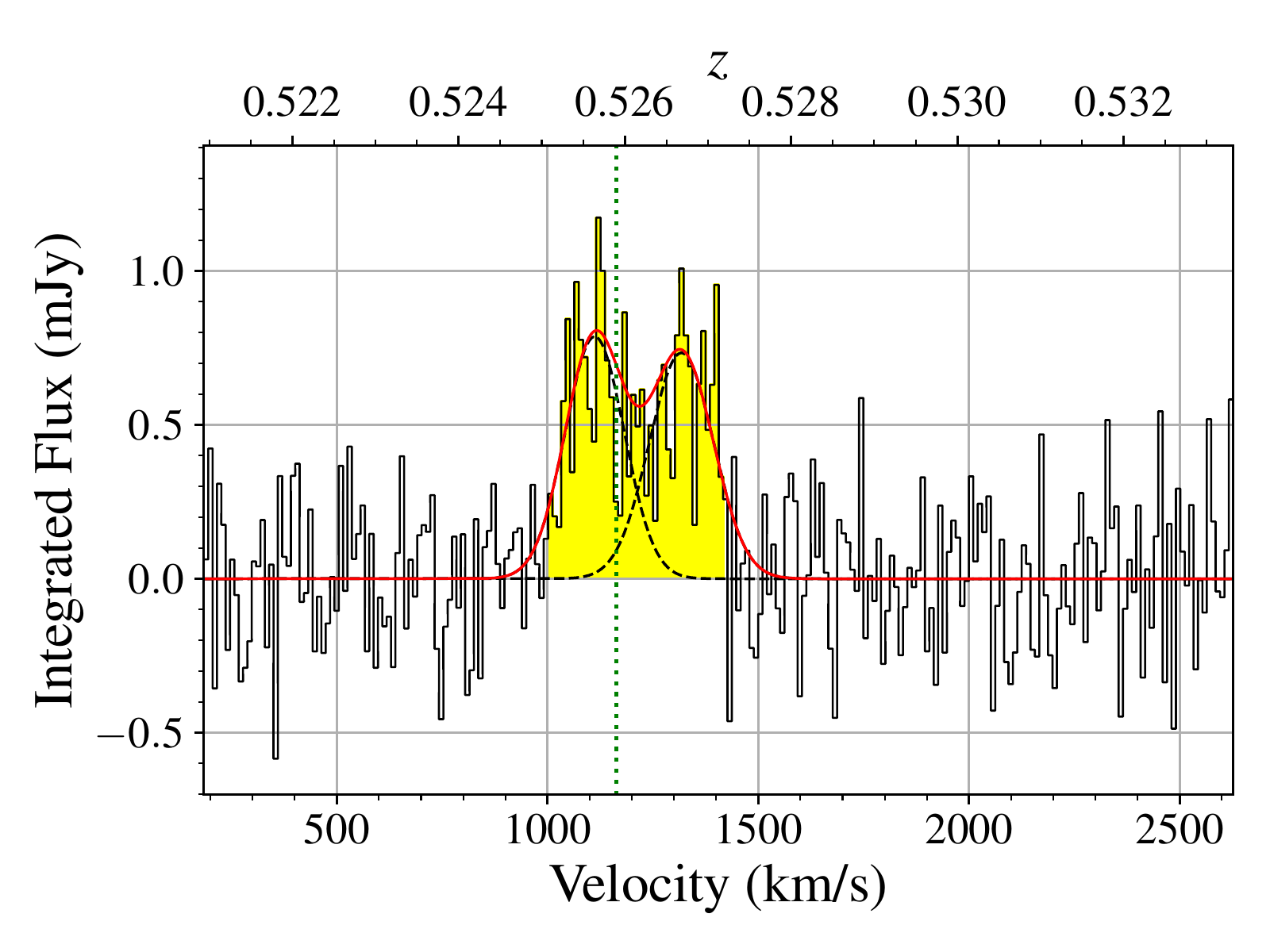}

	\includegraphics[scale=0.3]{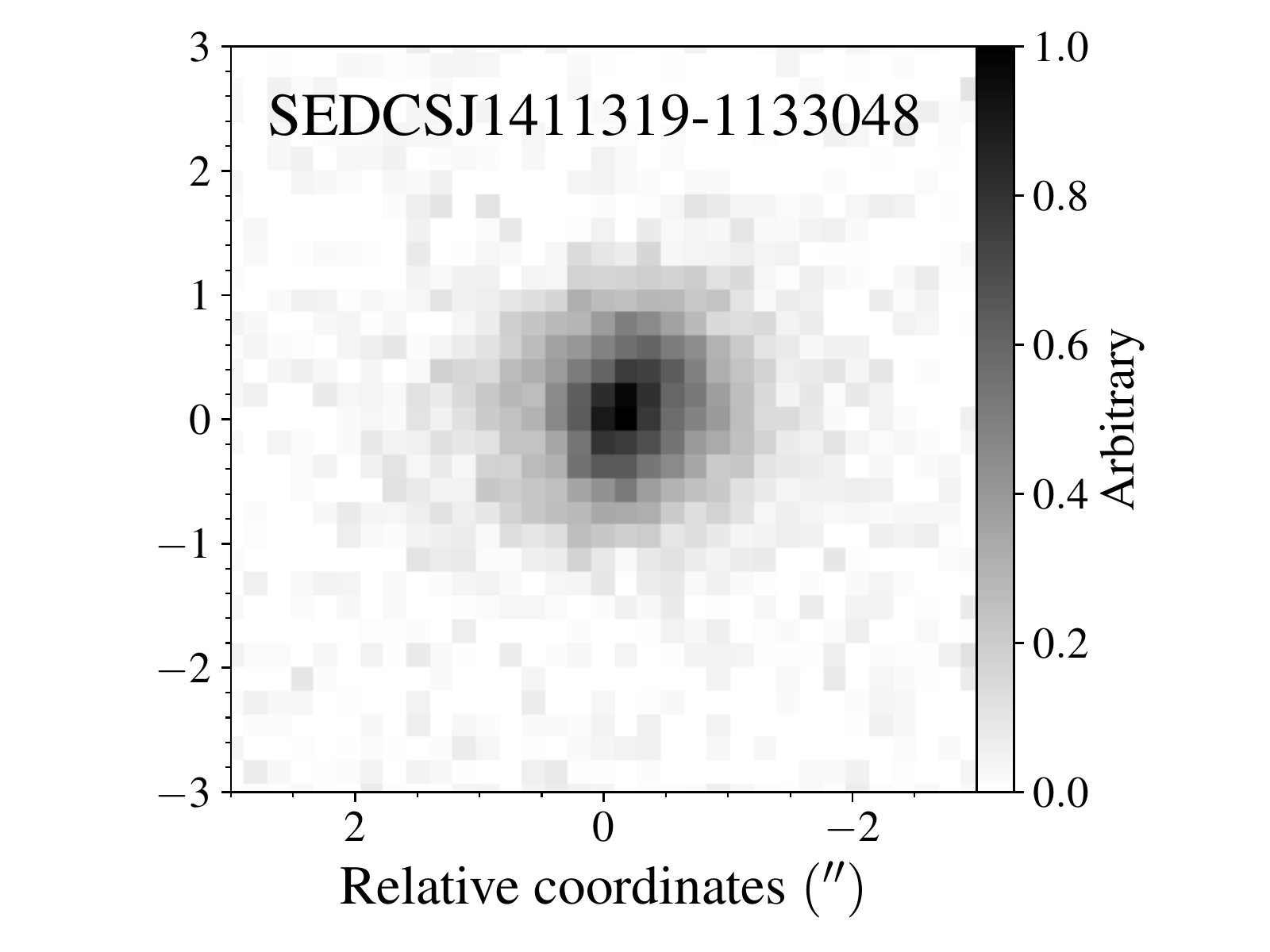}
	\includegraphics[scale=0.3]{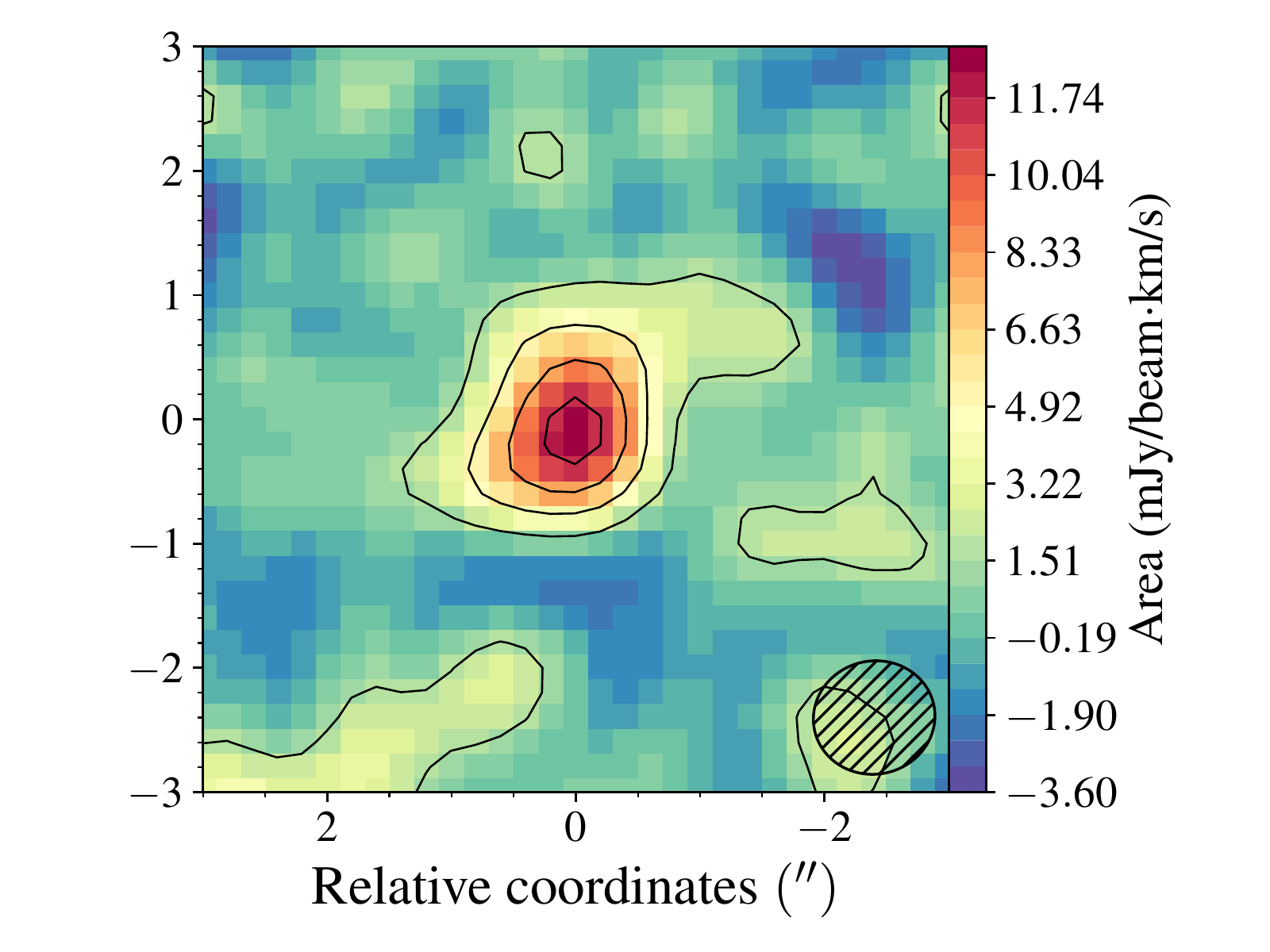}
	\includegraphics[scale=0.3]{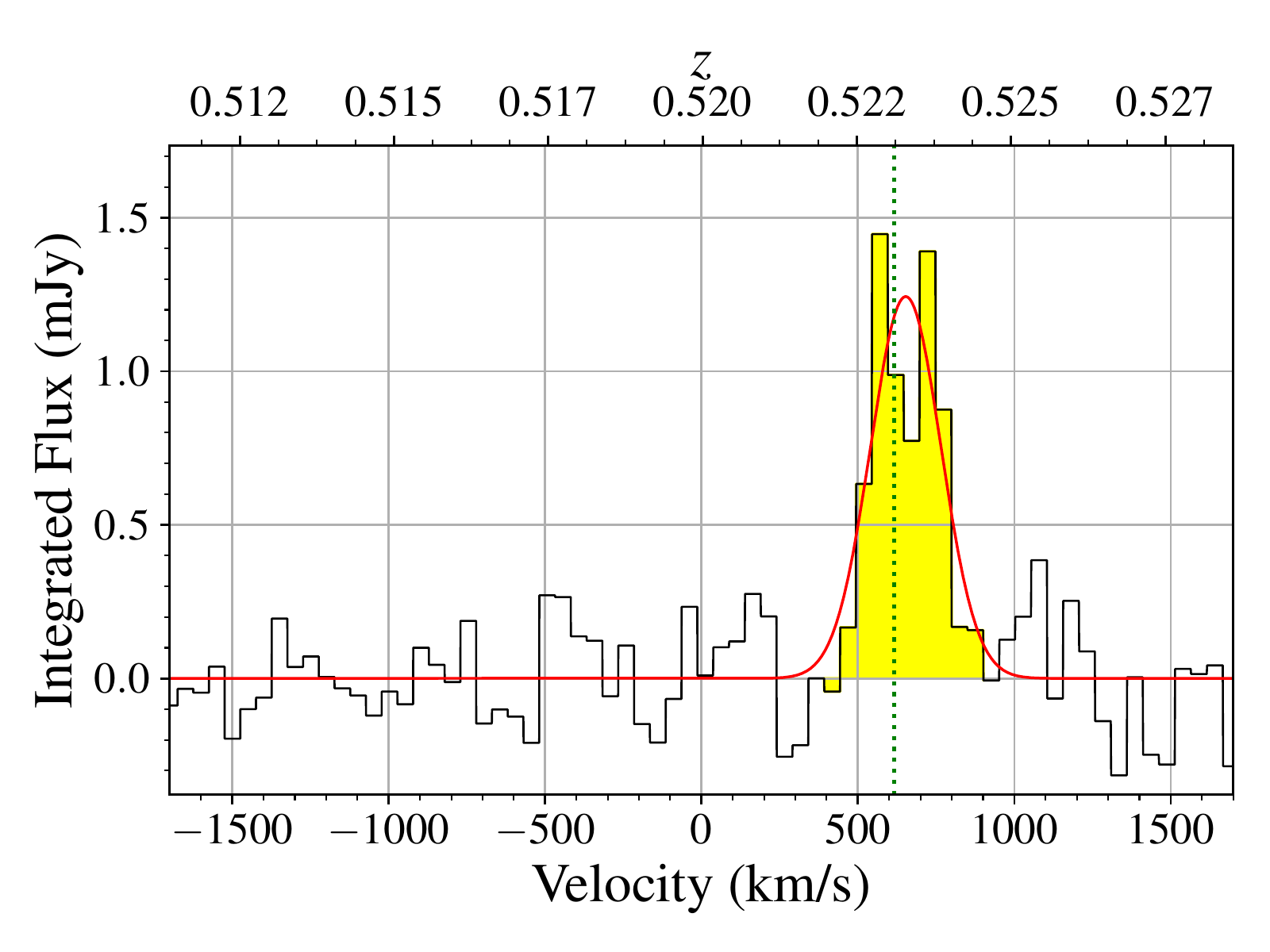}

	\includegraphics[scale=0.3]{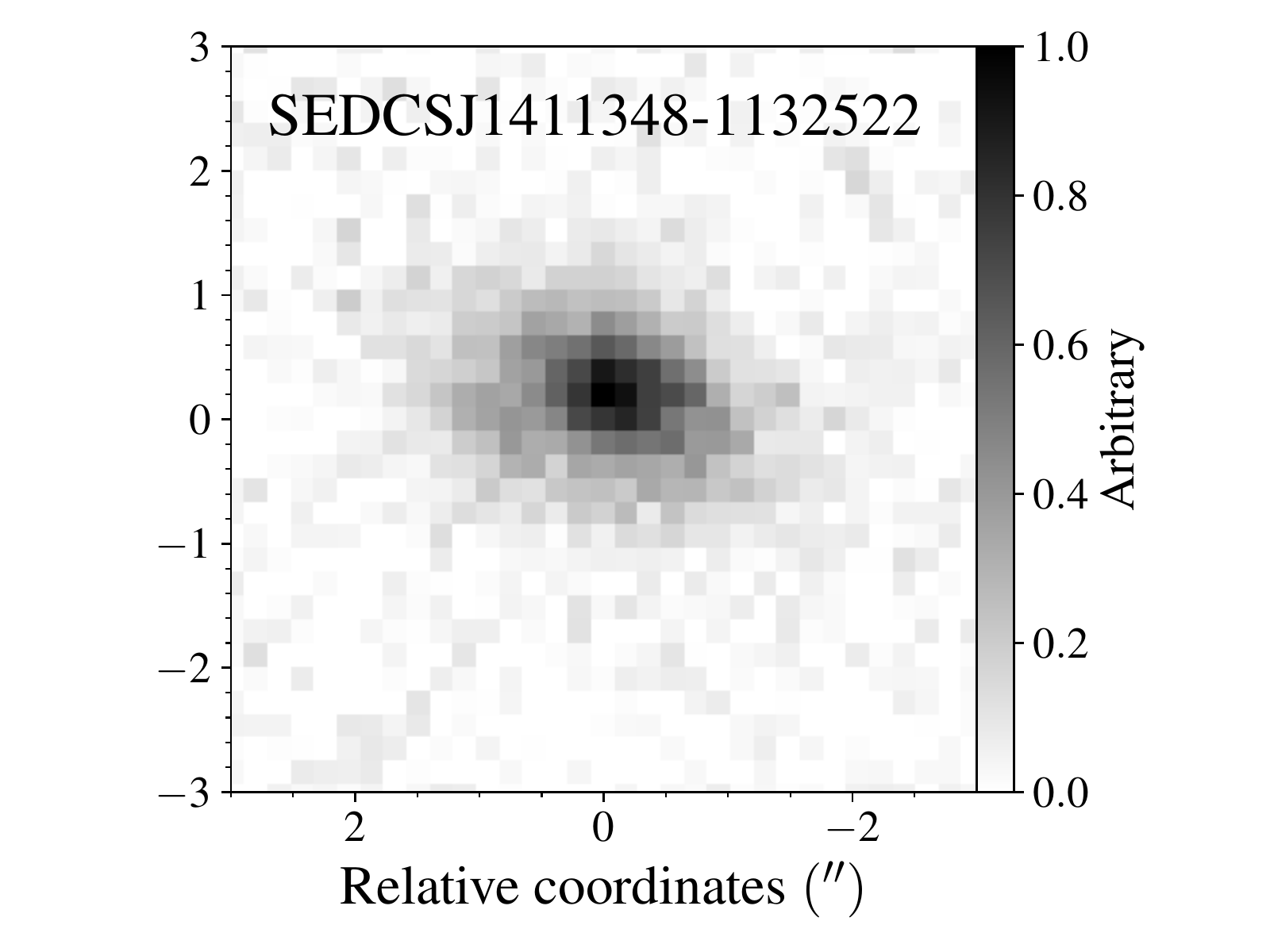}
	\includegraphics[scale=0.3]{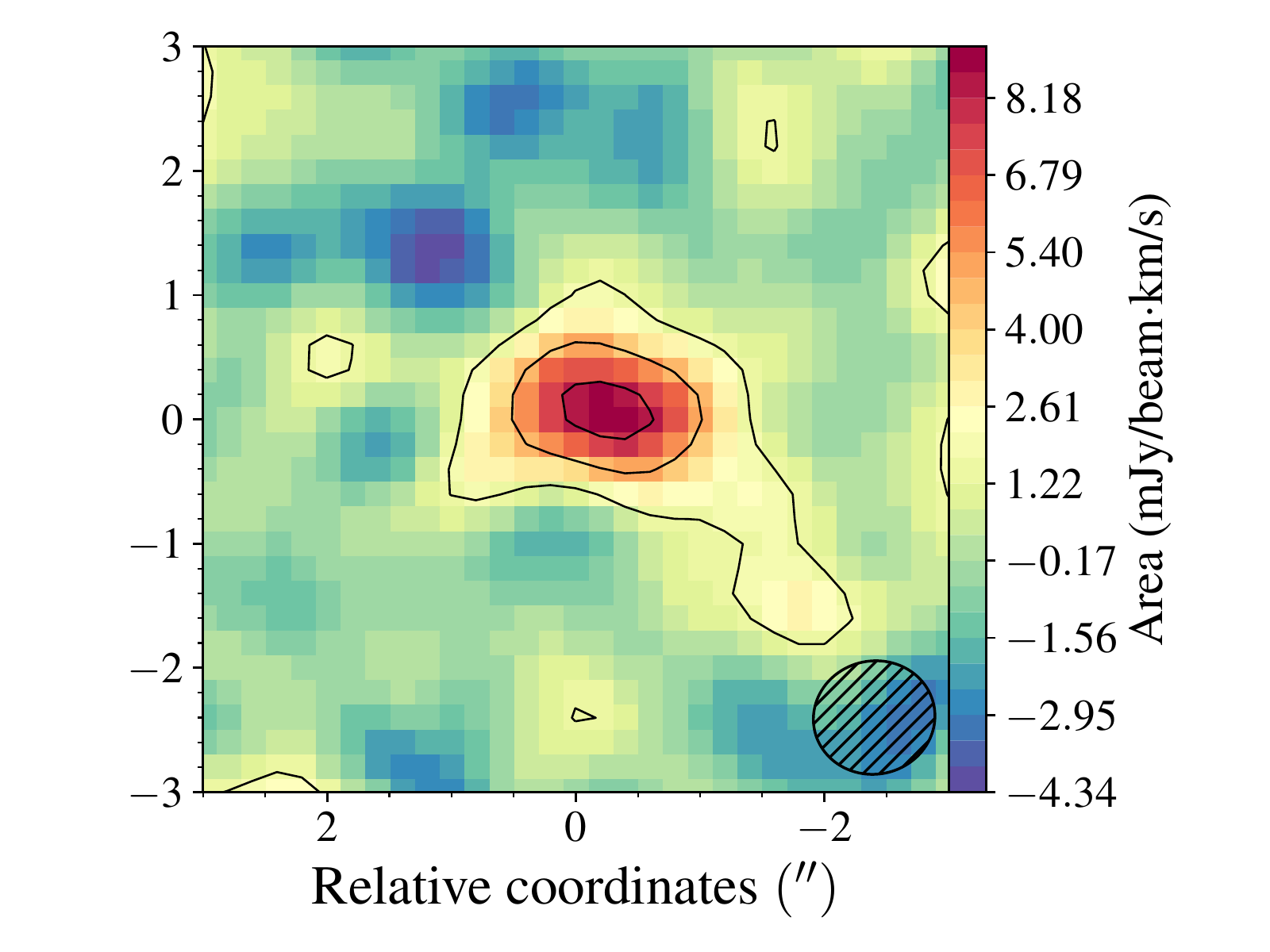}
	\includegraphics[scale=0.3]{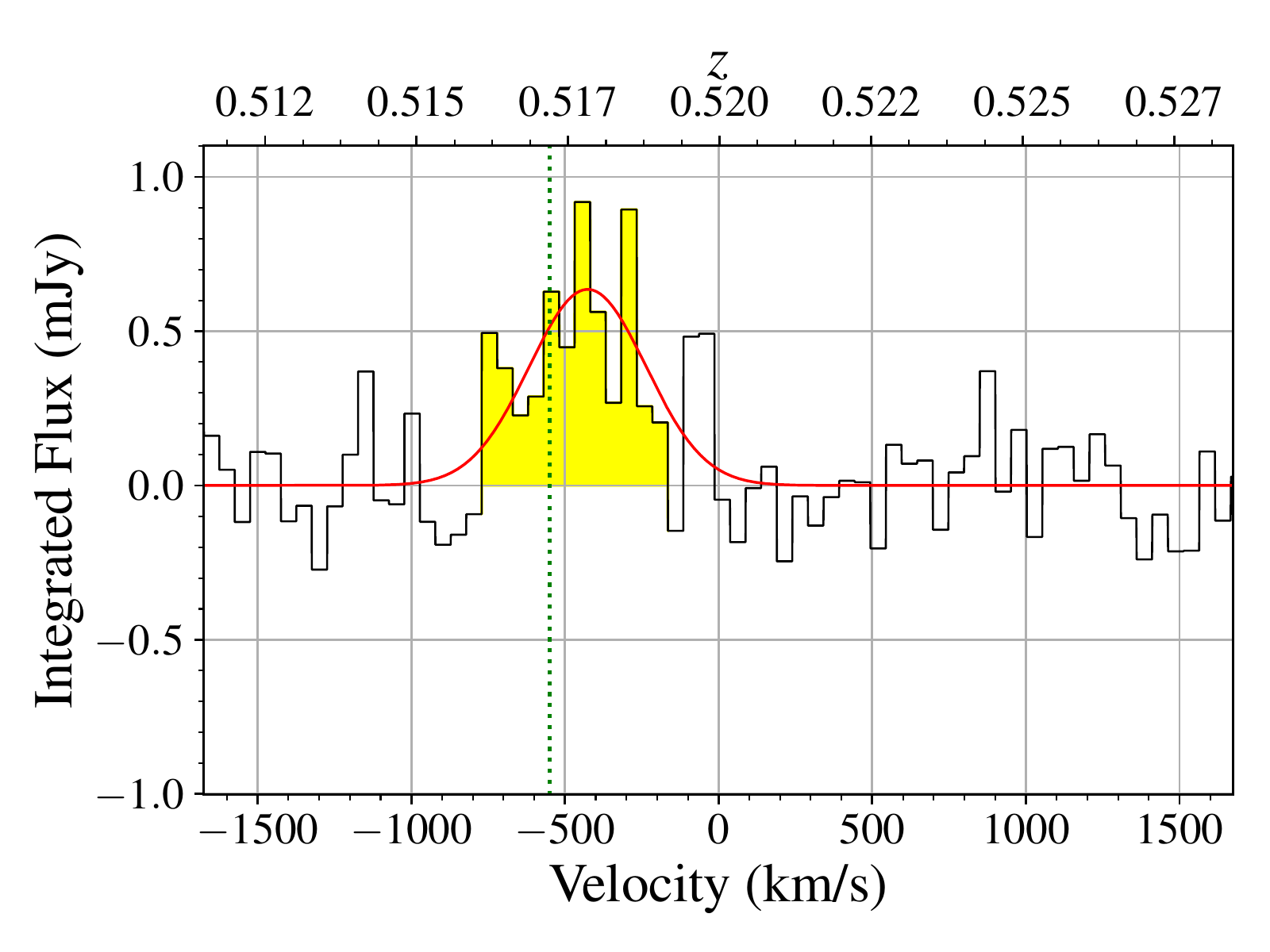}

	\includegraphics[scale=0.3]{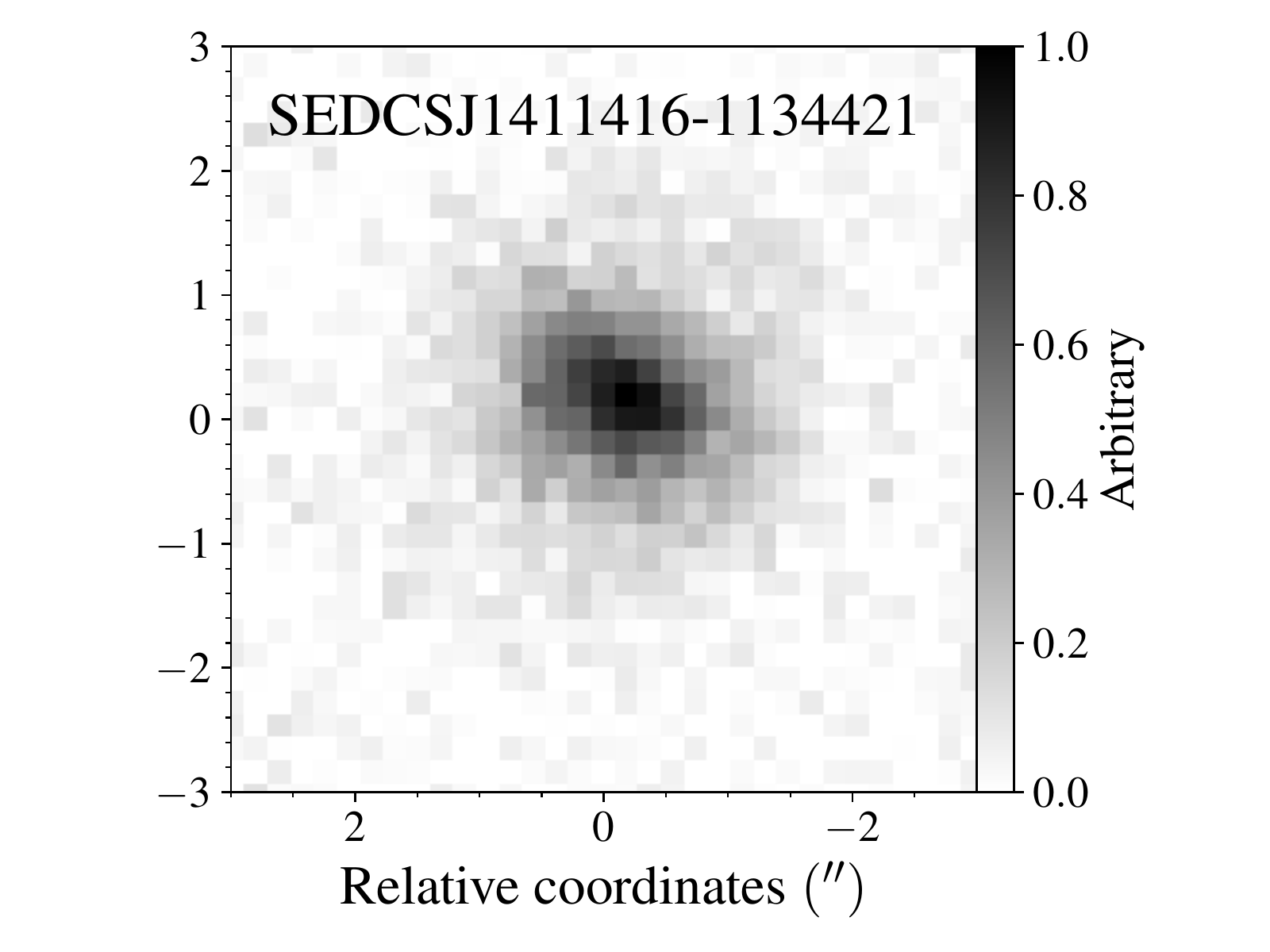}
	\includegraphics[scale=0.3]{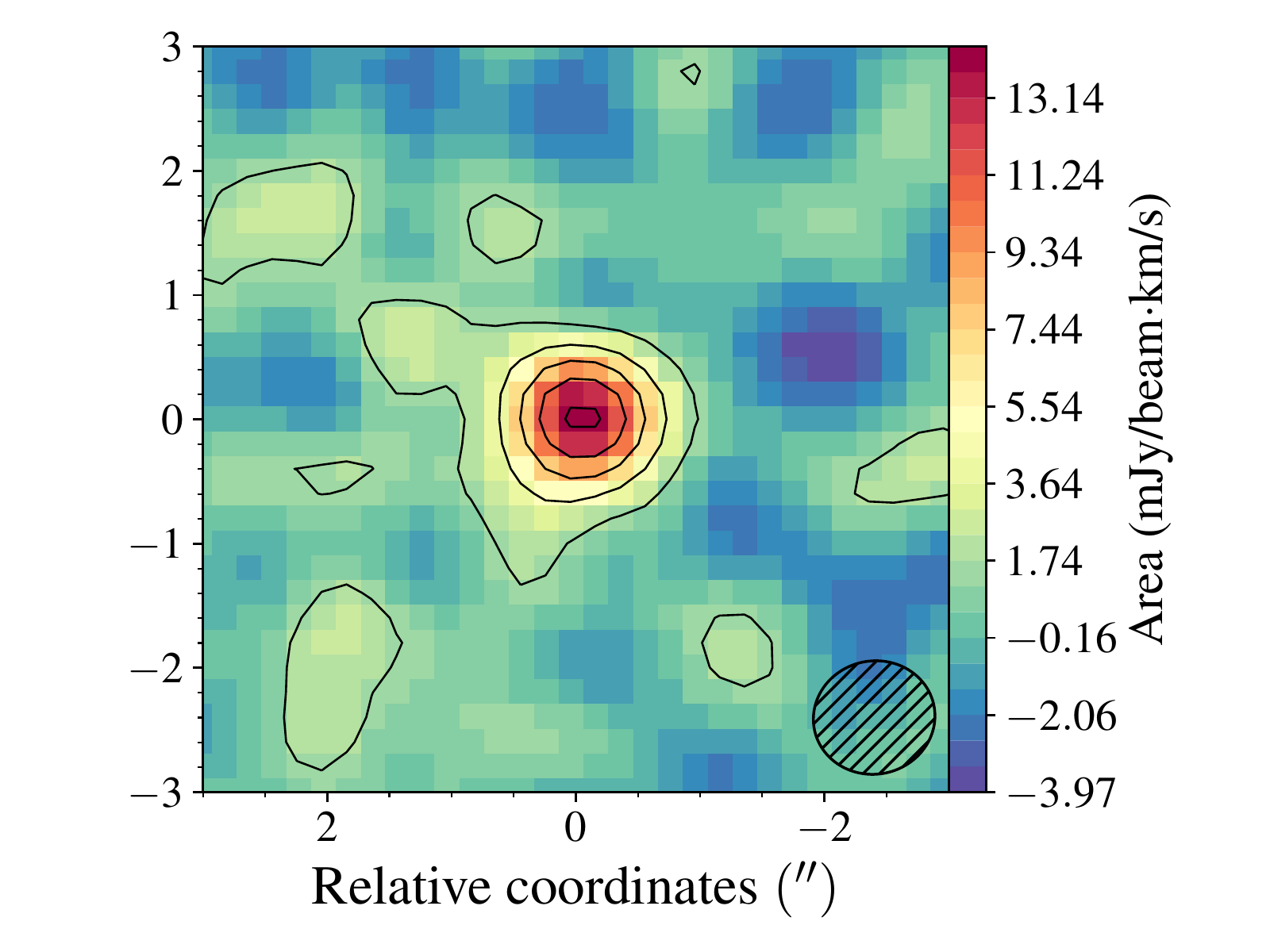}
	\includegraphics[scale=0.3]{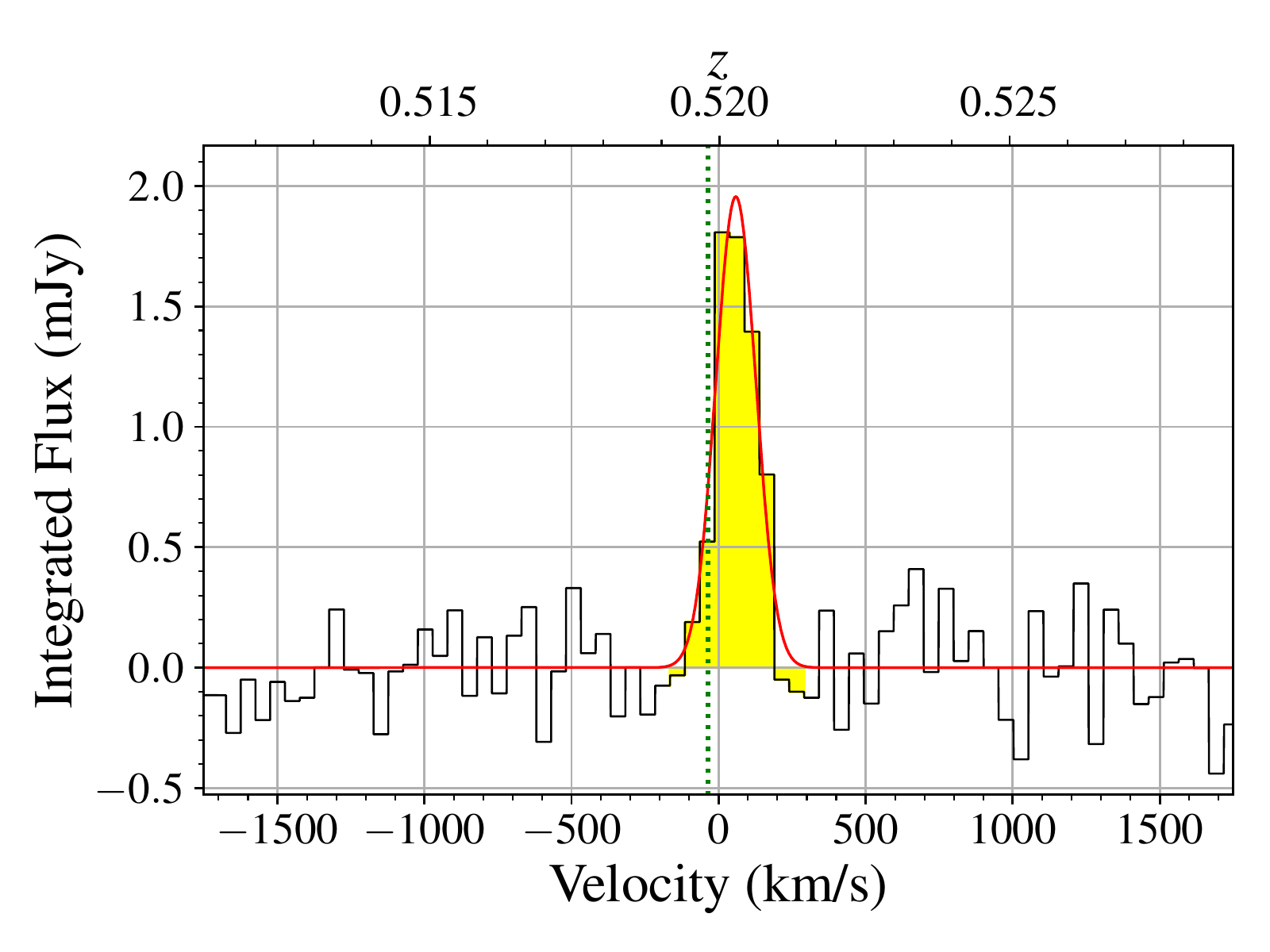}

	\includegraphics[scale=0.3]{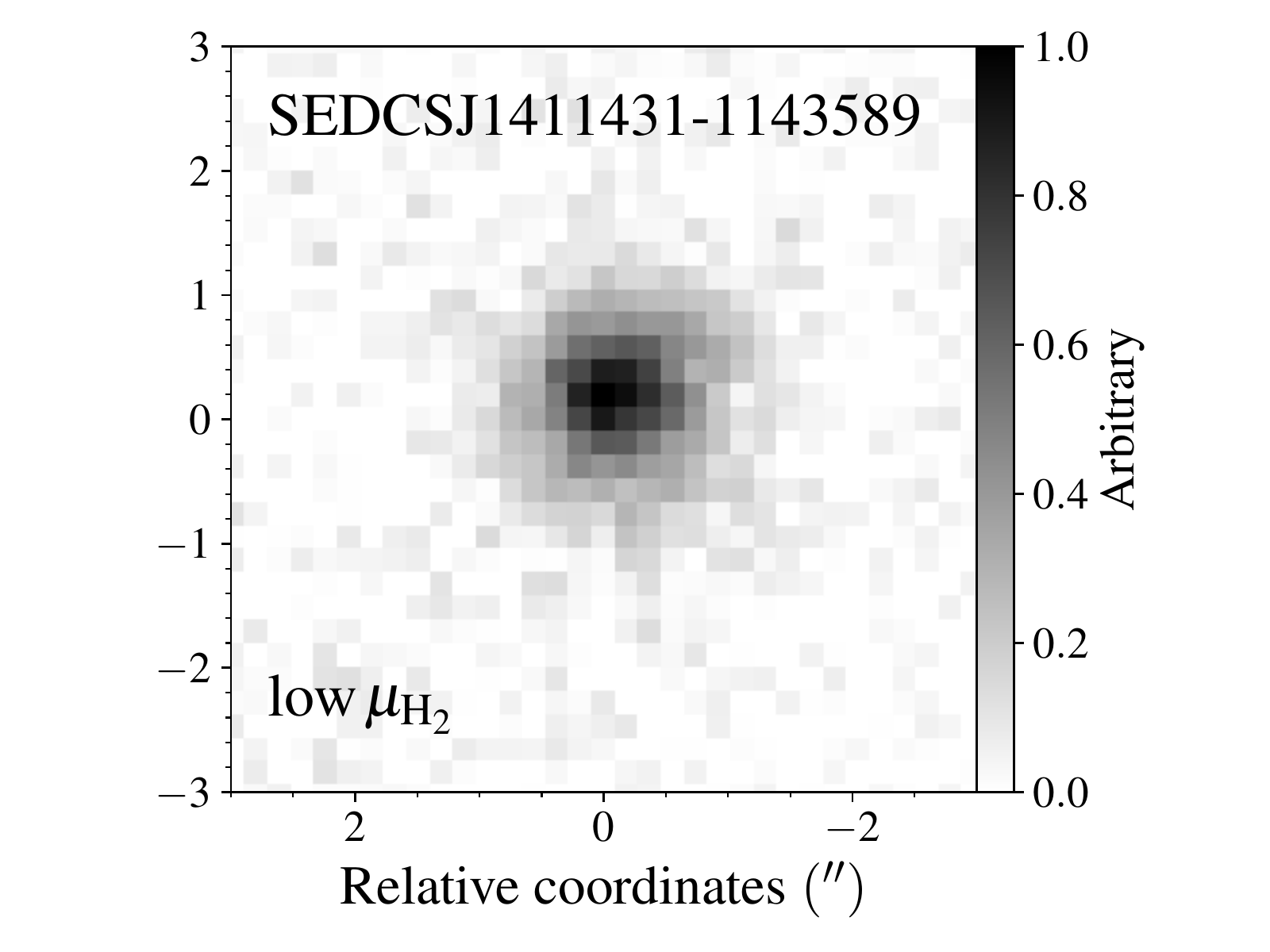}
	\includegraphics[scale=0.3]{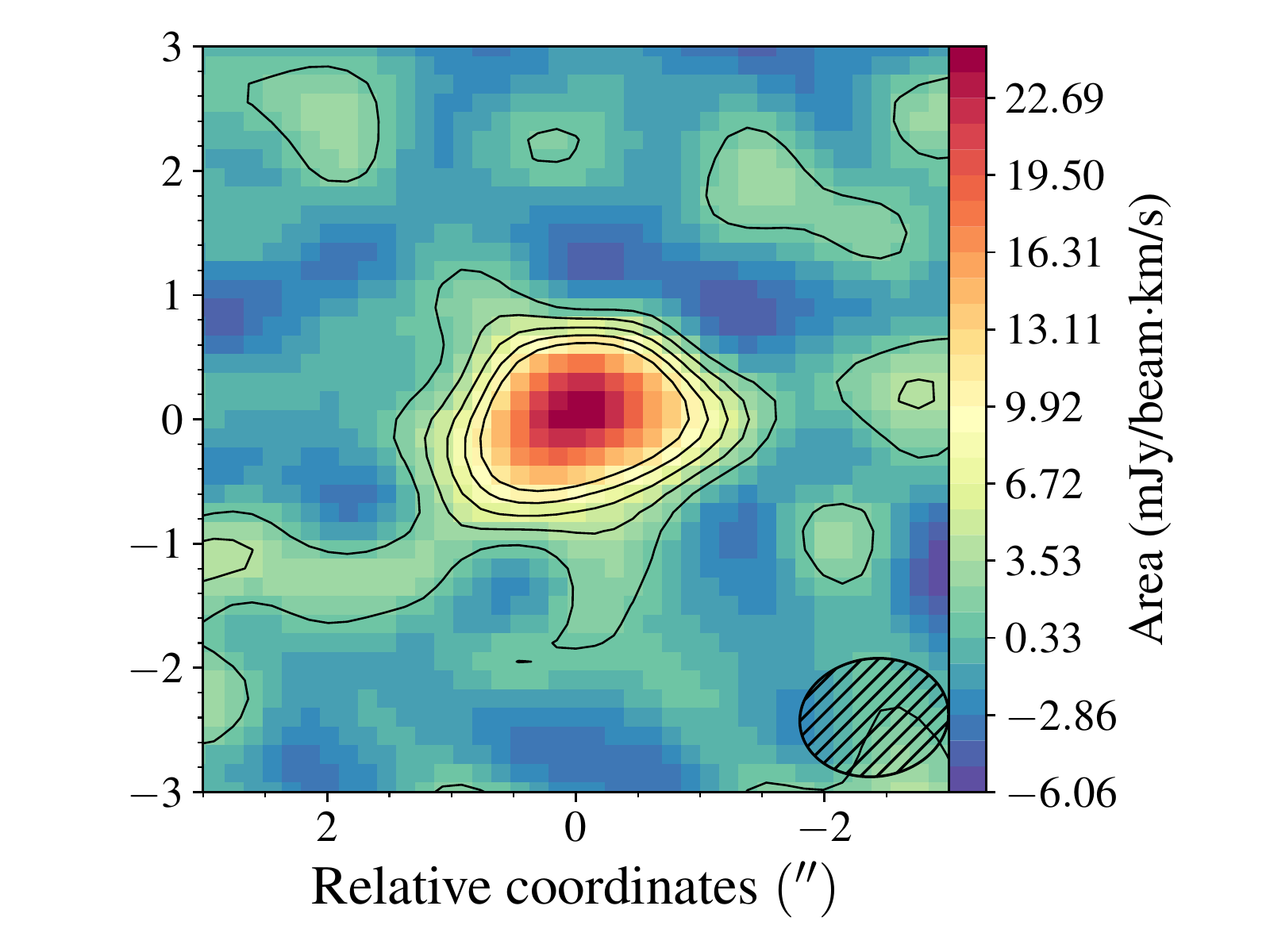}
	\includegraphics[scale=0.3]{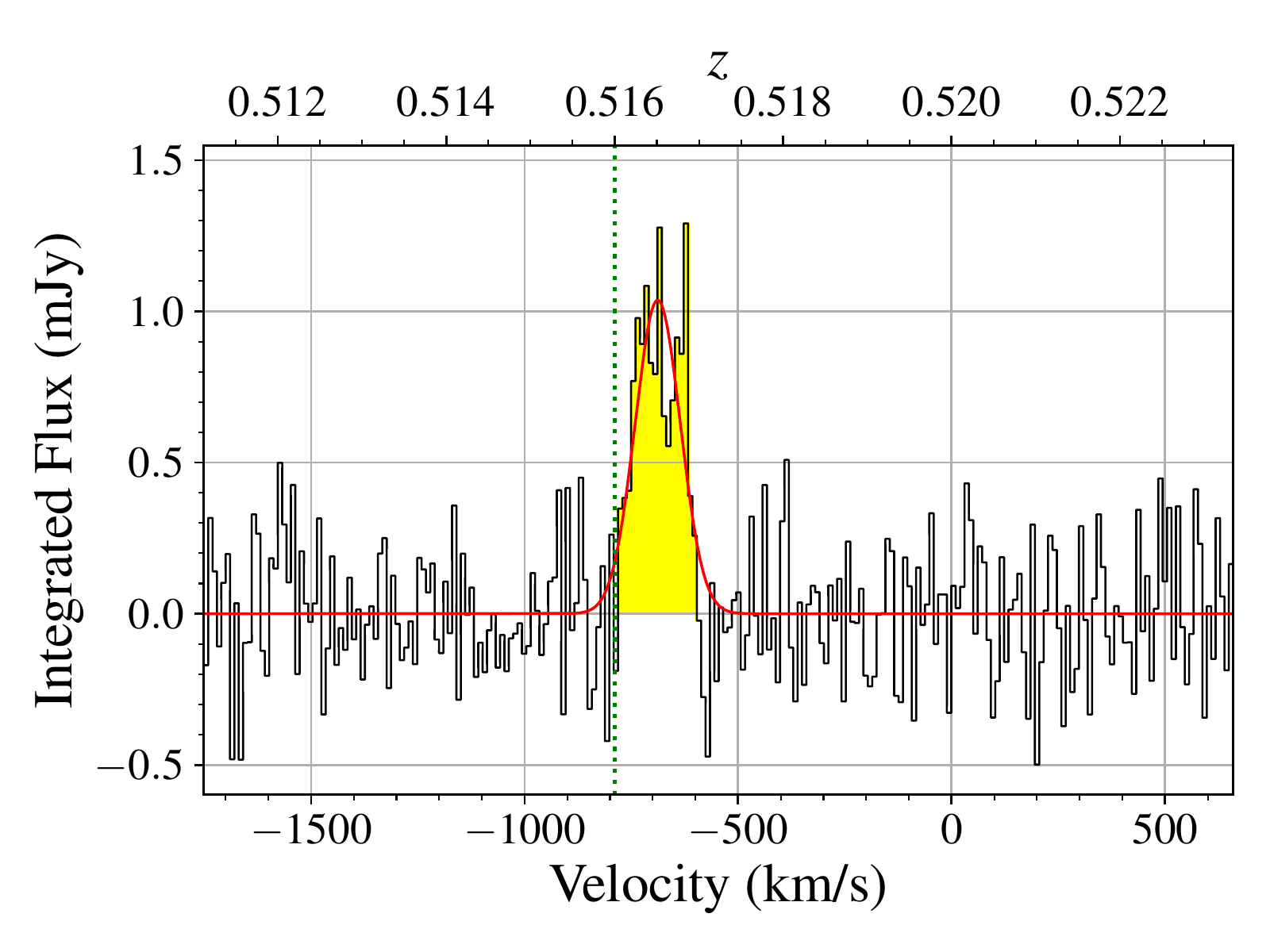}

    \caption{Continued.}
\end{figure*}

\begin{figure*}[htbp]\ContinuedFloat
\centering
	\includegraphics[scale=0.3]{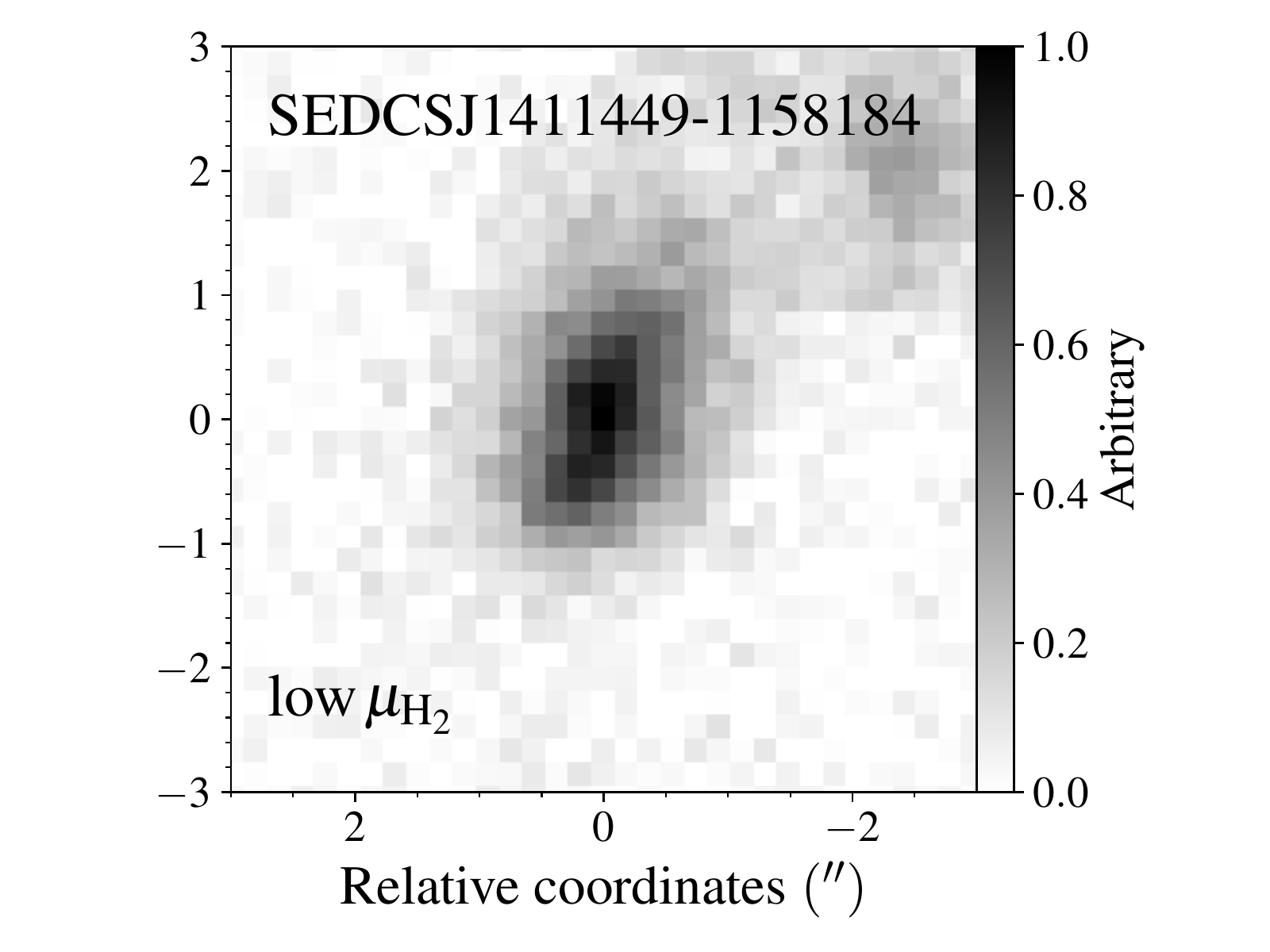}
	\includegraphics[scale=0.3]{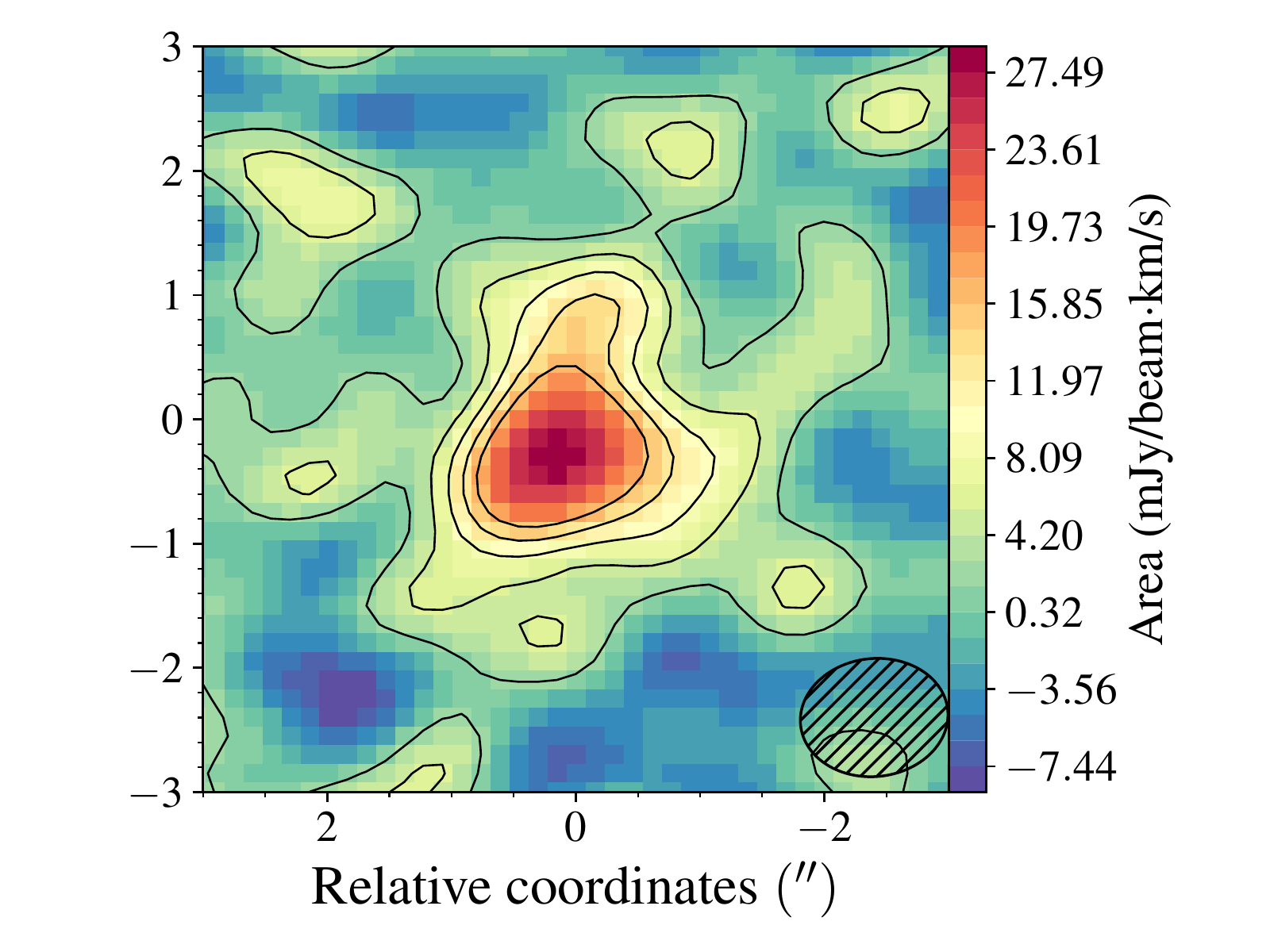}
	\includegraphics[scale=0.3]{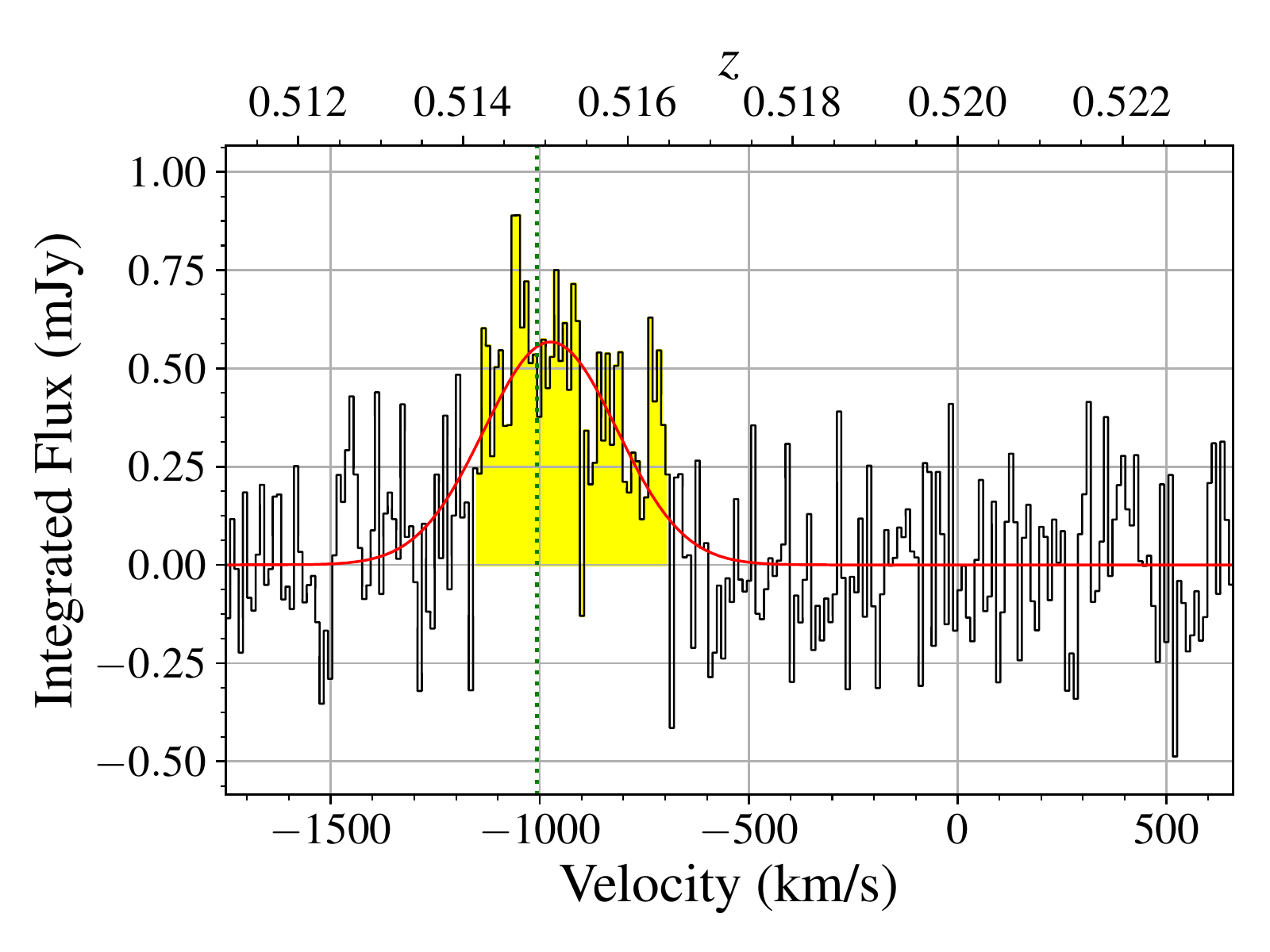}

	\includegraphics[scale=0.3]{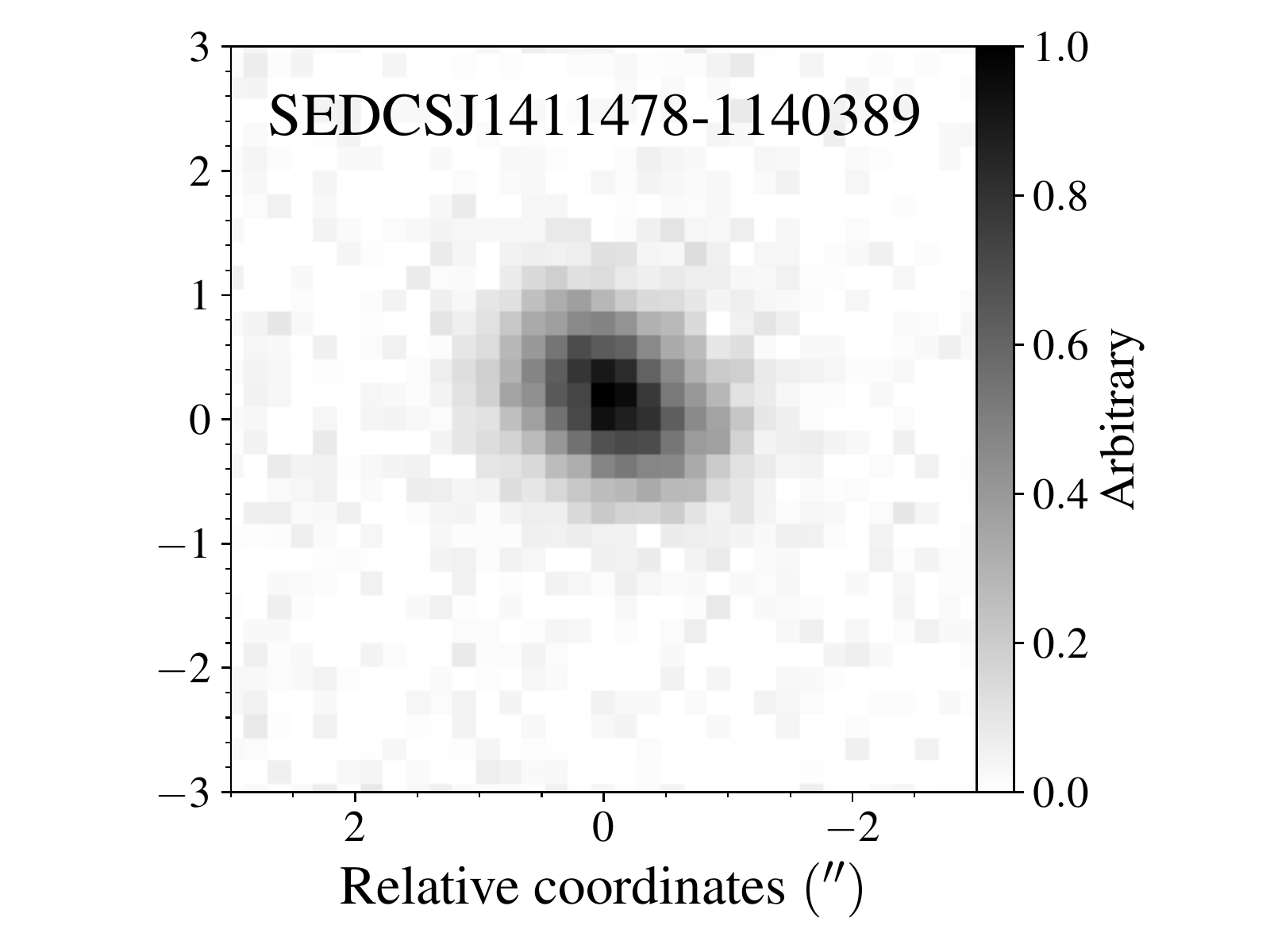}
	\includegraphics[scale=0.3]{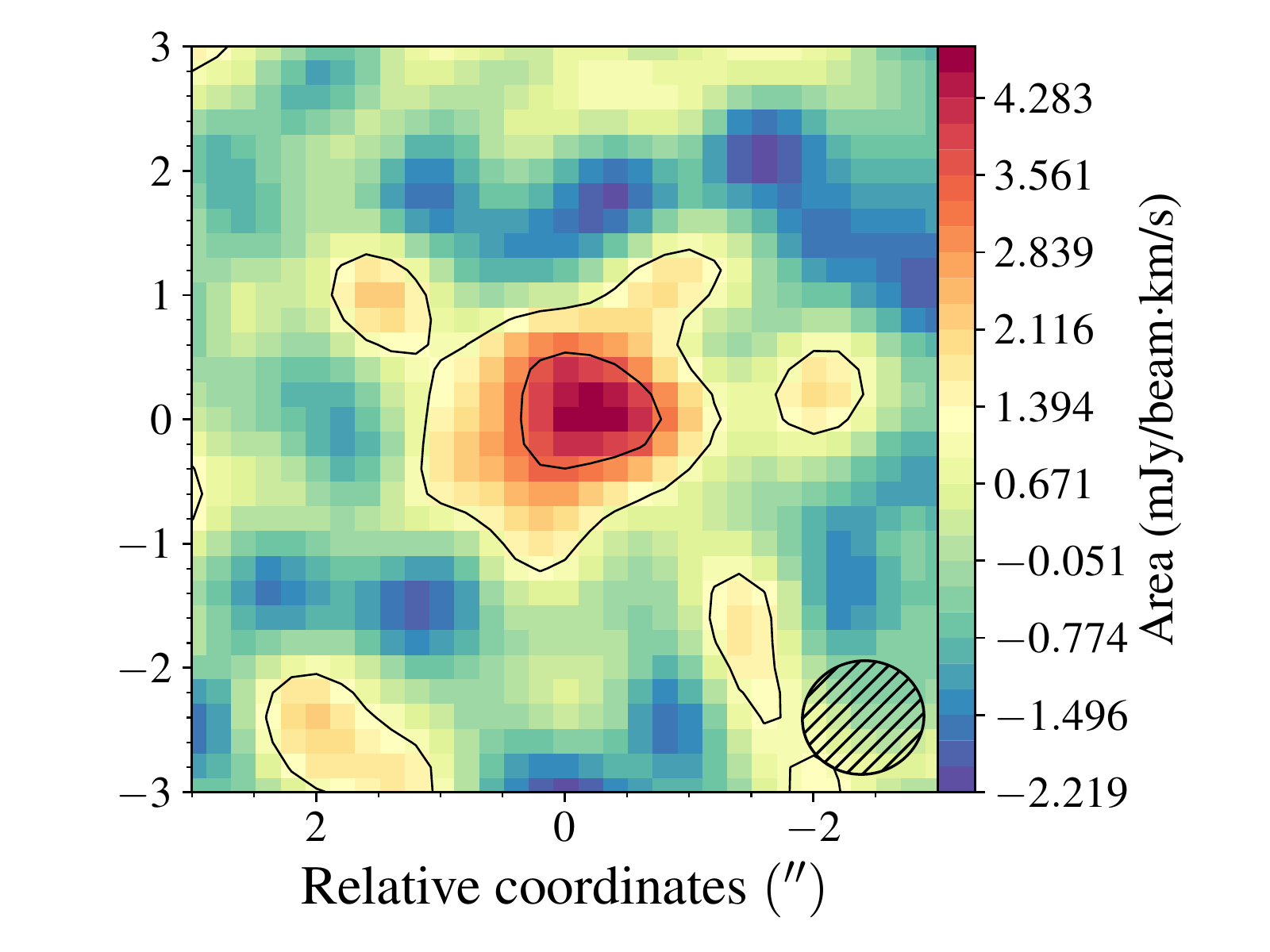}
	\includegraphics[scale=0.3]{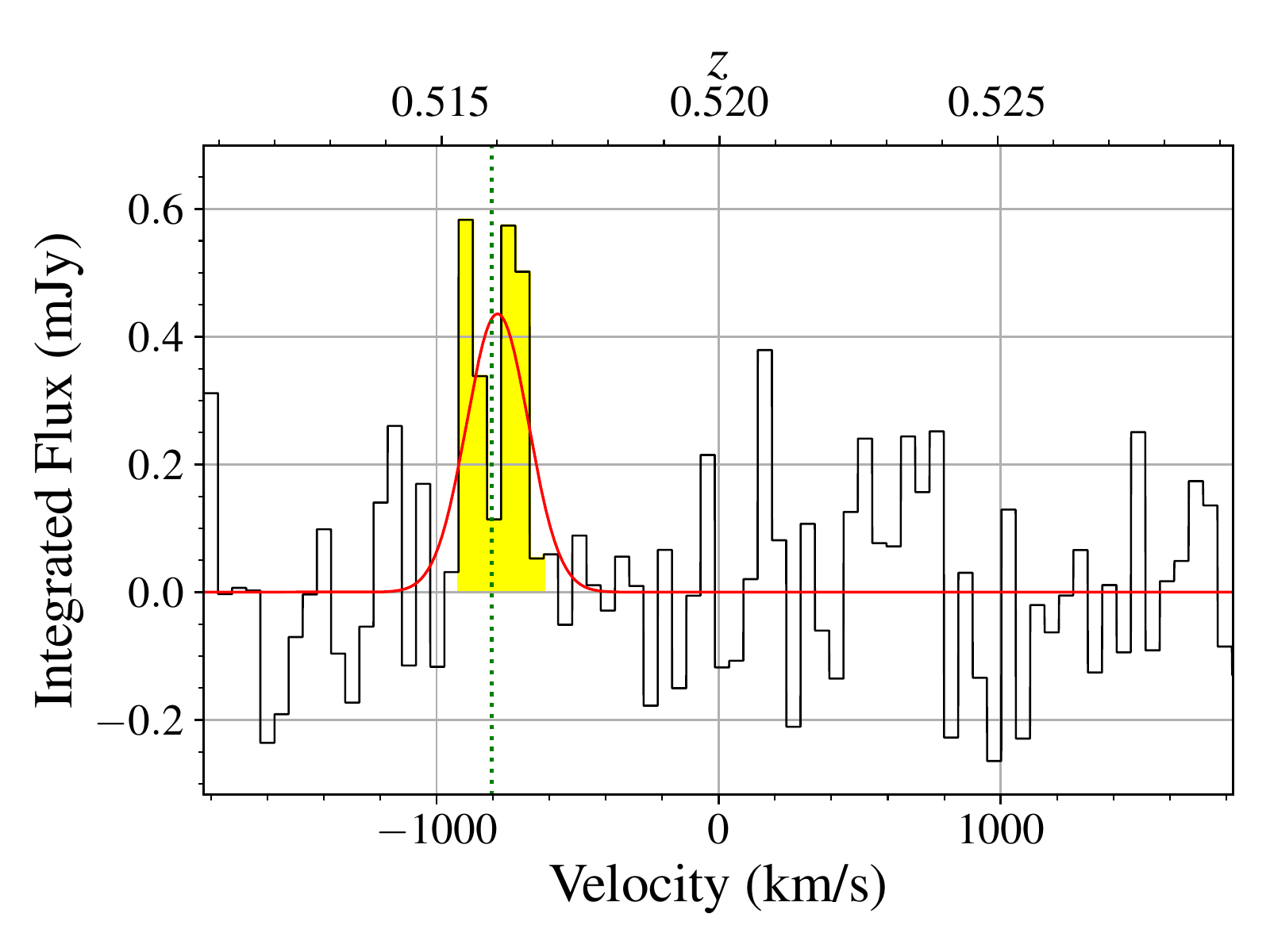}

	\includegraphics[scale=0.3]{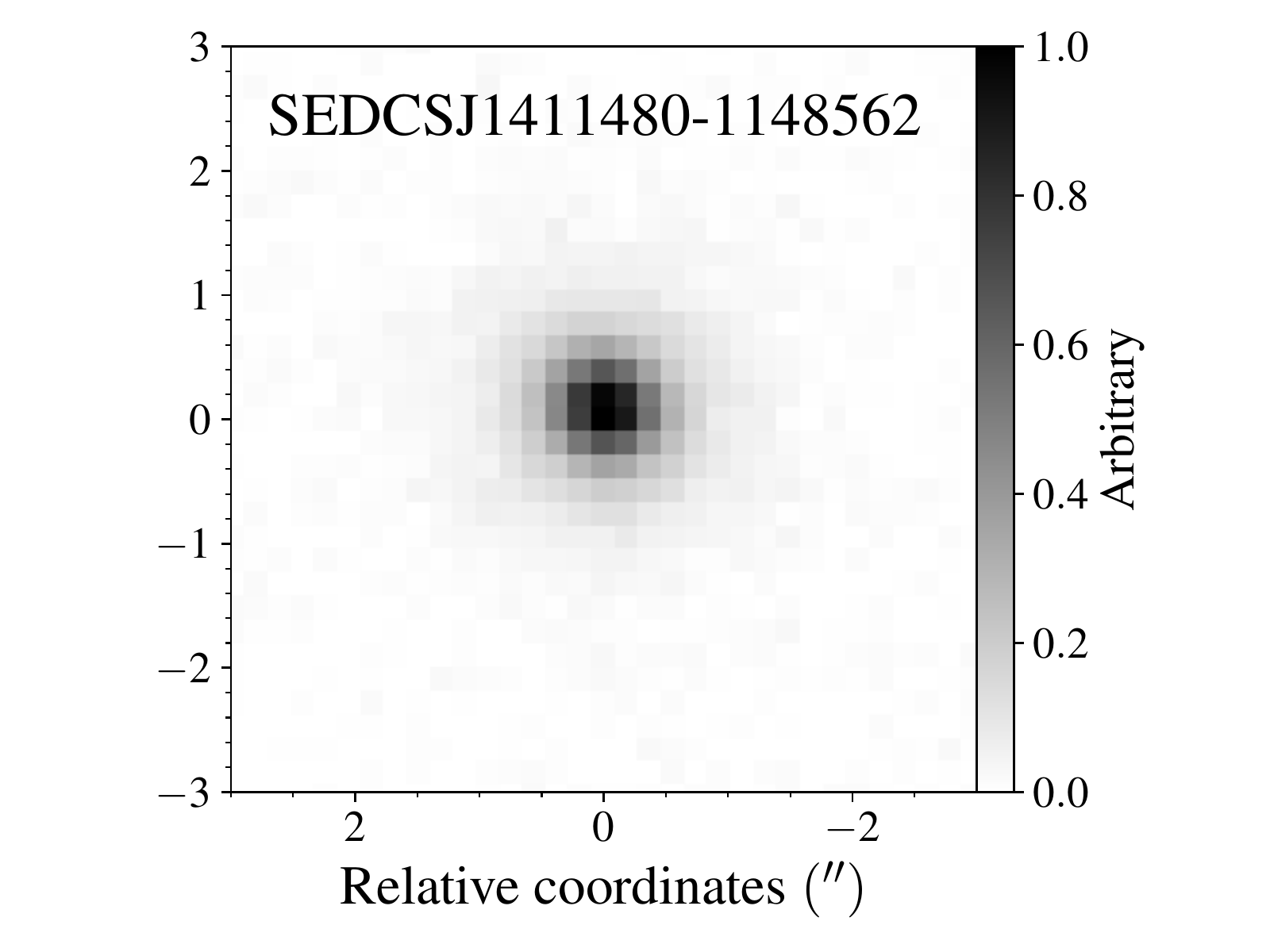}
	\includegraphics[scale=0.3]{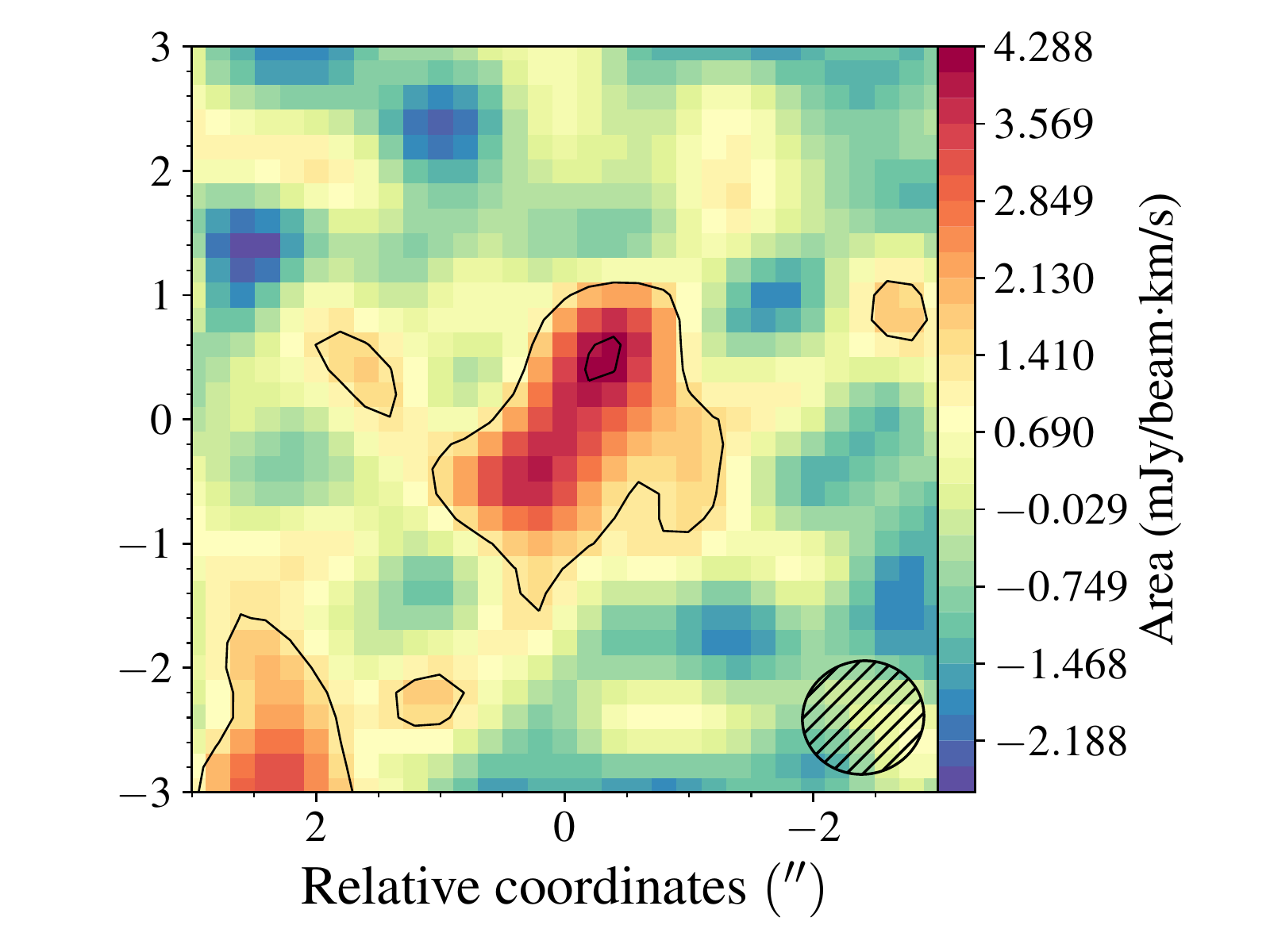}
	\includegraphics[scale=0.3]{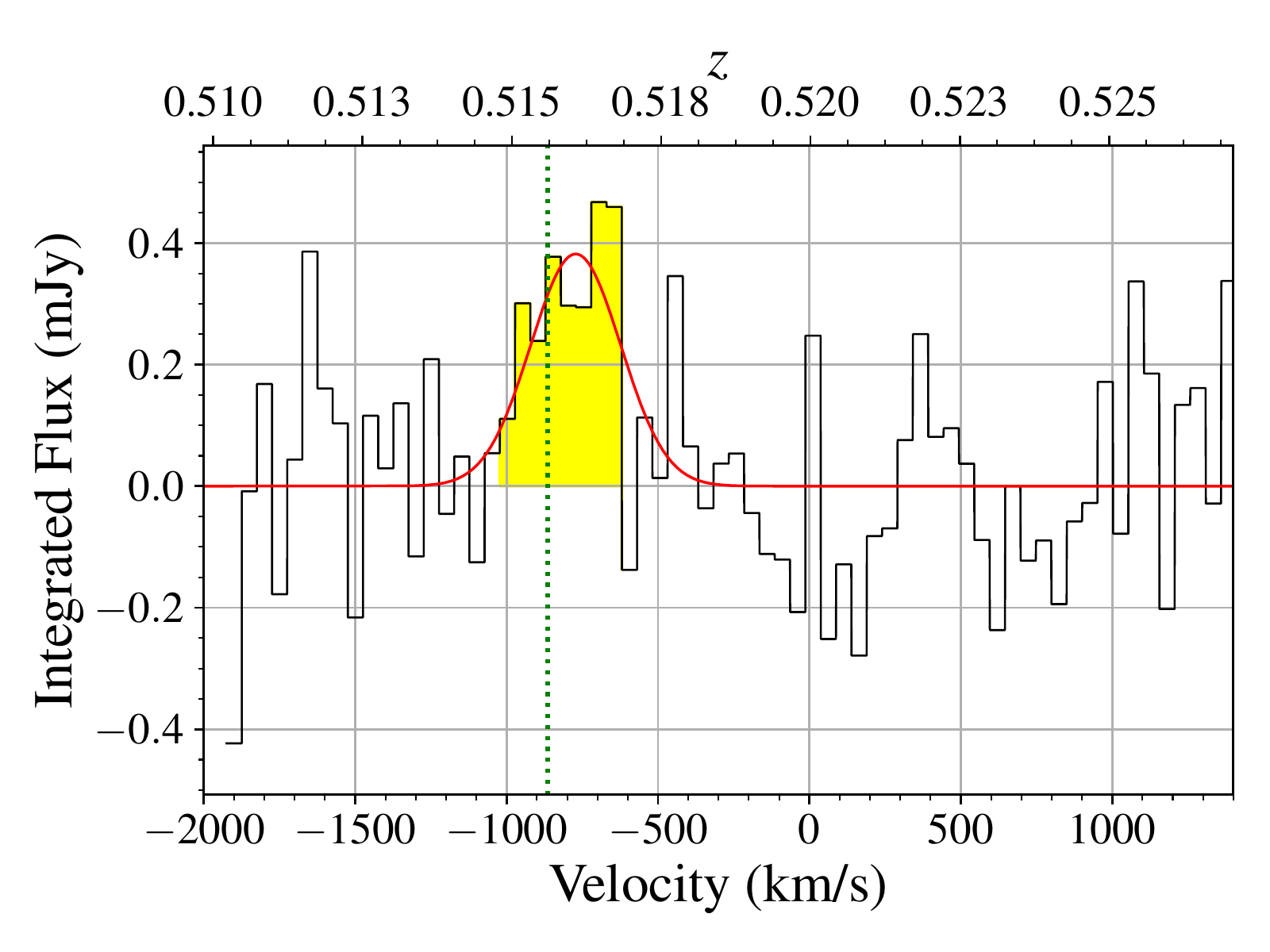}	

	\includegraphics[scale=0.3]{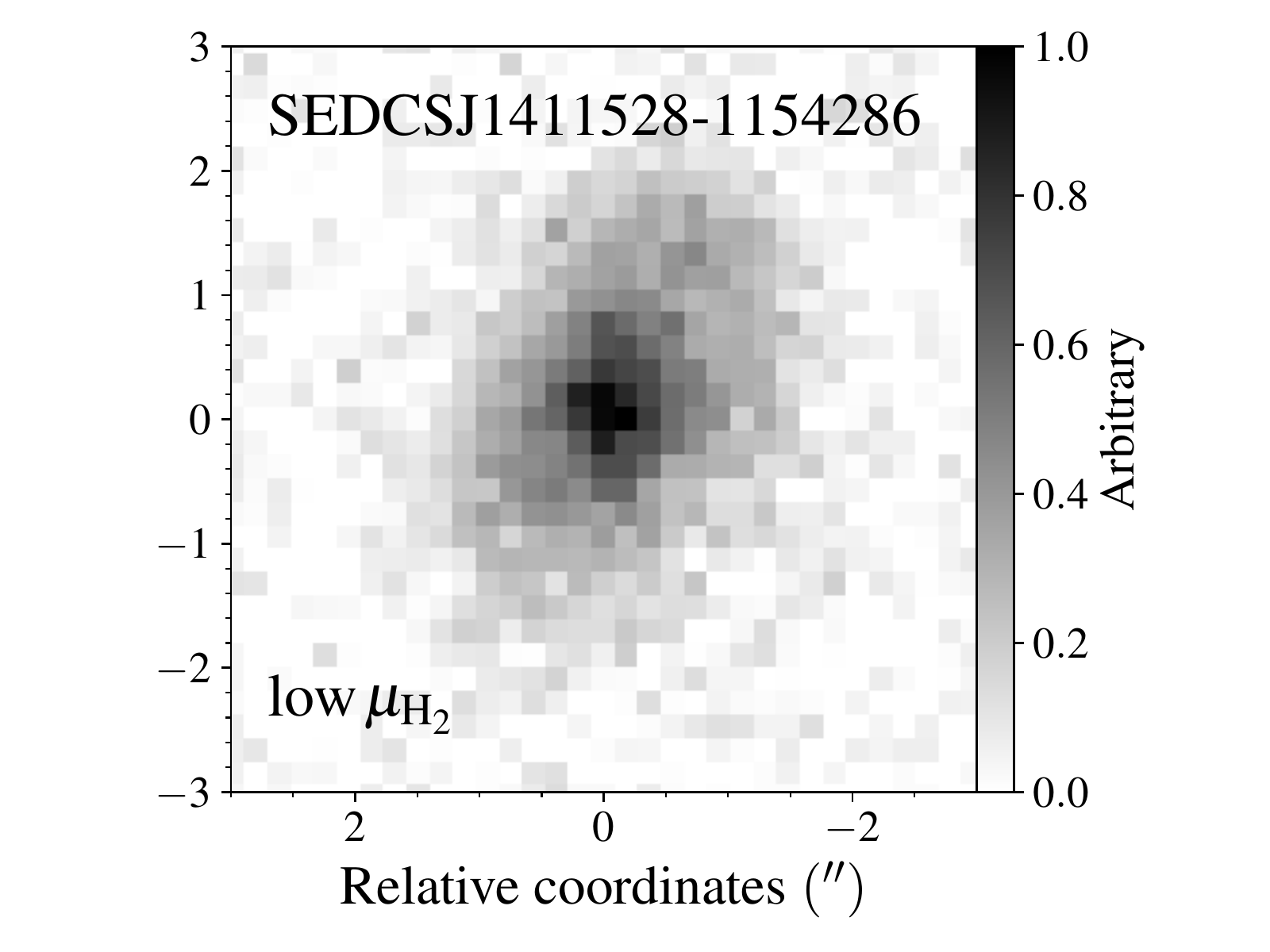}
	\includegraphics[scale=0.3]{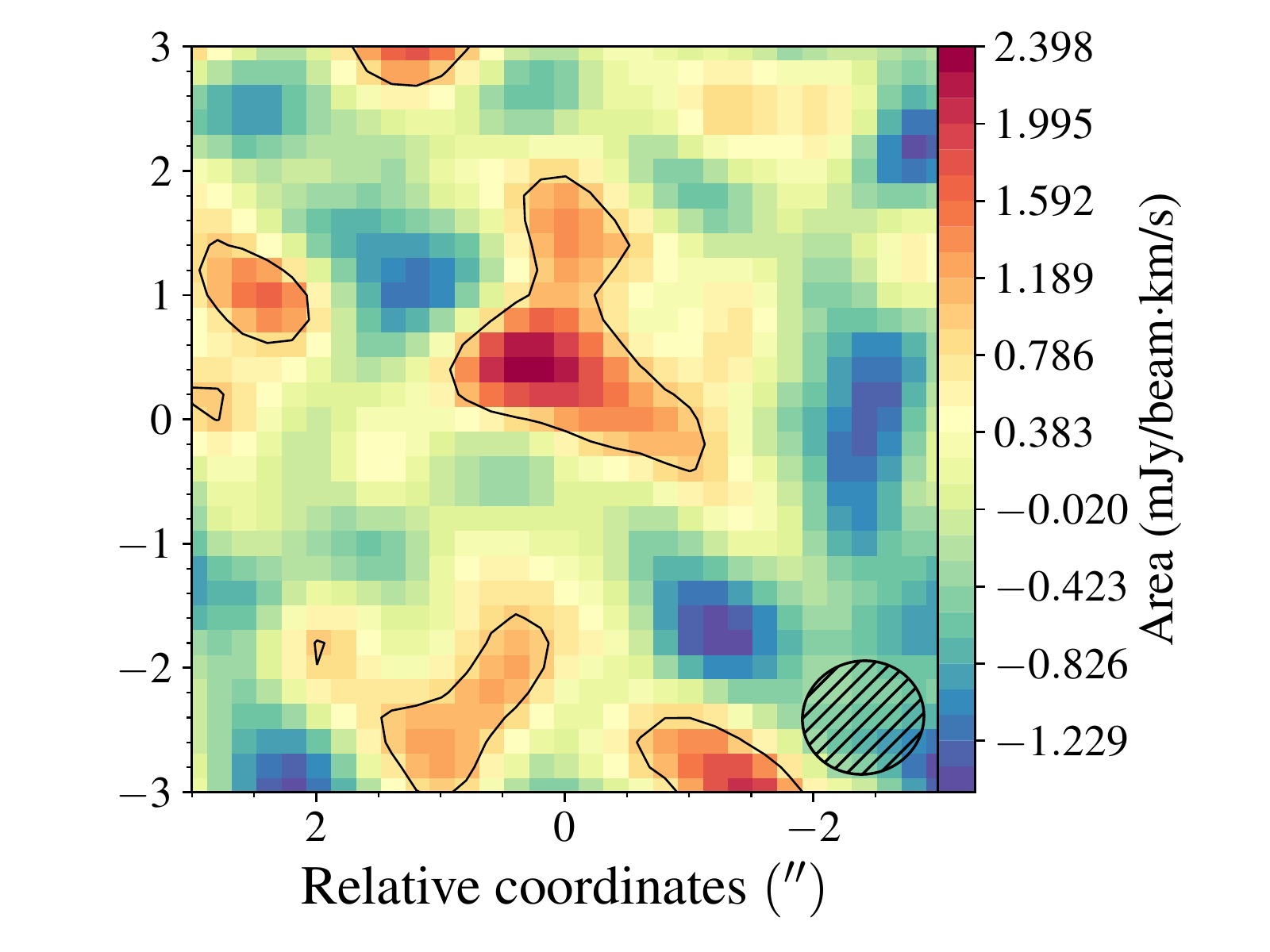}
	\includegraphics[scale=0.3]{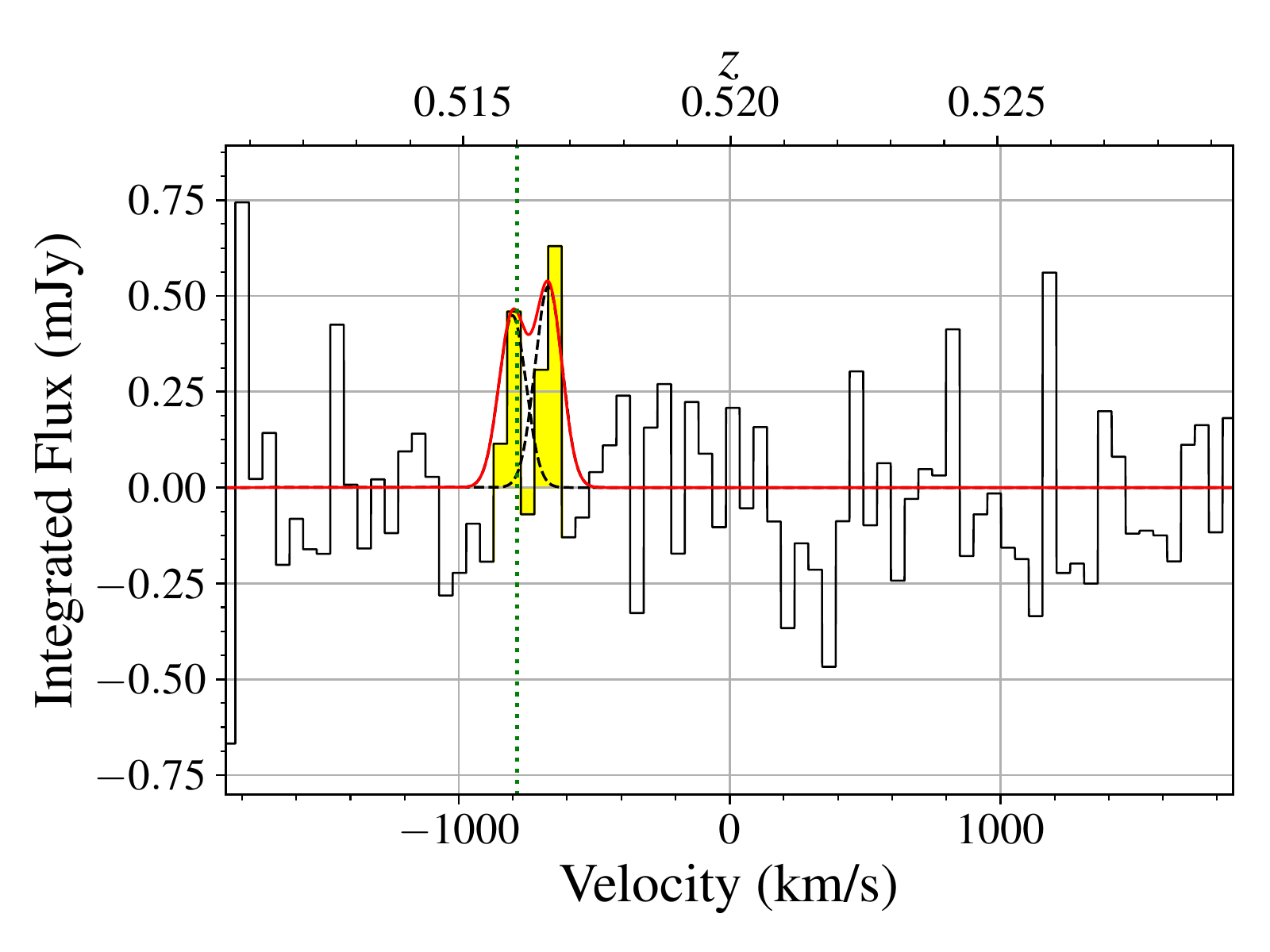}

    \caption{Continued.}
\end{figure*}

\end{appendix}

\end{document}